\documentclass[12paper]{elsart}
 
\usepackage[numbers]{natbib}
\usepackage{amsmath,amssymb}
\usepackage{algpseudocode,algorithm}
\usepackage{bm}
\usepackage{booktabs}
\usepackage{color}
\usepackage{graphicx,natbib}
\usepackage{threeparttable}
\usepackage{makecell}
\usepackage{hyperref}
\def\equationautorefname~#1\null{Eq. (#1)\null}
\def\figureautorefname~#1\null{Fig. #1\null}
\def\algorithmautorefname~#1\null{Algorithm #1\null}
\def\sectionautorefname~#1\null{Section #1\null}
\def\theoremautorefname~#1\null{Theorem #1\null}

\usepackage{subfigure,verbatim}

\renewcommand{\vec}[1]{\boldsymbol{#1}}  

\usepackage{tikz}
\newcommand{\tikzcircle}[2][red,fill=red]{\tikz[baseline=-0.5ex]\draw[#1,radius=#2] (0,0) circle ;}

\newcommand{\bmx}{\bm{x}}
\newcommand{\bmy}{\bm{y}}

\newcommand{\bxl}{\bar{\bm{x}}_{\bm{l}}}
\newcommand{\bym}{\bar{\bm{y}}_{\bm{m}}}
\newcommand{\by}{\bm{y}}

\usepackage{ntheorem}
\newtheorem{theorem}{Theorem}
\newtheorem*{theorem-non}{Theorem}
\newenvironment{proof}
{\textit{Proof:} }
{$\square$}

\begin{document}
\begin{frontmatter}



  \title{Fast algorithms for evaluating the stress field of dislocation lines in anisotropic elastic media} 
  
  \author{C. Chen$^1$, S. Aubry$^2$, T. Oppelstrup$^2$,}
  \author{A. Arsenlis$^2$ and E. Darve$^1$}
  
  \address{$^1$Stanford University, Stanford, CA}
  \address{$^2$ Lawrence Livermore National
    Laboratory, Livermore, CA.}  
  
  \date{\today}

\begin{abstract}
  In dislocation dynamics (DD) simulations, the most computationally
  intensive step is the evaluation of the elastic interaction forces
  among dislocation ensembles. Because the pair-wise interaction
  between dislocations is long-range, this force calculation step can
  be significantly accelerated by the fast multipole method (FMM). We
  implemented and compared four different methods in isotropic and
  anisotropic elastic media: one based on the Taylor series expansion
  (Taylor FMM), one based on the spherical harmonics expansion
  (Spherical FMM), one kernel-independent method based on the
  Chebyshev interpolation (Chebyshev FMM), and a new
  kernel-independent method that we call the Lagrange FMM. The Taylor
  FMM is an existing method, used in ParaDiS, one of the most popular
  DD simulation softwares. The Spherical FMM employs a more compact
  multipole representation than the Taylor FMM does and is thus more
  efficient. However, both the Taylor FMM and the Spherical FMM are
  difficult to derive in anisotropic elastic media because the
  interaction force is complex and has no closed analytical
  formula. The Chebyshev FMM requires only being able to evaluate the
  interaction between dislocations and thus can be applied easily in
  anisotropic elastic media. But it has a relatively large memory
  footprint, which limits its usage. The Lagrange FMM was designed to
  be a memory-efficient black-box method. Various numerical
  experiments are presented to demonstrate the convergence and the
  scalability of the four methods.
\end{abstract}


\end{frontmatter}



\section{Introduction}
Dislocations are line defects in crystalline materials,
and their presence strongly influences many material properties,
including the plastic deformability of crystals \cite{hirth1982theory}. The
dislocation dynamics (DD) method performs direct numerical simulations
of dislocation ensembles based on the fundamental physics of defect
motion, evolution, and interaction. It has been successfully used to
study the origins of strength and strain hardening for cubic crystals
\cite{barton2011multiscale}, the strength of micro-pillars~\cite{greer2008comparing},
and irradiated materials \cite{arsenlis2012dislocation}. However, the
computational cost for evaluating pair-wise elastic interactions of
dislocation ensembles in a naive way is prohibitive in large-scale DD simulations
\cite{kutka1998observations,schwarz1999simulation,weygand2002aspects}. 
In DD simulations, dislocation lines are
discretized and represented as straight segments, and evaluating the
pair-wise interaction among all segments directly requires $O(N_s^2)$
operations, where $N_s$ is the number of dislocation segments, which
can reach several millions in large-scale DD simulations.


Existing fast algorithms, which reduce the cost of computing pair-wise
dislocation interactions, fall into two categories. The first category,
cut-off methods, does not account for the interactions beyond a
predefined cut-off distance
\cite{kubin1992dislocation,devincre1996meso,devincre1997mesoscopic}.
However, employing a cut-off distance can lead to numerical artifacts
\cite{gulluoglu1989dislocation,wang1995n} due to the
slow decay of long-range dislocation interaction. The second category,
the fast multipole method (FMM) and its variants,
represents the elastic interactions of remote dislocations with
multipole expansions
\cite{zbib1998plastic,wang2004multipole,wang2006parallel,zhao2010new,lesar2002multipole,arsenlis2007enabling,yin2012computing}. As a
result, the calculation of elastic interaction forces requires only $O(N_s)$
operations. However, some of these methods have large pre-factors that
are not efficient for simulating practical dislocation systems.
Most existing FMMs are applicable only in isotropic
elastic media, and their extensions to be used in anisotropic elasticity are
 difficult to derive because the interaction elastic force does not have an analytic closed form 
 in anisotropic elasticity as it does in isotropic elasticity.

The goal of this study is to benchmark and compare different versions
of FMMs in DD simulations, and to present a new kernel-independent FMM that can be
applied efficiently in both isotropic and anisotropic elastic media. The
FMM was originally devised to evaluate long-range interactions among
large ensembles of particles through a kernel function such as the
Laplacian and the Stokes kernels~\cite{greengard1987fast}. To illustrate
an application of the FMM in DD simulations, consider a network of dislocation lines, of
which the elastic forces are to be evaluated at the ends of all dislocation segments.
The FMM first recursively divides the simulation
domain into small cells in a way that every cell at the
bottom of the hierarchy has a constant number of dislocation segments. Then,
an upward sweep populates cells at all hierarchical levels with the
so-called multipole expansions, which encode the information of
dislocation segments inside all cells. With multipole expansions, far-field
stress evaluations can be computed efficiently at every level in the
hierarchy, and the results are stored in the so-called local
expansions in all cells. After these multipole-to-local (M2L) translations,
a downward sweep, symmetric to the upward sweep, is executed to
translate local expansions hierarchically down to the smallest cells.
Last, for every cell at the bottom of the hierarchy, the far-field stress
encoded by its local expansion and the near-field stress induced by dislocation 
segments in neighboring cells sum up to be the stress field of this cell. 
This stress field is used to compute the force on the ends of every 
dislocation segment through Gaussian quadrature.

The first FMM used in DD simulations is based on the Taylor series
expansion of the Green's function in isotropic elastic media
\cite{lesar2002multipole,arsenlis2007enabling}. In isotropic elastic media,
the Green's function has a simple explicit expression, so the Taylor series expansion
is easy to derive. We refer to the method in \cite{arsenlis2007enabling} as
 the Taylor FMM. 
Unlike the Taylor FMM, the original FMM introduced by Greengard and Rokhlin \cite{greengard1987fast}
for particle simulations is based on spherical harmonics expansion. The spherical harmonics expansion 
 is not as general as the Taylor series expansion, but it leads to a more compact representation 
 of the Green's function in isotropic elastic media. We have implemented a variant of the original FMM 
 and call it the Spherical FMM. In the original FMM, the M2L translations are the most time-consuming steps.
While the original FMM takes $O(p^4)$ work for M2L translations, where $p$ is the order of the spherical harmonics expansion, our Spherical FMM needs only $O(p^3)$ work with a `point-and-shoot' translation scheme.
The authors of \cite{zhao2010new} implemented the optimized FMM
\cite{greengard1997new} for isotropic elastic media, and the optimization was to 
achieve faster M2L translations with $O(p^2)$ work via plane wave expansions and to
reduce the number of M2L translations by a `merge-and-translate' technique.

In anisotropic elastic media, however, methods based on either the Taylor series expansion 
or the spherical harmonics expansion are significantly more difficult to derive. 
The reason is that these methods 
rely on being able to differentiate the Green's function, 
but the Green's function does not have an analytic closed form in anisotropic elastic media, 
as it does in isotropic elastic media. In order to apply the FMM in anisotropic elastic media, 
the authors of \cite{yin2012computing} proposed approximating the derivatives of the Green's function 
in anisotropic elasticity with truncated spherical harmonics series. 
However, the calculation is quite cumbersome, 
and the resulting method is not efficient for large-scale simulations.

To overcome the limitations of kernel-specific FMMs, such as the Taylor FMM 
and the Spherical FMM, several kernel-independent FMMs
\cite{gimbutas2003generalized, ying2004kernel, martinsson2007accelerated, fong2009black} 
have been developed, the formulation of which does not depend on the kernel function 
(the Green's functions in isotropic and anisotropic elastic media). These kernel-independent
methods require only being able to numerically evaluate the kernel function; 
thus, they can be applied with the Green's function in
both isotropic and anisotropic elastic media. The differences among these
kernel-independent methods are detailed in \autoref{sec:FMM}. In this work, we have implemented
the kernel-independent method in \cite{fong2009black} for DD simulations, 
which we call the Chebyshev FMM.
The Chebyshev FMM uses Chebyshev interpolants of the kernel function to approximate far-field interactions between dislocations, and the Chebyshev approximation is known to be nearly optimal \cite{trefethen2013approximation}.
In the Chebyshev FMM, M2L operators are pre-computed and the corresponding storage increases empirically as $O(p^5)$,
where $p$ is the degree of the Chebyshev interpolant. 
This memory cost is further exacerbated by the fact that the kernel function in DD is a tensor with 54 components, and it reached $1.7$ GB 
when $p=7$.
This intensive memory consumption limits the use of the Chebyshev FMM in large-scale DD simulations.

To reduce the memory cost of the Chebyshev FMM, we introduce
a new kernel-independent FMM based on the Lagrange interpolation on equally spaced grids, 
which we call the Lagrange FMM. The Lagrange FMM has three
advantages when it is applied in DD simulations. First, like the Chebyshev FMM, it is a kernel-independent method 
and thus can be applied in both isotropic and anisotropic elastic
media. Second, the M2L operators in the Lagrange FMM are block-Toeplitz
matrices, and the memory cost of storing them increases as $O(p^3)$, where $p$ is the degree of the Lagrange interpolant. 
This cost is much smaller than that of the Chebyshev FMM, e.g., when $p=7$, the memory 
usage of the Lagrange FMM is only about $300$ MB. Third, the Lagrange FMM 
takes advantage of the structured M2L operators to achieve fast M2L translations using the fast 
Fourier transform (FFT).

To summarize, the main contributions of this paper are
\begin{enumerate}
\item Introduction of a new kernel-independent method, namely the Lagrange FMM, for use in both isotropic and anisotropic elastic media. This algorithm has a reduced computational cost and memory footprint compared to the Chebyshev FMM.
\item Benchmarks and comparisons of four different FMMs, namely the Taylor FMM, the Spherical FMM, the Chebyshev FMM and the Lagrange FMM, in isotropic elastic media.
\item Benchmarks and comparisons of two kernel-independent FMMs, namely the Chebyshev FMM and the Lagrange FMM, in anisotropic elastic media.
\item Benchmarks on sequential and parallel computers (up to 4,096 cores).
\end{enumerate}

The remainder of this paper details our new method and comparison of the four methods,
and it is organized as follows. In \autoref{sec:DD}, the stress field of
dislocation ensembles is formulated in terms of the Green's function in isotropic and anisotropic elastic media.
In \autoref{sec:FMM}, we review the three existing methods: the Taylor FMM, the Spherical
FMM and the Chebyshev FMM. In \autoref{sec:new_fmm}, the
Lagrange FMM algorithm is introduced, and the computational costs and memory footprint of the four FMMs are analyzed and compared.
\autoref{sec:results} presents numerical results of our benchmark and comparisons.



\section{Dislocation dynamics}\label{sec:DD}

The motion of dislocations is driven by the Peach-Koehler force:
\[
{\bm f} = {\bm \sigma} \cdot {\bm b} \times {\bm \xi}
\]
where ${\bm \sigma}$ is the stress tensor, $\bm b$ is the dislocation Burgers vector and $\bm \xi$ is the local unit tangent vector to the dislocation line. More details can be found in Hirth and Lothe's book \cite{hirth1982theory} and the references therein.

Here we formulate the stress field of dislocation ensembles using the Green's function in isotropic and anisotropic elastic media. In a linear elastic medium, the stress field $\bm \sigma$ at $\bm x$ induced by a closed dislocation loop ${\cal C}$ can be defined with Mura's formula~\cite{mura1971green} as
\begin{equation}
\sigma_{\alpha\beta}(\bm{x}) = b_w \oint_{{\cal C}} \epsilon_{ngr} C_{\alpha\beta vg} C_{pqwn} \frac{\partial G_{vp}}{\partial x_q}( \bm{x}-\bm{y} ) dy_r
\label{eq:stress}
\end{equation}
where $\bm b$ is the Burgers vector, $\bm{\epsilon}$ is the permutation tensor, $\bm C$ is the elastic stiffness tensor, and $G_{vp}(\bm{x}-\bm{y})$ is the Green's function defined as the displacement in the $x_v$-direction at $\bm{x}$ in response to a unit point force applied in the $y_p$-direction at $\bm{y}$.

In isotropic elasticity, the Green's function $\bm G$ has a simple expression \cite{cai2006non}
\begin{equation}
G_{vp}( \bm{x}-\bm{y} ) = {1\over 8 \pi \mu} \bigg[ \delta_{vp} {\partial^2 R \over \partial x_i \partial x_i}
- {1\over 2(1-\nu)} {\partial^2 R \over \partial x_v \partial x_p} \bigg] 
\label{eq:k_iso}
\end{equation}
where $\mu$ is the shear modulus, $\nu$ is the Poisson's ratio, $\bm{\delta}$ is the identity matrix in three dimensions, $R = | \bm{x}-\bm{y} |$, and repeated indices imply summation. Note that \autoref{eq:k_iso} involves only the derivatives of $R$, and expanding these derivatives in Taylor series (spherical harmonics series) leads to the Taylor FMM (the Spherical FMM).

In anisotropic elasticity, the Green's function $\bm G$ does not have an analytic closed form, which makes it difficult to derive its Taylor series expansion or its spherical harmonics series expansion. Although an analytic closed form is not available, the Green's function can be evaluated numerically \cite{aubry2013use,aubry2013methods,yin2010efficient}. Here, we use the scheme in \cite{aubry2013use}, which expands the derivatives of $\bm G$ with the spherical harmonics series
\begin{equation}
  \frac{\partial G_{vp}}{\partial x_q}( \bm{x}-\bm{y} ) = \frac{1}{R^2}
  \sum\limits_{d=0}^{\infty} 
  \sum\limits_{m=0}^{2d+1}  
  \Re \left( S^{dm}_{vpq} (\bm{T} \cdot \bm{e_{12}})^m (\bm{T} \cdot \bm{e_3})^{2d+1-m}
  \right) 
  \label{eq:k_aniso}
\end{equation}
where $\bm{T} = (\bm{x}-\bm{y})/R$, $(\bm{e}_1, \bm{e}_2, \bm{e}_3)$ are the canonical bases, $\bm{e}_{12} = \bm{e}_1 + i \, \bm{e}_2$, $\Re()$ is the real part of a complex number, and $\bm{S}$ is a pre-computed coefficient tensor defined in~\cite{aubry2013use}. In practice, only the first $q_{\max}$ terms are kept for a prescribed accuracy. The choice of $q_{\max}$ is studied in detail in \cite{aubry2013use}. 

To evaluate \autoref{eq:stress} for a dislocation loop $\cal C$ with an arbitrary shape, the dislocation loop ${\cal C}$ is first discretized as a polygon, i.e., a set of nodes connected by straight segments. Assume the dislocation loop ${\cal C}$ is approximated by a polygonal loop with $m$ straight segments ${{\cal S}^i}, i=1,2,\ldots,m$. The contour integral in \autoref{eq:stress} can be approximated by the sum of $m$ line integrals along ${{\cal S}^i}$, and line integrals are evaluated numerically with Gaussian quadrature. We also define the kernel function in DD as
\begin{equation}
K_{\alpha\beta wr}(\bm{x} - \bm{y}) \overset{\text{def}}{=} \epsilon_{ngr} C_{\alpha\beta vg} C_{pqwn} \frac{\partial G_{vp}}{\partial x_q}( \bm{x}-\bm{y} )
\label{eq:kernel}
\end{equation}
which is a tensor of size $6 \cdot 3 \cdot 3 = 54$ ($\bm K$ is symmetric in terms of $\alpha$ and $\beta$, i.e., $K_{\alpha\beta wr} = K_{\beta\alpha wr}$).

With the definition of the kernel function, \autoref{eq:stress} becomes 
\begin{align}
\sigma_{\alpha\beta}(\bm{x}) 
&= b_w \oint_{{\cal C}} K_{\alpha\beta wr}(\bm{x} - \bm{y}) dy_r 
&& \text{[plug in kernel function definition]} \nonumber \\
&\approx b_w \sum_{i=1}^{m} \int_{{\cal S}^i} K_{\alpha\beta wr}(\bm{x} - \bm{y}^{i}) dy_r
&& \text{[discretize the dislocation loop]} \nonumber \\
&\approx {1\over 2} b_w \sum_{i=1}^{m} \sum_{j=1}^n \omega^j K_{\alpha\beta wr}(\bm{x} - \bm{y}^{ij}) {{\cal S}^i_r}  
&& \text{[use $n$-point Gaussian quadrature]} \nonumber \\
& = \sum_{i=1}^{m} \sum_{j=1}^n K_{\alpha\beta wr}(\bm{x} - \bm{y}^{ij})  q^{ij}_{wr}
&& [q^{ij}_{wr} \overset{\text{def}}{=} {1\over 2} b_w \omega^j {{\cal S}^i_r}]
\label{eq:stress_discrete}
\end{align}
where $w^j$ and $\bm{y}^{ij}$ $(j=1,2,\dots, n)$ are quadrature weights and quadrature points on dislocation segment ${\cal S}^i$, ${{\cal S}^i_r}$ is the length of segment ${\cal S}^i$ projected in the $x_r$-direction, and $q^{ij}_{wr}$ is defined to be ${1\over 2} b_w \omega^j {{\cal S}^i_r}$.

In linear elasticity, the stress field induced by multiple dislocation loops ${\cal C}_l, l=1,2,\ldots,r$ can be obtained with the principle of superposition. In other words, the definition of the stress field, i.e., \autoref{eq:stress}, can be extended to the summation of multiple contour integrals with respect to all dislocation loops. Correspondingly, the stress field discretization, i.e., \autoref{eq:stress_discrete}, is also extended to the summation over all dislocation loops as
\begin{align*}
\sigma_{\alpha\beta}(\bm{x}) 
&\approx \sum_{l=1}^r \sum_{i=1}^{m} \sum_{j=1}^n K_{\alpha\beta wr}(\bm{x} - \bm{y}^{ijl})  q^{ijl}_{wr}
&& \text{[sum over all dislocation loops]} \nonumber \\
&= \sum_{k=1}^{N} K_{\alpha\beta wr}(\bm{x} - \bm{y}^{k})  q^{k}_{wr}
&& \text{[combine three summations]}
\end{align*}
where $N=m n r$ is the total number of Gaussian quadrature points over all dislocation loops. In a DD simulation, the stress field needs to be evaluated at multiple positions along all dislocation lines, e.g., all Gaussian quadrature points used in the discretization. As a result, a direct evaluation of \autoref{eq:stress3} costs $O(N^2)$ operations, which is prohibitive for large-scale simulations. Note that the total number of dislocation segments is $N_s =  m r$, so given a fixed quadrature rule ($n$ fixed), the cost is also $O(N_s^2)$.

In summary, evaluating the stress field of dislocation ensembles at $N$ points $\{ {\bm x}_i \}_{i=1}^N$ is equivalent to evaluating the sum 
\begin{equation} \label{eq:stress3}
{\bm s}({\bm x}_i) = \sum_{j=1}^N {\bm K}({\bm x}_i - \bm{y}_j)  {\bm q}_j \quad \text{for} \quad i=1,2,\ldots, N 
\end{equation}
where  $\{ {\bm y}_i \}_{i=1}^N$ and $\{ {\bm q}_i \}_{i=1}^N$ are $N$ points and the associated strength, respectively, $\bm K$ is the kernel function in DD, and ${\bm s}({\bm x}_i)$ is the computed stress field.

\section{Existing fast multipole methods\label{sec:FMM}}

In this section, we review existing FMMs, with a particular focus on the three methods we have implemented, namely the Taylor FMM, the Spherical FMM, and the Chebyshev FMM. The FMM was originally introduced by Greengard and Rokhlin to speed up the evaluation of pair-wise long-range interactions for particles ensembles. It has since been successfully used for many different types of kernels including the Laplace kernel \cite{greengard1987fast, greengard1997new}, the Helmholtz kernel \cite{rokhlin1990rapid, rokhlin1992diagonal, darve2000fast, darve2000siam, engquist2007fast}, and the Stokes kernel \cite{fu2000fast}, among others. 

The FMM follows five steps. First, the dislocation ensemble region is recursively partitioned until every leaf cell at the bottom of the hierarchy has a small number of segments. Second, an upward sweep populates cells at all levels with the so-called multipole expansion, which encodes the information of dislocation segments inside the cells. Third, the multipole expansions are used to compute far-field interactions of every cell, and the results are stored in the so-called local expansions in every cell. This M2L translation step is the most computationally intensive step in the FMM.
Fourth, a downward sweep, symmetric to the upward sweep, is executed to gather local expansions onto leaf cells. Last, the stress fields in all cells are the sum of far-field interactions and the near-field interactions, which are calculated exactly using \autoref{eq:stress3} with dislocation segments in neighboring cells. A technical analysis on how errors propagate through all translations can be found in \cite{darve2000siam}.

The first FMM used in DD is based on the Taylor series expansion \cite{lesar2002multipole, arsenlis2007enabling}. Since the Green's function in isotropic elasticity, i.e., \autoref{eq:k_iso}, has a simple explicit expression that involves only the derivatives of $R$, the Taylor FMM uses the Taylor series expansions of those derivatives to approximate far-field dislocation interactions. The implementation details of the Taylor FMM can be found in the appendix of \cite{arsenlis2007enabling}. In the Taylor FMM, using the $p$-th order Taylor series expansion leads to a number of $O(p^3)$ multipole and local coefficients. As a result, an M2L translation (applying an $O(p^3)$ by $O(p^3)$ matrix to a vector) takes as much as $O(p^6)$ work. The cost of the Taylor FMM is further exacerbated by relatively slow convergence of approximating the Green's function with the Taylor series expansion at the center of an FMM cell. The calculation error of the Taylor FMM decreases as $O((\sqrt{3}/3)^{p})$, where $\sqrt{3}/3$ is the ratio between the radius of the circumscribed sphere an FMM cell and the distance from the center of the cell to the nearest cell in its far-field.

Greengard and Rokhlin's original FMM~\cite{greengard1987fast} is based on using the spherical harmonics 
expansion of the kernel $1/R$. To use the same machinery, we derived the spherical 
harmonics expansion of the Green's function in isotropic elasticity from that of $1/R$. The observation is that
the Green's function in \autoref{eq:k_iso} involves only the derivatives of $R$, and the spherical harmonics
expansion of $R$ can be derived from that of $1/R$ using the relationship $R=|\bmx - \bmy|=({|\bmx|^{2}}+{|\bmy|^{2}}-2{\bmx \cdot \bmy})/{R}$. Therefore, we were able to implement the Spherical FMM with five different fields of ``charges:'' $1$, $|\bmy|^2$, $y_1$, $y_2$ and $y_3$, following the original FMM's machinery. The convergence rate of the Spherical FMM is the same as that of the Taylor FMM~\cite{greengard1987fast}. 

Although the spherical harmonics expansion is not as general as the Taylor series expansion, it leads to a more compact representation of the Green's function in isotropic elasticity with only $O(p^2)$ multipole and local coefficients. This leads to the $O(p^4)$ cost of M2L translation in the original FMM. To further reduce the M2L translation cost, our Spherical FMM employs the `point-and-shoot' scheme, which takes only $O(p^3)$ work, as follows. Consider the translation between two boxes $A$ and $B$. First, the coordinate system can be rotated in a way that the $z$-axis is aligned with the direction from the center of box $A$ to the center of box $B$. Second, the subsequent M2L translation operators apply only along the $z$-axis, which are effectively sparse in the original FMM. Finally, the coordinate system is rotated back to the original. The details can be found in \cite{greengard1997new}.
 
Both the Taylor FMM and the Spherical FMM rely on using the derivatives of the Green's function in isotropic elasticity, which were obtained from the simple explicit expression in \autoref{eq:k_iso}. However, the Green's function does not have an analytic closed form in anisotropic elasticity; the Taylor FMM and the Spherical FMM are thus difficult to derive. In contrast, kernel-independent FMMs, which require only numerical evaluations of the kernel function, can be applied directly in anisotropic elastic media using \autoref{eq:k_aniso} to evaluate the corresponding Green's function.



Several ways to construct a kernel-independent FMM are as follows.
The first one is to approximate the kernel function with polynomials. This approach leads to algorithms that can be applied to a variety of non-oscillatory kernels that are sufficiently smooth away from their singularities. The singular value decomposition (SVD) is typically used to compress the resulting representation for achieving the optimal running time, but the pre-computation with the SVD can be costly. Examples of such kernel-independent FMMs are methods in \cite{gimbutas2003generalized, dutt1996fast, fong2009black}. The method in \cite{gimbutas2003generalized} uses the Legendre polynomials and uses the SVD to compute the optimal representations adaptively \cite{yarvin1998generalized, yarvin1999improved}. The method in \cite{dutt1996fast} uses the Chebyshev polynomials, and it was developed for the purpose of fast interpolation, integration and differentiation. The method in \cite{fong2009black} also uses the Chebyshev polynomials, and its advantage is that the pre-computation is independent of particle positions.



The second way to construct a kernel-independent FMM is to replace analytical expansions of the kernel function by equivalent densities \cite{anderson1992implementation, makino1999yet}. One such method in \cite{ying2004kernel} computes equivalent densities numerically through solving local Dirichlet-type boundary value problems. Since the solution of an elliptic partial differential equation in a sufficiently well-behaved region is characterized by its values on the boundary, the method in \cite{ying2004kernel} works for various kernels associated with fundamental solutions of elliptical PDEs, such as the Laplacian, the Stokes, and the Navier operators. Like other kernel-independent methods, the method in \cite{ying2004kernel} also uses the SVD to speed up M2L translations (the FFT is used when the equivalent densities are put on regular girds).

The third way to construct a kernel-independent FMM utilizes numerical techniques including the interpolative decomposition (ID) \cite{cheng2005compression}, and Cauchy's integral formula. For instance, the method in \cite{martinsson2007accelerated} uses ID to choose a subset of particles in every cell as the skeletons and passes information through the FMM tree via these skeletons. The ID serves the same purpose as the SVD in previous methods, and the advantage of using ID is that interactions between neighboring cells can also be compressed.
Another example of kernel-independent FMM is the method in \cite{letourneau2014cauchy}, which is based on the Cauchy's integral formula and the Laplace transform. The method employs diagonal M2L operators by design and thus has a relatively small computational cost, especially when high accuracy is needed.

Among the existing kernel-independent methods, we have implemented the method in \cite{fong2009black}, namely the Chebyshev FMM, for DD simulations. The Chebyshev FMM uses Chebyshev polynomials to approximate far-field interactions between dislocations, which is nearly optimal in minimizing the maximum error of polynomial approximation \cite{trefethen2013approximation}. The calculation error of Chebyshev approximation depends on the size of the corresponding Bernstein ellipse \cite{trefethen2013approximation}, and it is empirically $O((1/3)^p)$ for DD simulations in isotropic elastic media, where $p$ is the degree of Chebyshev polynomial. In addition, the Chebyshev FMM uses the SVD to compress M2L operators and thus uses the minimal number of multipole and local coefficients. In the Chebyshev FMM, suppose $\alpha_p$ is the compression ratio from using the SVD. The number of multipole and local coefficients is $O(\alpha_p \, p^3)$, and an M2L translation (applying an $O(\alpha_p \, p^3)$ by $O(\alpha_p \, p^3)$ matrix to a vector) takes $O(\alpha_p^2 \, p^6)$ work. The memory footprint of the Chebyshev FMM is dominated by the cost of storing pre-computed M2L operators, which is
\begin{equation}
{\cal M}_{C-FMM} = 54 \cdot 316 \cdot p^6 \cdot \alpha_p^2 \cdot f
\label{eq:mem_cheb}
\end{equation}
where $54$ is the number of components in the kernel function, $316$ is the total number of pre-computed M2L operators in three dimensions\footnote{Since the kernel function in DD is homogeneous, i.e., $\bm K(\alpha \bm x, \alpha \bm y) = \alpha^{-2} \bm K(\bm x, \bm y) $, the M2L operators only need to be computed for one level in the FMM hierarchical tree.}, and $f$ is the memory cost to store a floating-point number ($f=4$ bytes for single-precision and $f=8$ bytes for double-precision).

\begin{figure}[htb]
\begin{center}
\includegraphics[width=7.2cm]{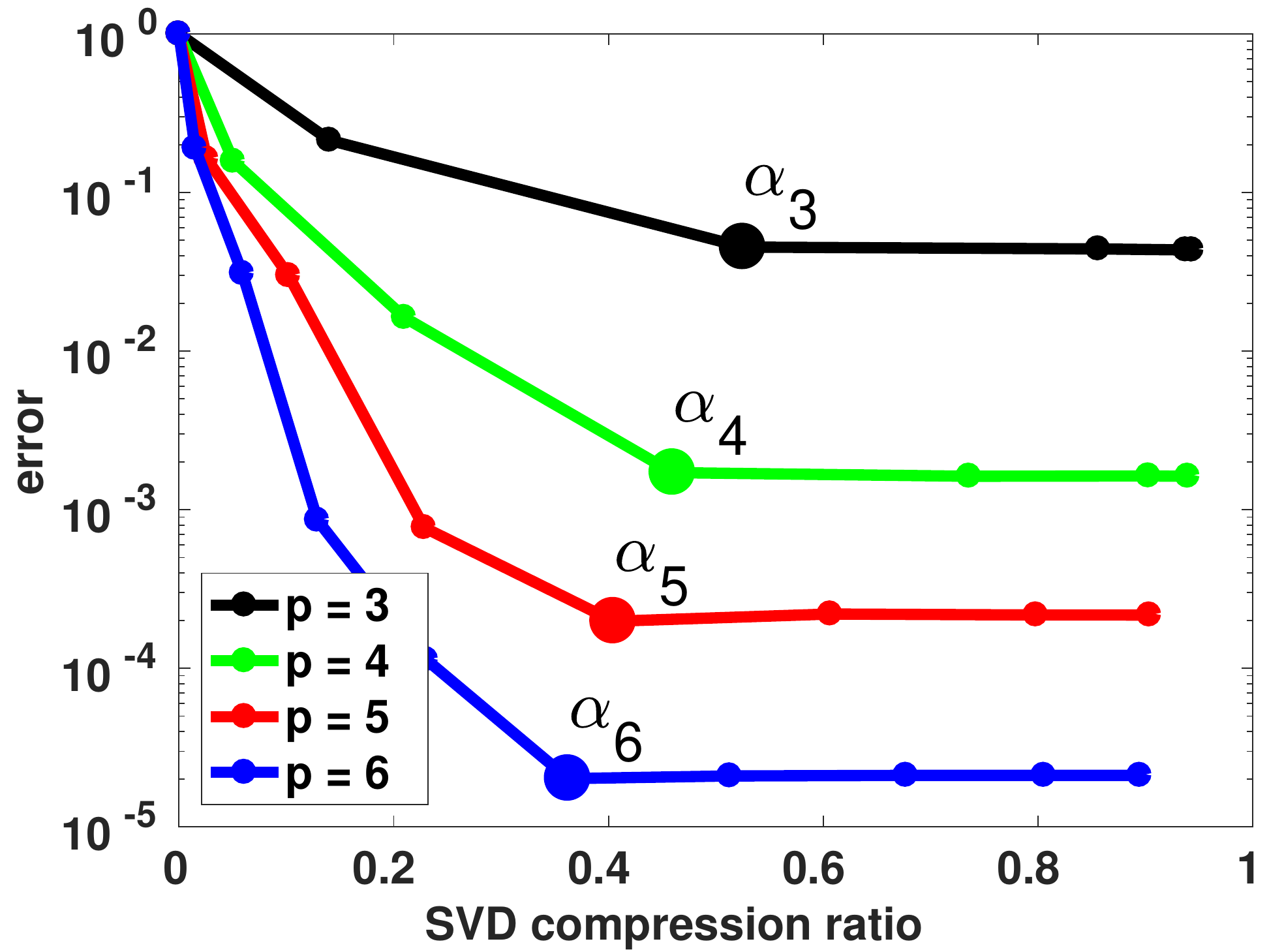}
\caption{The calculation error of the stress field in an isotropic elastic media using the Chebyshev FMM with different SVD compression ratios. In the Chebyshev FMM, $p$ is the Chebyshev interpolation order. A compression ratio of $1$ means no compression.} 
\label{fig:SVD}
\end{center}
\end{figure}

In the Chebyshev FMM, the optimal SVD compression ratio $\alpha_p$ is chosen to obtain the largest compression on the number of multiple and local coefficients while keeping the calculation error unchanged. \autoref{fig:SVD} shows an example of $\alpha_p$ ($p=3$, $4$, $5$, and $6$) when the Chebyshev FMM is used in an isotropic elastic media. As the figure shows, when the Chebyshev interpolation order $p$ is fixed, the calculation error does not deteriorate as long as the SVD compression ratio is larger than $\alpha_p$.

A summary of the comparisons among the Taylor FMM, the Spherical FMM, the Chebyshev FMM and the new method we are going to introduce is given in \autoref{subsec:compare}.

\section{The Lagrange FMM\label{sec:new_fmm}}

In this section, we introduce a new kernel-independent method that we call the Lagrange FMM. We designed the Lagrange FMM such that its M2L translation operators are block-Toeplitz matrices, which leads to a smaller memory footprint compared with the Chebyshev FMM. The kernel function in DD, i.e., \autoref{eq:kernel} can be viewed as a translation-invariant function of two variables, i.e., ${\bm K}({\bm x}, {\bm y}) \overset{\text{def}}{=} {\bm K}({\bm x} - {\bm y})$; we thus use ${\bm K}({\bm x} - {\bm y})$ and ${\bm K}({\bm x}, {\bm y})$ interchangeably in this section.

\subsection{Reduced memory footprint}\label{subsec:mem_unif}

The amount of computer memory required for storing the pre-computed M2L translation operators dominates the memory footprint in the Lagrange FMM and the Chebyshev FMM. In the following, we derive explicit formulas of M2L translation operators in the Lagrange FMM and calculate their memory consumption.

Like the Chebyshev FMM, the Lagrange FMM also uses Lagrange interpolating polynomials to approximate far-field dislocation interactions. Consider the interaction ${\bm K}({\bm x}, {\bm y})$ between two points $\bm x = (x_1, x_2, x_3)$ and $\bm y = (y_1, y_2, y_3)$ lying in two distant boxes $A$ and $B$, respectively. This far-field interaction can be approximated using the $p$-th order Lagrange interpolating polynomial as below 
\begin{align}
{\bm K}(\bm{x}, \bm{y}) &\approx \sum_{\bm{l}} {\bm K}(\bar{\bm{x}}_{\bm{l}}, \bm{y}) L_p(\bar{\bm{x}}_{\bm{l}}, \bm{x}) 
&& \text{[interpolate on $\bm x$]} \nonumber \\
&\approx \underbrace{ \sum_{\bm{l}} \sum_{\bm{m}} {\bm K}(\bar{\bm{x}}_{\bm{l}}, \bar{\bm{y}}_{\bm{m}}) 
L_p(\bar{\bm{x}}_{\bm{l}}, \bm{x}) L_p(\bar{\bm{y}}_{\bm{m}}, \bm{y}) }_{\text{Lagrange interpolant of $\bm K(\bm x, \bm y)$}}
&& \text{[interpolate on $\bm y$]}
\label{eq:interpolant}
\end{align}
where $\bar{\bm{x}}_{\bm{l}} = (\bar{x}_{l_1}, \bar{x}_{l_2}, \bar{x}_{l_3})$, $l_i = 1, 2, \ldots, p$  and $\bar{\bm{y}}_{\bm{m}} = (\bar{y}_{m_1}, \bar{y}_{m_2}, \bar{y}_{m_3}), m_i = 1, 2, \ldots, p$ are two sets of interpolation nodes in box $A$ and box $B$, respectively, and 
\begin{align*}
L_p(\bar{\bm{x}}_{\bm l}, \bm{x}) &= \ell_p(\bar{x}_{l_1}, x_1) \ell_p(\bar{x}_{l_2}, x_2) \ell_p(\bar{x}_{l_3}, x_3) \\
L_p(\bar{\bm{y}}_{\bm m}, \bm{y}) &= \ell_p(\bar{y}_{m_1}, y_1) \ell_p(\bar{y}_{m_2}, y_2) \ell_p(\bar{y}_{m_3}, y_3)
\end{align*}
 are products of three Lagrange basis polynomials,\footnote{A $p$-th order Lagrange basis polynomial $\ell_p(\bar{x}_i, x) = \prod_{1\le k \le p, k\not=i} (x-\bar{x}_k)/(\bar{x}_i - \bar{x}_k)$, where $\bar{x}_1, \bar{x}_2, \ldots, \bar{x}_p$ are interpolation nodes.} one for each dimension. 
 
In other words, the M2L translation operator $\bm{T}_{A \rightarrow B}$ between box $A$ and box $B$ takes the form
\begin{equation}
\bm{T}_{A \rightarrow B} = \big( {\bm K}(\bxl - \bym) \big)_{p^3 \times p^3}
\label{eq:m2l}
\end{equation}
which can be viewed as a $p^3$ by $p^3$ matrix with entry ${\bm K}(\bxl - \bym)$ at the $\bm l$-th row and $\bm m$-th column.
The difference between the Chebyshev FMM and the Lagrange FMM is the choice of interpolation nodes. In the Chebyshev FMM, the interpolation nodes are Chebyshev nodes, which guarantees that the approximation in \autoref{eq:interpolant} is nearly optimal \cite{trefethen2013approximation}. In the Lagrange FMM, the interpolation nodes lie on equally spaced grids in box $A$ and box $B$. For example, if box $A$ is the unit cube: $[0,1]^3$, then $x_{l_i} = (i-1)/(p-1)$ for $l_i = 1, 2, \ldots, p$. This choice of interpolation nodes in the Lagrange FMM leads to the block-Toeplitz structure of the M2L translation operator $\bm{T}_{A \rightarrow B}$, as stated in the following theorem (its proof is given in the Appendix A).

\begin{center}
\begin{theorem} \label{thm1}
In the $p$-th order Lagrange FMM, every M2L operator $\bm{T}_{A \rightarrow B}$ is a three-level block-Toeplitz matrix\footnote{A one-level block-Toeplitz matrix is simply a Toeplitz matrix having constant entries along its diagonals, and a $k$-level block-Toeplitz matrix is a matrix that has constant blocks along the diagonals and in which every block is a $(k-1)$-level block-Toeplitz matrix.} with at most $(2p-1)^3$ unique entries, and applying $\bm{T}_{A \rightarrow B}$ to a vector takes $O(p^3\log(p))$ work. 
\end{theorem}
\end{center}

Therefore, the memory footprint of the Lagrange FMM, which is dominated by the cost of storing all M2L operator, is
\begin{equation}
{\cal M}_{L-FMM} = 54 \cdot 316 \cdot (2p-1)^3 \cdot f
\label{eq:mem_unif}
\end{equation}
where the two constants 54 and 316 are respectively the number of components in the Green's function and the total number of pre-computed M2L operators in three dimensions,\!\footnote{Since the kernel function in DD is homogeneous, i.e., $\bm K(\alpha \bm x, \alpha \bm y) = \alpha^{-2} \bm K(\bm x, \bm y) $, the M2L operators only need to be computed for one level in the FMM hierarchical tree.} $p$ is the degree of interpolating polynomial, and $f$ is the memory cost to store a floating-point number ($f=4$ bytes for single-precision and $f=8$ bytes for double-precision). 

Note that the M2L operators in the Chebyshev FMM and the Lagrange FMM depend only on the kernel function $\bm K$, not on the problem configuration such as the number and location of dislocation segments. Therefore, these operators can be pre-computed one time and then used in multiple DD simulations.

\subsection{Comparison of four FMMs}\label{subsec:compare}

Assume the expansion order is $p$ in the Taylor FMM and the Spherical FMM, and, correspondingly, the $p$-th order polynomial interpolation is used in the Chebyshev FMM and the Lagrange FMM. For a fixed problem size, we compare the four methods: the Taylor FMM, the Spherical FMM, the Chebyshev FMM, and the Lagrange FMM, in terms of (1) their capability in isotropic and anisotropic elastic media, (2) the number of multipole and local coefficients, (3) the running time, (4) the convergence rate, (5) the memory footprint, and (6) the pre-computation time. Since the FMM running time is dominated by the cost of the M2L translation step, we analyze the M2L translation time for each method. The M2L translation operators are pre-computed in the Chebyshev FMM and the Lagrange FMM, and the cost for storing the pre-computed M2L translation operators dominates the memory footprint of both methods. 

The Taylor FMM applies in only isotropic elastic media. In the Taylor FMM, the number of multipole and local coefficients is $O(p^3)$, which is roughly the number of terms in the $p$-th order Taylor series expansion of the Green's function in DD. As a result, every M2L translation (applying a $O(p^3)$ by $O(p^3)$ matrix to a vector) requires $O(p^6)$ work. The calculation error decays as $O((\sqrt{3}/3)^p)$, the same as that of the Taylor series expansion used. The calculations of the derivatives of $R$ are pre-computed with $O(p^3)$ CPU time and computer memory.

The Spherical FMM applies in only isotropic elastic media. In the Spherical FMM, the number of multipole and local coefficients is $O(p^2)$, which is roughly the number of terms in the $p$-th order spherical harmonics expansion of the Green's function in DD. The M2L translation step requires $O(p^3)$ work using the `point-and-shoot' scheme. In the Spherical FMM, the rotation operators in the `point-and-shoot' scheme, which rotates the coordinate system, can be pre-computed. These operators typically require only a few megabytes of memory and can be pre-computed in a couple minutes. Since the pre-computation in the Spherical FMM is not for M2L translation operators, the costs are not comparable to that of the Chebyshev FMM and the Lagrange FMM. The convergence rate of the Spherical FMM is about the same as that of the Taylor FMM. For the Spherical FMM, the rotation operators, which first rotates the multipole expansion around the $z$-axis and then around the $y$-axis, can be pre-computed. These operators have $O(p^{3})$ floating point numbers, and require $O(p^{4})$ floating point operations to compute. But they typically require only a few mega bytes and can be computed in a couple minutes. 

The Chebyshev FMM applies in both isotropic and anisotropic elastic media. In the Chebyshev FMM, the number of multipole and local coefficients is $\alpha_p \, p^3$ with $\alpha_p$ being the SVD compression ratio. As a result, every M2L translation (applying an $\alpha_p \, p^3$ by $\alpha_p \, p^3$ matrix to a vector) requires $O(\alpha_p^2 \, p^6)$ work. Unlike the Taylor series expansion, which approximates a function at one point, the Chebyshev approximation minimizes the maximum error over a whole region, and the calculation error is empirically $O((1/3)^p)$. The memory footprint of the Chebyshev FMM is $O(\alpha^2_p \, p^6)$, as in \autoref{eq:mem_cheb}. In the pre-computation stage of the Chebyshev FMM, an SVD of the $p^3$ by $p^3$ full M2L translation operator needs to be computed, which leads to a huge cost of $O(p^9)$.

The Lagrange FMM applies in both isotropic and anisotropic elastic media. In the Lagrange FMM, the multipole and local coefficients in three dimensions are tensor products of three sets of $p$ coefficients in each dimension, and the number of multipole and local coefficients is thus $p^3$. The M2L translation cost is $O(p^3 \log(p))$ work according to \autoref{thm1}. The convergence rate of the Lagrange FMM is about the same as that of the Chebyshev FMM in practice. The memory footprint is $O(p^3)$ as in  \autoref{eq:mem_unif}. Since every M2L operator has at most $(2p-1)^3$ entries, the cost of pre-computing all M2L operators is $O(p^3)$.

The comparison of four methods is summarized in the following table.

\begin{table}[!htb]
\caption{Comparisons of four methods, where `Taylor,' `Spherical,' `Chebyshev,' and `Lagrange' indicate the corresponding FMMs. Notation: $p$ is the expansion order in the Taylor FMM and the Spherical FMM; correspondingly, $p$ is the degree of interpolating polynomial in the Chebyshev FMM and the Lagrange FMM. In the Chebyshev FMM, the SVD compression ratio is $\alpha_p$ ($\alpha^2_p$ ranges from $p^{-1}$ to $p^{-2}$ in isotropic elastic media).}   
\label{table:compare}
\centering
\begin{tabular}{c c c c c} \toprule
                          & Taylor   & Spherical   & Chebyshev   & Lagrange   \\  \toprule
Capability	& isotropic & isotropic & \thead{isotropic \\ \& anisotropic}  & \thead{isotropic \\ \& anisotropic} \\ 
\# of coefficients          & $O(p^3)$     & $O(p^2)$        & $O(\alpha_p \, p^3)$            & $O(p^3)$  \\ 
Running time$\dagger$ & $O(p^6)$     & $O(p^3)$        & $O(\alpha^2_p \, p^6)$        & $O(p^3 \log(p))$ \\
Calculation error         & $O(3^{-\frac{p}{2}})$     & $O(3^{-\frac{p}{2}})$        & $O(3^{-p})^\diamond$      & $O(3^{-p})^\diamond$ \\ 
Precomputation memory* & $O(p^3)$    & $O(p^3)$        & $O(\alpha^2_p \, p^6)$         & $O(p^3)$ \\  
Precomputation time*  & $O(p^3)$     & $O(p^4)$        & $O(p^9)$                              & $O(p^3)$ \\    
\bottomrule
\end{tabular}
\begin{tablenotes}
\footnotesize \item $\dagger$ FMM running time (M2L translation cost) with respect to $p$, assuming the number of FMM cells is fixed. See Appendix B for the results of optimal running time with respect to the number of FMM cells.
\footnotesize \item *\ In practice, the pre-computation memory and pre-computation time of the Taylor FMM and the Spherical FMM are much smaller than those of the Chebyshev FMM and the Lagrange FMM.
\footnotesize \item $\diamond$ This bound was found empirically.
\end{tablenotes}
\label{table:fmm_cost}
\end{table}

\subsection{Lagrange FMM algorithm}

We present the Lagrange FMM, which evaluates the following sum
\begin{equation} \label{eq:nbody}
{\bm s}(\bm{x}_i) = \sum_{j=1}^N  {\bm K}(\bm{x}_i, \bm{y}_j) \, {\bm q}_j \quad \text{for} \quad i=1,2,\ldots,N
\end{equation}
where $ {\bm K}(\bm{x}_i, \bm{y}_j) \overset{\text{def}}{=}  {\bm K}( \bm{x}_i - \bm{y}_j )$ is a translation-invariant kernel function and may be singular when $\bm{x}_i = \bm{y}_j$. Note that the Lagrange FMM still works when $\{ \bm{x}_i \}_{i=1}^N$ and  $\{ \bm{y}_i \}_{i=1}^N$ are the same set of points. 

Before introducing the entire algorithm, we first focus on the core component in the Lagrange FMM---evaluating far-field interactions. Consider the far-field interaction between two sets of points $\{ \bm{x}_i \}_{i=1}^N$ and  $\{ \bm{y}_i \}_{i=1}^N$, lying in two distant FMM cells. With the Lagrange interpolant in \autoref{eq:interpolant}, the stress field at $\bm{x}_i$ can be approximated as
\begin{align} \label{eqn:far-field}
{\bm s}(\bm{x}_i) 
& \approx \sum_{j=1}^N 
\bigg( \sum_{\bm{l}} \sum_{\bm{m}} {\bm K}(\bar{\bm{x}}_{\bm{l}}, \bar{\bm{y}}_{\bm{m}}) 
L_p(\bar{\bm{x}}_{\bm{l}}, \bm{x}_i) L_p(\bar{\bm{y}}_{\bm{m}}, \bm{y}_j) \bigg)  \, {\bm q}_j \nonumber \\
& =   \sum_{\bm{l}}  L_p(\bar{\bm{x}}_{\bm{l}}, \bm{x}_i)  
\sum_{\bm{m}}  {\bm K}(\bar{\bm{x}}_{\bm{l}}, \bar{\bm{y}}_{\bm{m}})
\sum_{j=1}^N  L_p(\bar{\bm{y}}_{\bm{m}}, \bm{y}_j) \, {\bm q}_j . 
\end{align}

A fast algorithm for evaluating far-field interaction follows immediately from \autoref{eqn:far-field} as below. 

\begin{algorithm} [htb]
  \begin{enumerate}
    \item
    Compute weights at $\bar{\bm{y}}_{\bm{m}}$ by anterpolation:
    $${\bm W}_{\bm{m}} = \sum_{j=1}^N L_p(\bar{\bm{y}}_{\bm{m}}, \bm{y}_j) {\bm q}_j$$
    \item
    Compute ${\bm s}(\bm{x})$ at $\bar{\bm{x}}_{\bm{l}}$:
    $${\bm s}(\bxl) = \sum_{\bm{m}} {\bm K}(\bxl, \bym)  {\bm W}_{\bm{m}}$$
    \item
    Compute ${\bm s}(\bm{x})$ at $\bm{x}_i$ by interpolation:
    $${\bm s}(\bm{x}_i) = \sum_{\bm{l}} L_p(\bxl, \bm{x}_i) {\bm s}(\bxl)$$
  \end{enumerate}
  \caption{Fast algorithm for evaluating far-field interaction}
  \label{alg:fast_sum}
where $\{ \bm{x}_i \}_{i=1}^N$, $\{ \bm{y}_i \}_{i=1}^N$ belong to two distant FMM cells, and $\bar{\bm{x}}_{\bm{l}} = (x_{l_1}, x_{l_2}, x_{l_3}), \bar{\bm{y}}_{\bm{m}} = (y_{m_1}, y_{m_2}, y_{m_3})$ are 3-vectors of interpolation nodes with $l_i, m_i =1, 2, \ldots, p$.
\end{algorithm}

In the above algorithm, both step (1) and step (3) require $O(pN)$ work, and step (2) needs $O(p^2)$ work. When $p \ll N$, \autoref{alg:fast_sum} is significantly faster than a naive evaluation of \autoref{eq:nbody}, which costs $O(N^2)$ work. 

Note that the Lagrange interpolant of high degree over equally spaced interpolation nodes does not always converge, even for smooth functions, which is known as Runge's phenomenon; however, we find the low-order approximations sufficiently accurate for DD simulations in both isotropic and anisotropic media, as shown in our numerical experiments. The scheme of Lagrange interpolation on uniform nodes can be stabilized; for example, it could be stabilized by fitting a polynomial of degree $d < 2\sqrt{p}$ using least-squares, where $p$ is the number of equidistant points \cite{dahlquist1974numeriska}, or by using spline curves \cite{prenter2008splines}, which are piecewise polynomials.

\begin{algorithm}[b] 
\begin{itemize}
\item Particle to moment (P2M)\\
Step (1.a) for every leaf cell $I$:
$$
{\bm W}_{\bm{m}} = \sum_{\by_j \in I}  L_p(\bym, \bm{y}_j) \, {\bm q}_j
$$
\item Moment to moment (M2M)\\
Step (1.b) (Go up the tree) for every non-leaf cell $I$:
$$
{\bm W}_{\bm{m}'} = \sum_{J\in \mathcal{C}(I)} \sum_{\bym \in J} L_p(\bar{\bm{y}}_{\bm{m}'}, \bym) \, {\bm W}_{\bm{m}} 
$$
\item Moment to local (M2L)\\
Step (2) for every cell $I$:
$$
{\bm s}(\bxl) = \sum_{J\in \mathcal{I}(I)} \sum_{\bym \in J} {\bm K} (\bxl, \bym) \, {\bm W}_{\bm{m}} 
$$
\item Local to local (L2L)\\
Step (3.a) (Go down the tree) for every cell $I$:
$$
{\bm s}(\bar{\bm{x}}_{\bm{l}'}) \mathrel{{+}{=}} \sum_{\bxl \in \mathcal{P}(I)}  L_p(\bar{\bm{x}}_{\bm{l}'}, \bxl) \, {\bm s}(\bxl) 
$$
\item Local to particle (L2P)\\
Step (3.b) for every leaf cell $I$:
$$
{\bm s}(\bm{x}_i) = \sum_{\bxl \in I}  L_p(\bm{x}_i, \bxl) \, {\bm s}(\bxl)  + \sum_{\bm{y}_j \in \mathcal{N}(I)} {\bm K}(\bm{x}_i, \bm{y}_j) \, {\bm q}_j
$$
\end{itemize}
\caption{Lagrange FMM algorithm}
\label{alg:lagrange}
\end{algorithm}

With \autoref{alg:fast_sum} for evaluating far-field interactions, we can derive the Lagrange FMM using the standard FMM machinery. The Lagrange FMM is based on the standard FMM tree structure \cite{greengard1987fast}, which defines the set of children cells $\mathcal{C}(\cdot)$, the interaction list $\mathcal{I}(\cdot)$, the parent cell $\mathcal{P}(\cdot)$, and the set of neighboring cells $\mathcal{N}(\cdot)$ for every FMM cell. The Lagrange FMM, which evaluates the sum in \autoref{eq:nbody} is shown in \autoref{alg:lagrange}.

\section{Numerical results\label{sec:results}}

In this section, results of benchmarking and comparing the four methods introduced: the Taylor FMM, the Spherical FMM, the Chebyshev FMM and the Lagrange FMM are presented. In \autoref{subsec:precompute}, memory footprint and pre-computation time of the Chebyshev FMM and the Lagrange FMM are presented. In \autoref{subsec:convergence} and \autoref{subsec:timing}, convergence rates and running time of the four methods are presented, respectively. In \autoref{subsec:linear}, the minimum running time and linear scalability of the four methods are presented. In \autoref{subsec:parallel}, parallel scalability is presented. 

We refer to the expansion order in the Taylor FMM and the Spherical FMM and the interpolation order in the Chebyshev FMM and the Lagrange FMM as the FMM order. In the Chebyshev FMM, the optimal SVD compression ratios are used. To measure the relative error of FMM calculations, we use the stress fields calculated by evaluating the summation in \autoref{eq:stress3} directly as the reference. For simulation in anisotropic elastic media, the anisotropic kernel function in \autoref{eq:k_aniso} was calculated through a series of spherical harmonics with an expansion order of $q_{\max}$. As shown in \autoref{table:qmax}, for different anisotropic media, we chose $q_{\max}$ according to the anisotropic ratio $A=2C_{44}/(C_{11}-C_{12})$, where $\bm C$ is the elastic stiffness tensor, so that the error from the truncated spherical harmonics series is smaller than that from FMM calculations. 

\begin{table}[h]
\centering
\caption{Truncation term $q_{\max}$, relative cost and accuracy of evaluating the anisotropic kernel function in \autoref{eq:k_aniso}.}
\begin{tabular}{l c c c} \toprule
 Anisotropic ratio      & $q_{\max}$            & relative cost            & accuracy  \\ \toprule
$A=1$                    & 1            &           1      &  $10^{-16}$ \\
$A=0.31$, $3.16$  & 10          &           16    &  $10^{-6}$ \\
$A=0.1$, $10$       & 20          &            52   &  $10^{-4}$ \\
 \bottomrule
\end{tabular}
\label{table:qmax}
\end{table}

We implemented the four methods in the dislocation dynamics simulation code ParaDiS \cite{arsenlis2007enabling}. All results were obtained from running ParaDiS on the Quartz machine\footnote{\url{https://hpc.llnl.gov/hardware/platforms/Quartz}} at the Lawrence Livermore National Laboratory.

\subsection{Memory footprint and pre-computation time}\label{subsec:precompute}

In this section, we compare the memory footprint and pre-computation time of the four methods. In practice, the memory footprint and pre-computation time of the Taylor FMM and the Spherical FMM are much smaller than those of the Chebyshev FMM and the Lagrange FMM. Typically, the memory footprint of the Taylor FMM and the Spherical FMM are around a few megabytes, and the pre-computation time is at most a few minutes. Therefore, we show results only for the Chebyshev FMM and the Lagrange FMM.

In the Chebyshev FMM and the Lagrange FMM, M2L translation operators are pre-computed, which speeds up applying both methods but results in large memory footprint. The memory cost for storing these pre-computed matrices and the pre-computation time are summarized in \autoref{table:fmm_cost}. 

\begin{figure}[htb]
\begin{center}
\hspace*{\fill}
\subfigure[Memory footprint]{\includegraphics[width=6.7cm]{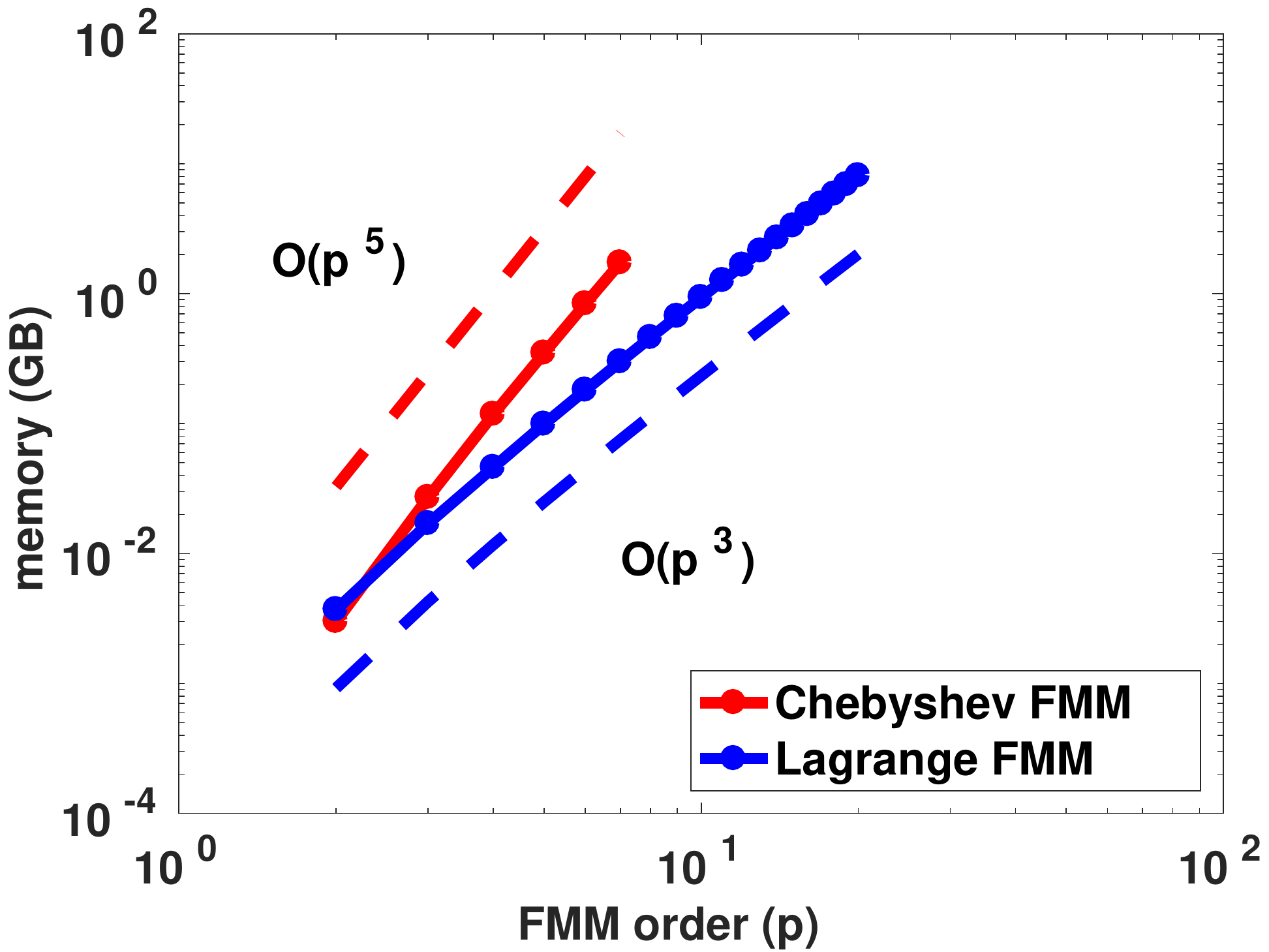}}
\hfill
\subfigure[Pre-computation time]{\includegraphics[width=6.7cm]{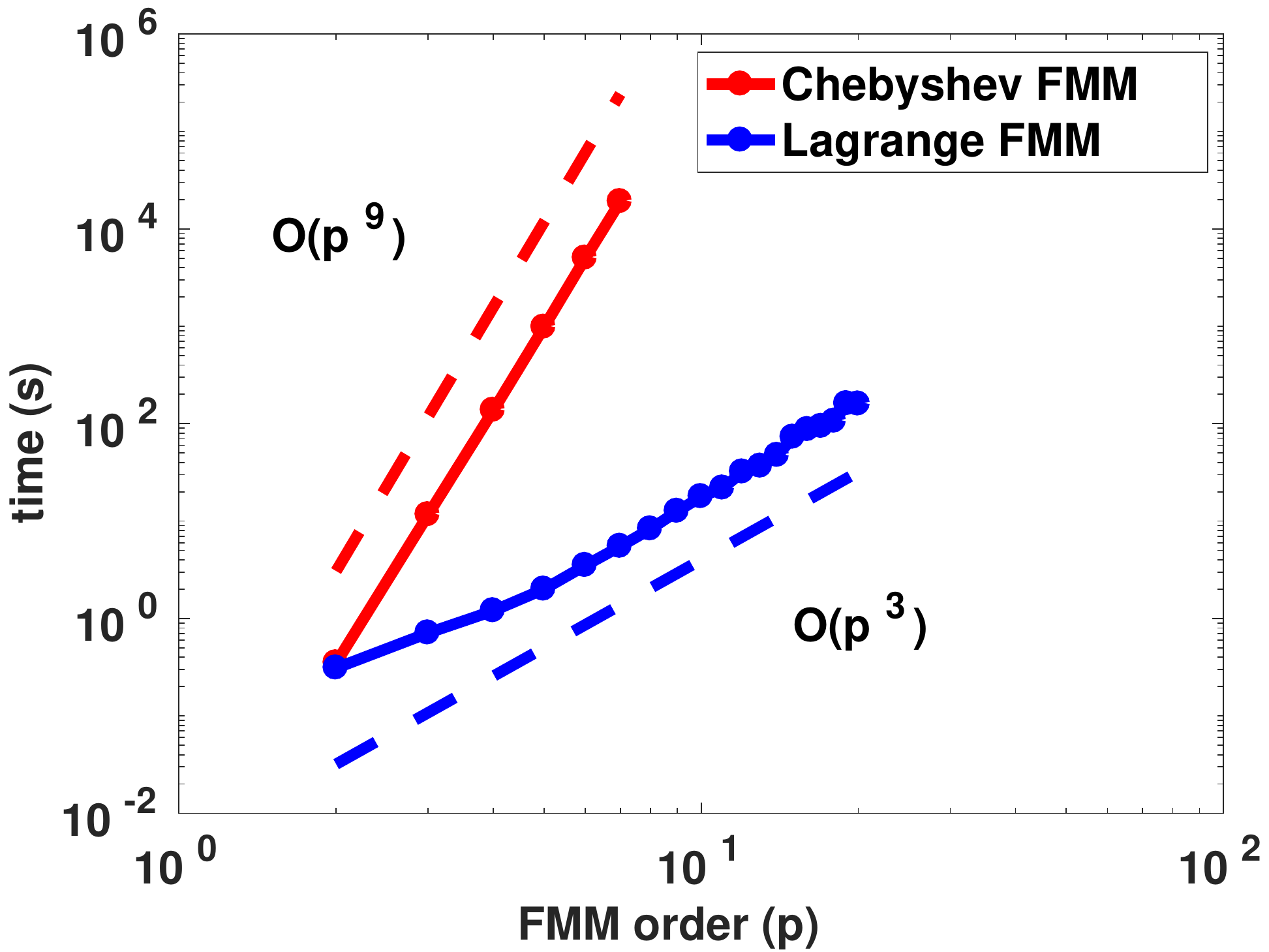}}
\hspace*{\fill}
\caption{Comparison of memory footprint and pre-computation time between the Chebyshev FMM and the Lagrange FMM in isotropic elastic media.}
\label{fig:precomputation}
\end{center}
\end{figure}

We focus on the results in isotropic elastic media, and results in anisotropic elastic media follow similar trends. The memory footprint and pre-computation time of the Chebyshev FMM and the Lagrange FMM are shown in \autoref{fig:precomputation}. As the figure shows, the Lagrange FMM has lower memory footprint and requires significantly smaller amount of pre-computation time than the Chebyshev FMM. For example, the memory footprint of the Lagrange FMM, compared with that of the Chebyshev FMM, is reduced by a factor of $5.7$ when $p=7$. In this case, the Chebyshev FMM consumes about $1.7$ GB memory, while the same memory footprint is reached when $p=12$ in the Lagrange FMM ($1.6$ GB). 

Due to the large memory footprint and pre-computation time of the Chebyshev FMM, we present results only for $p \le 7$ for the Chebyshev FMM in this section.

\subsection{FMM Convergence}\label{subsec:convergence}
\begin{figure}[b]
\centering
\ifx\JPicScale\undefined\def\JPicScale{0.25}\fi
\unitlength \JPicScale mm
\begin{picture}(180,160)(0,0)
\linethickness{0.3mm}
\put(20,0){\framebox(160,160){}}
\linethickness{0.3mm}
\put(20,0){\framebox(160,120){}}
\linethickness{0.3mm}
\put(20,0){\framebox(160,80){}}
\linethickness{0.3mm}
\put(20,0){\framebox(160,40){}}
\linethickness{0.3mm}
\put(20,0){\framebox(120,160){}}
\linethickness{0.3mm}
\put(20,0){\framebox(80,160){}}
\linethickness{0.3mm}
\put(20,0){\framebox(40,160){}}
\linethickness{0.5mm}
\multiput(110,95)(0.36,-0.12){42}{\line(1,0){0.36}}
\linethickness{0.5mm}
\put(125,90){\line(0,1){15}}
\linethickness{0.5mm}
\multiput(110,95)(0.18,0.12){83}{\line(1,0){0.18}}
\linethickness{0.5mm}
\put(27.5,12.5){\tikzcircle[fill=black]{1pt}}
\put(37,15){\makebox(0,0)[cc]{$x$}}
\end{picture}
\caption{Single-loop configuration (projected on the plane). The dislocation loop is shown as a triangle, and the stress field is evaluated at a point denoted by $x$. The cubic domain is divided into $4 \times 4 \times 4$ FMM cells. }
\label{fig:SimpleLoop} 
\begin{center}
\hspace*{\fill}
\subfigure[210 dislocation loops]{\includegraphics[clip, width=6cm, trim= 3cm 6cm 3cm 6cm]{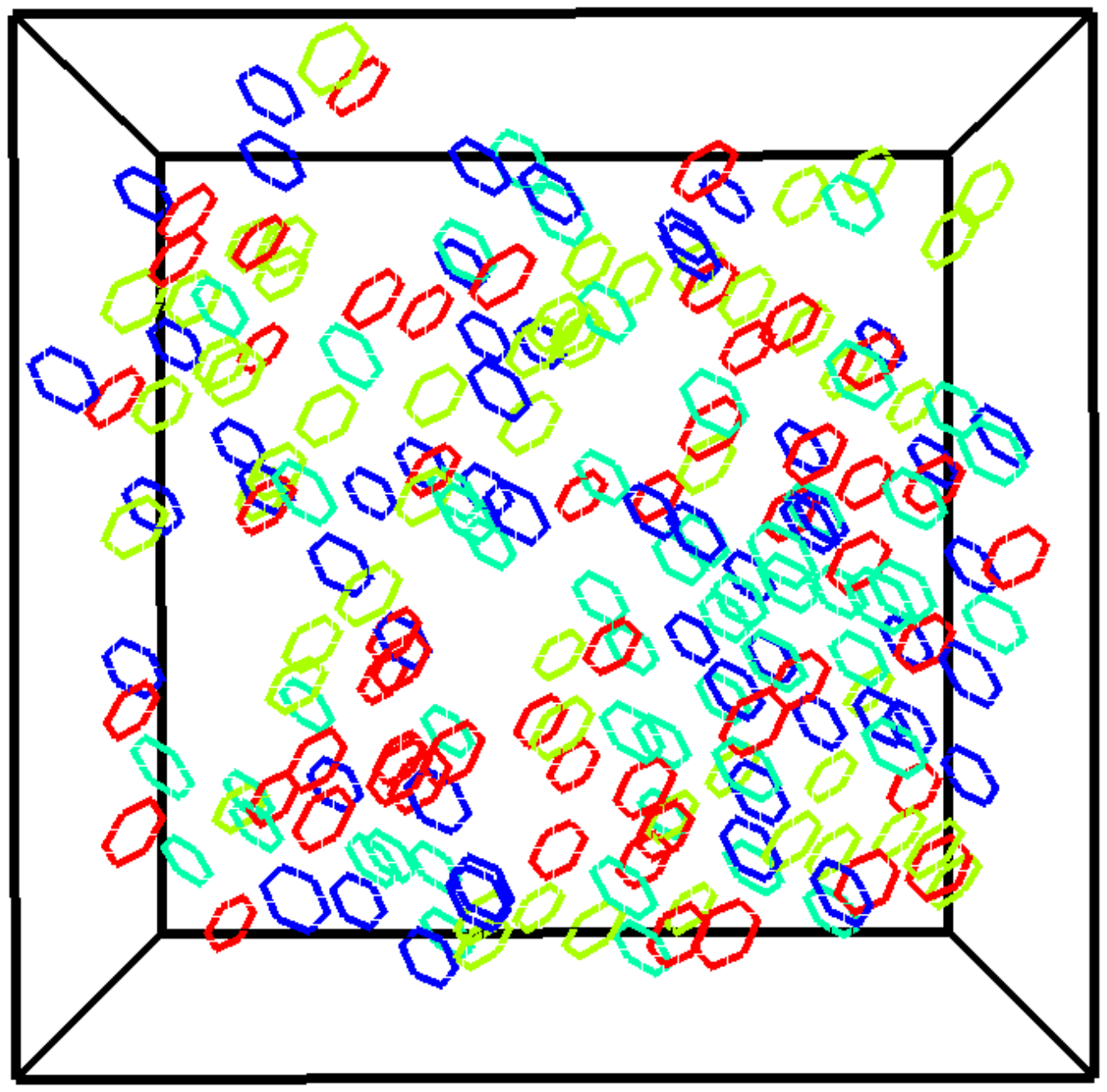}}
\hfill
\subfigure[Evaluation points (blue)]{\includegraphics[clip, width=6cm, trim= 4.5cm 7.5cm 4.5cm 7.5cm]{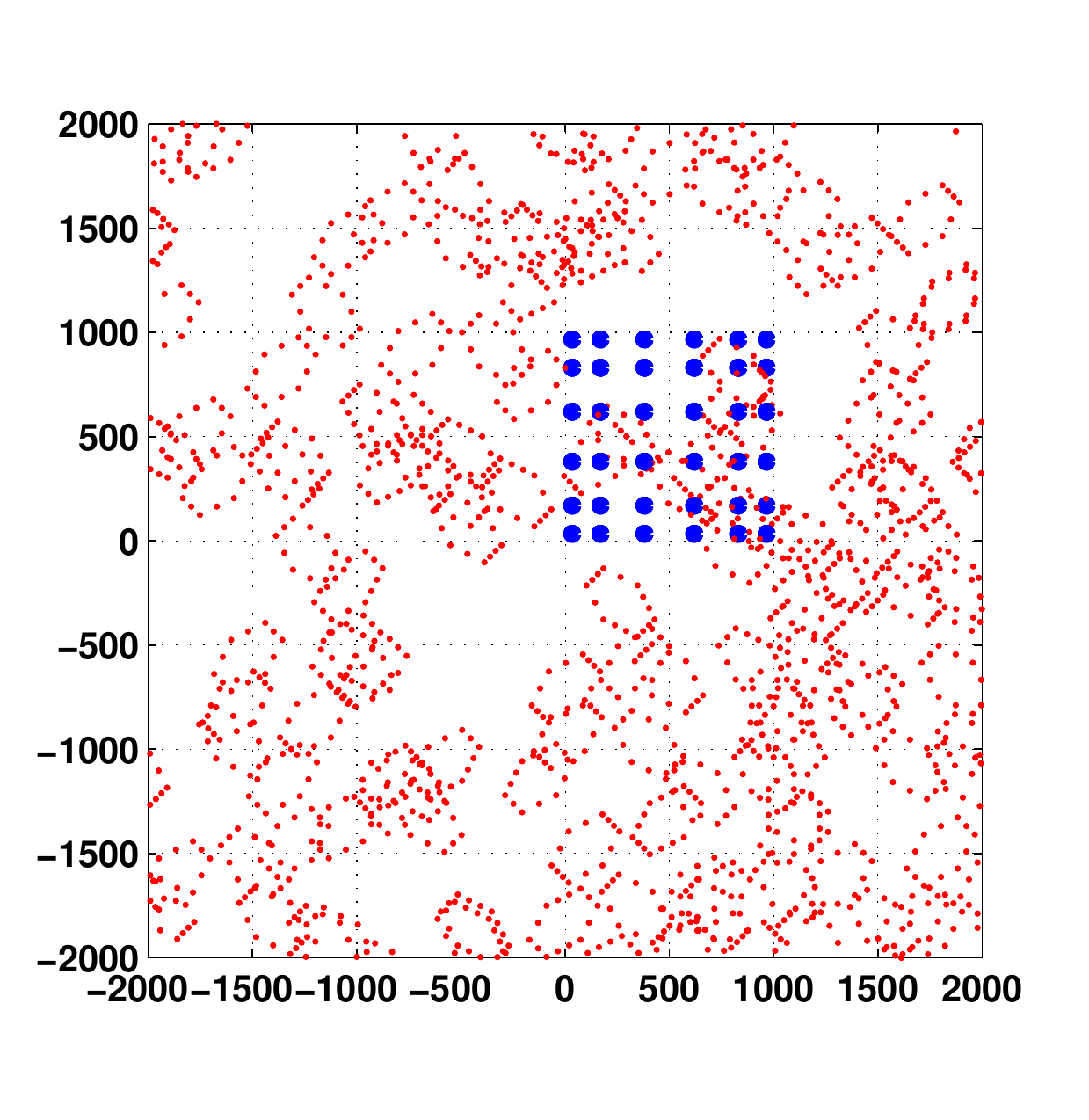}}
\hspace*{\fill}
\caption{Multiple-loops configuration. (a) shows a cubic domain of size $(4,000 b)^3$ containing $210$ closed dislocation loops, which are discretized into about $3,000$ segments. (b) shows the points (blue) where the stress field is evaluated at and the dislocation loops (red), projected on the plane. The cubic domain is divided into $8 \times 8 \times 8$ FMM cells, and some dislocation loops cross multiple FMM cells.}
\label{fig:BigLoop}
\end{center}
\end{figure}

To show the convergence of the four FMM methods, we present the
results for a single-loop configure and a multiple-loops
configuration, as shown in \autoref{fig:SimpleLoop} and
\autoref{fig:BigLoop}, respectively. The single-loop configuration has
one dislocation loop completely lying inside an FMM cell. The
multiple-loops configuration represents a realistic initial
dislocation dynamics simulation configuration, where some dislocation
loops cross multiple FMM cells. We experiment using both
configurations with isotropic elasticity and five anisotropic ratios
($A=0.1, 0.31, 1.0, 3.1, 10$).

\begin{figure}[htb]
\begin{center}
\subfigure[single-loop configuration]{\includegraphics[width=6.7cm]{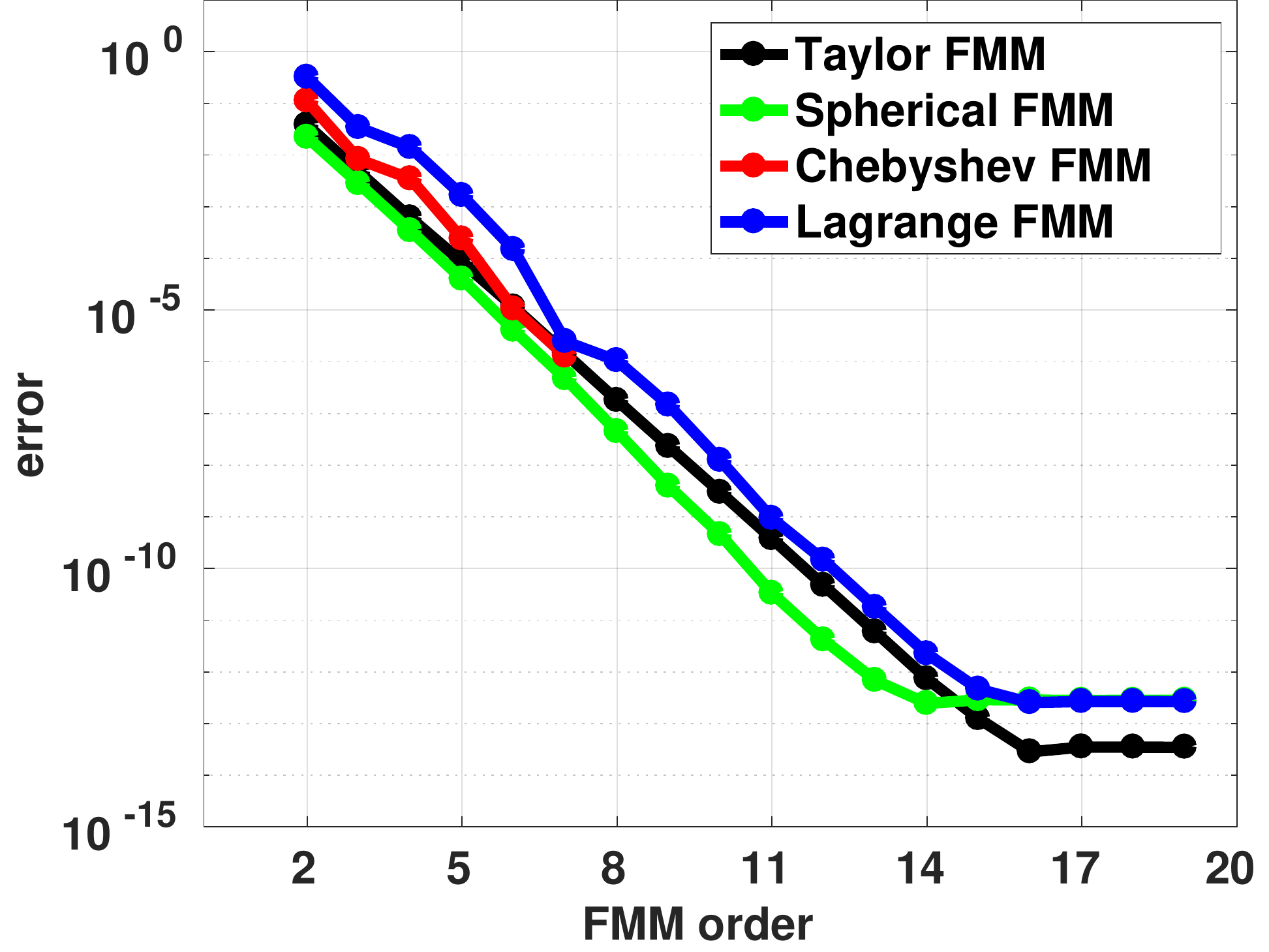}}
\subfigure[multiple-loops configuration]{\includegraphics[width=6.7cm]{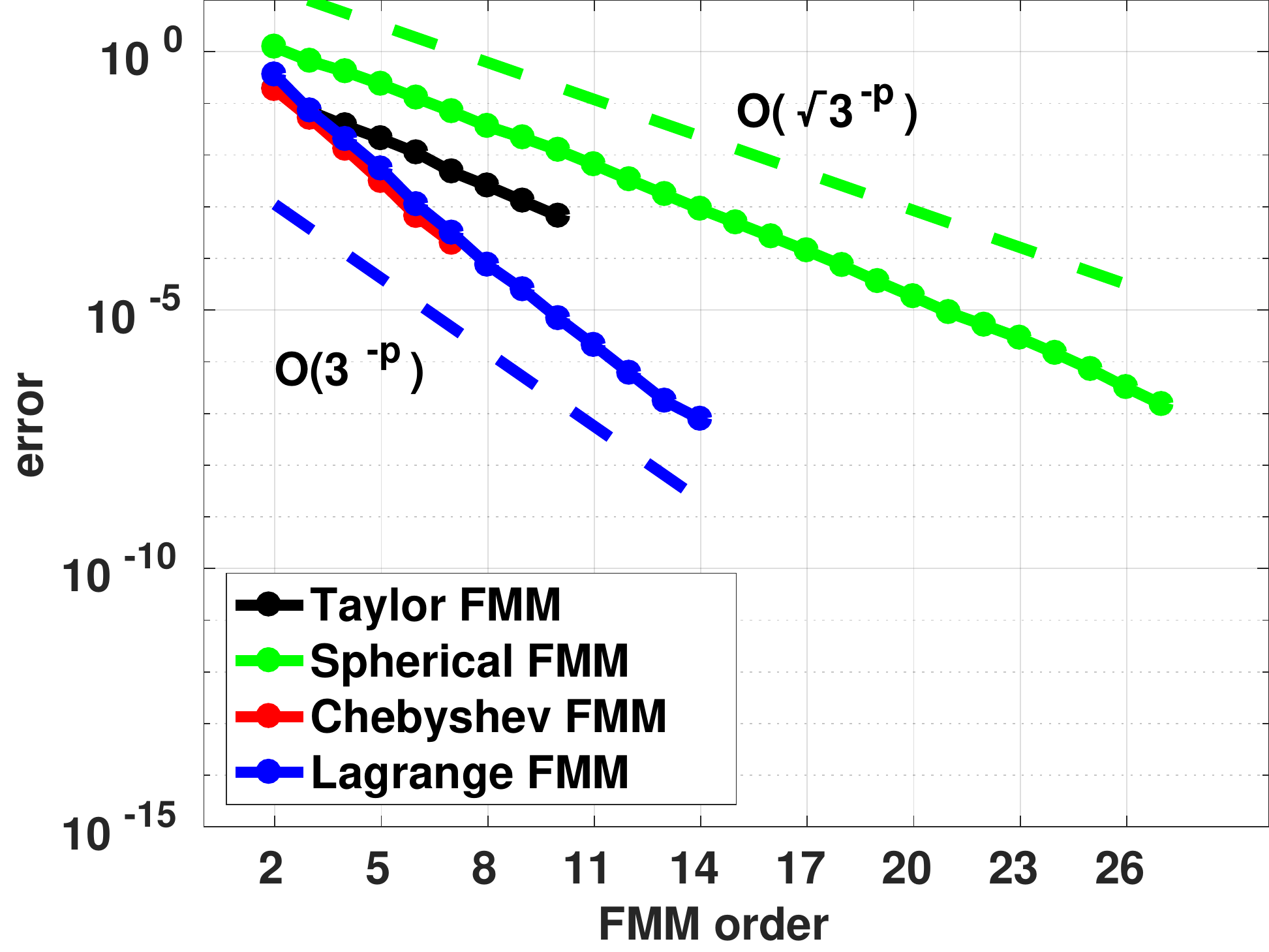}}
\caption{FMM convergence in isotropic elastic media. In (a) and (b), the Chebyshev FMM was stopped at order 7 due to its large pre-computation storage (1.7 GB at order 7). In (b), the Taylor FMM was stopped at order 10 due to its large running time (40 min/run at order 10), and the Lagrange FMM was stopped at order 14 because the error stopped decreasing.}
\label{fig:convergence_iso}
\end{center}
\end{figure}

In isotropic elastic media, convergence results with the single-loop configuration and with the multiple-loops configuration are shown in \autoref{fig:convergence_iso}(a) and \autoref{fig:convergence_iso}(b), respectively. As \autoref{fig:convergence_iso}(a) shows, three of the four methods converged to an accuracy of $10^{-12}$, with the exception of the Chebyshev FMM, which was stopped at order seven due to its large memory footprint. These results verify the correctness of our implementation.

Convergence results for the one and multiple-loops configurations are
shown in \autoref{fig:convergence_iso}(b). The figure shows two
differences among the four methods. First, the Chebyshev FMM and the
Lagrange FMM converged faster than the Taylor FMM and the Spherical
FMM. The convergence rate of the Chebyshev FMM and the Lagrange FMM is
about $3^{-p}$, and the convergence rate of the Taylor FMM and the
Spherical FMM is about $\sqrt{3}^{-p}$. Second,
with a limited memory of 3 GB, the Lagrange FMM was able to achieve a
smaller error than what the Chebyshev FMM reaches. As the figure
shows, the Lagrange FMM converges to an error of $10^{-7}$ whereas the
Chebyshev FMM converges to $10^{-4}$.


\begin{figure}[htb]
\centering
\subfigure[`Chebyshev' for single-loop]{\includegraphics[width=6.7cm]{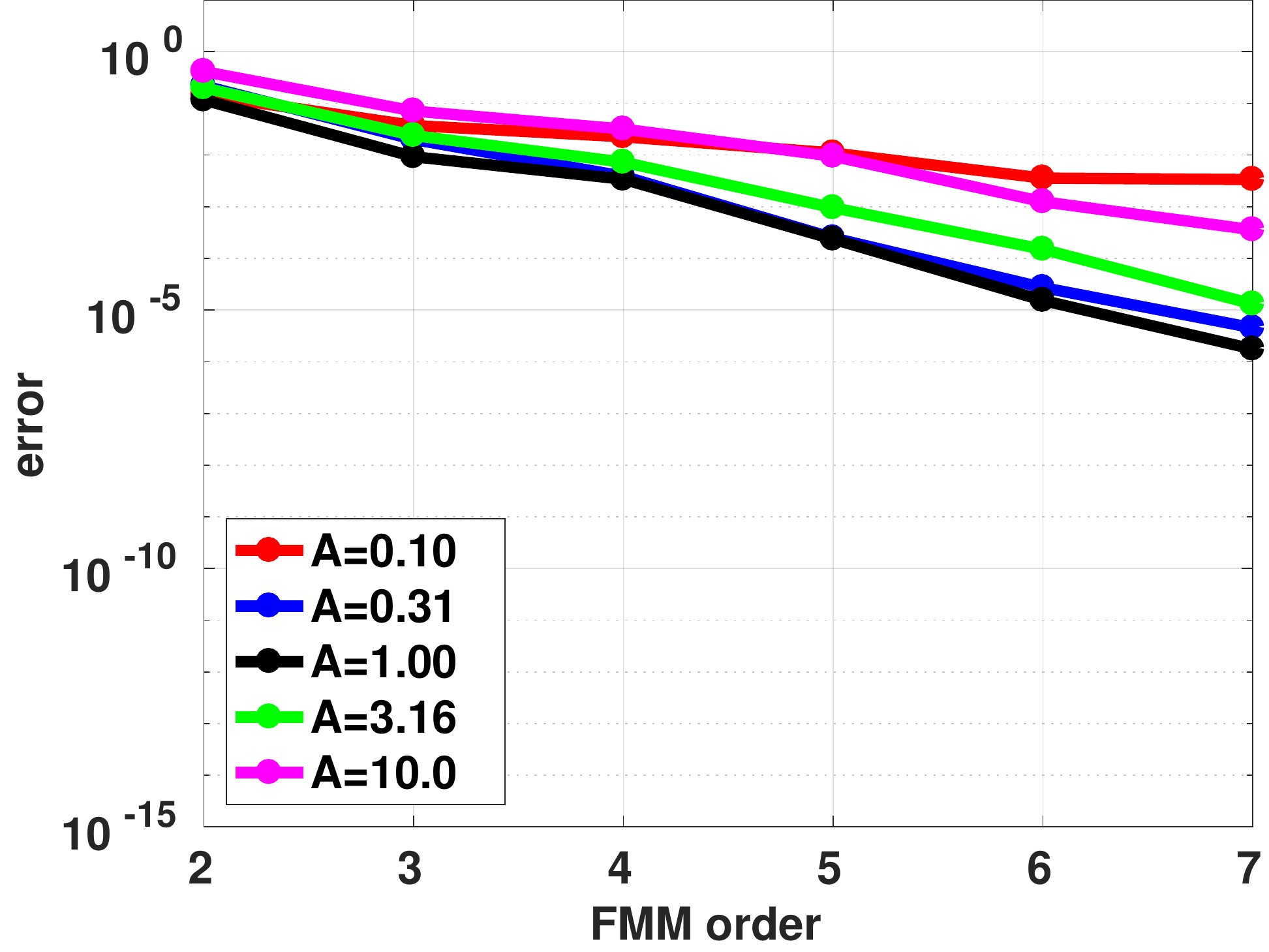}}
\subfigure[`Lagrange' for single-loop]{\includegraphics[width=6.7cm]{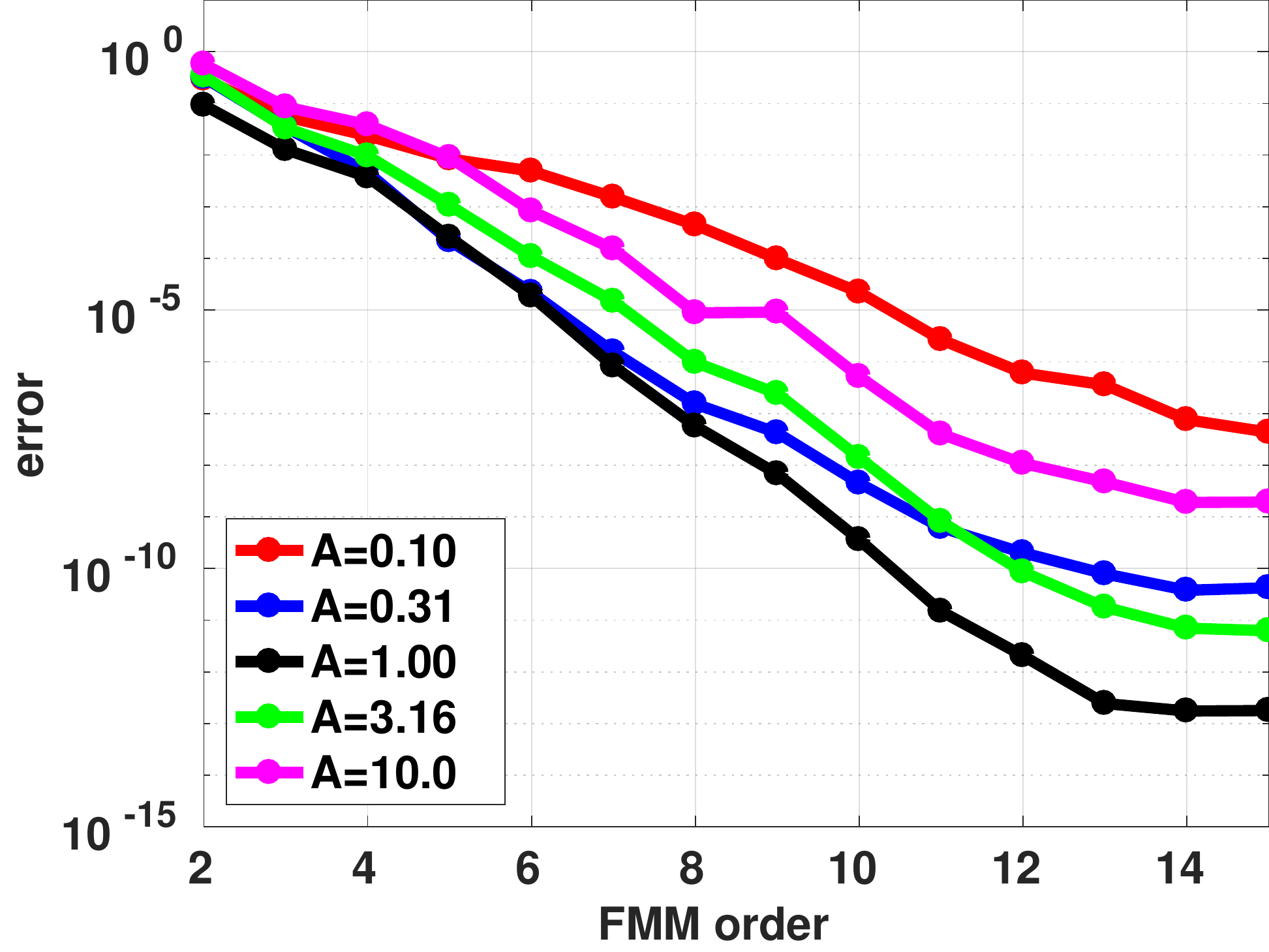}}
\subfigure[`Chebyshev' for multiple-loops]{\includegraphics[width=6.7cm]{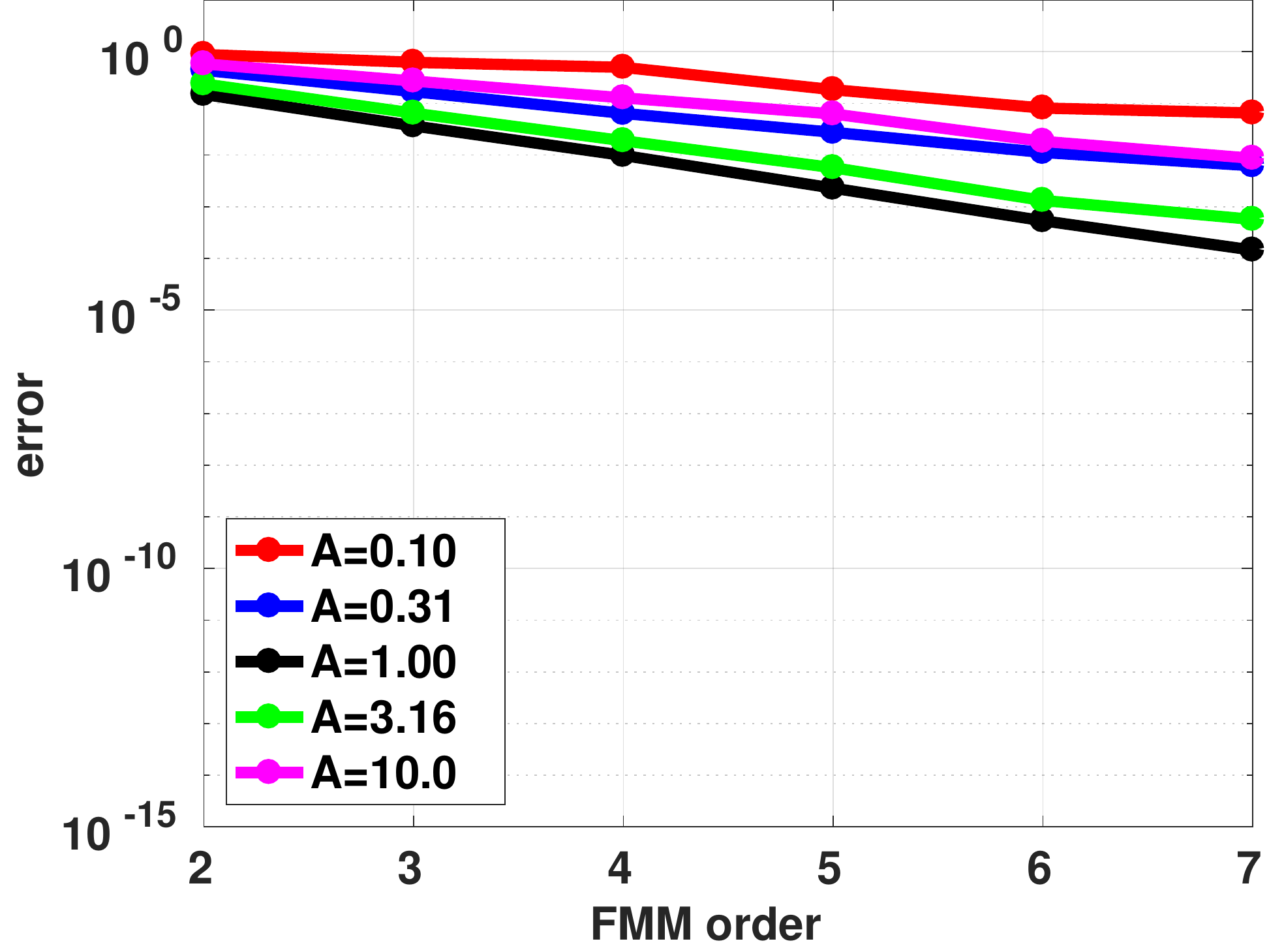}}
\subfigure[`Lagrange' for multiple-loops]{\includegraphics[width=6.7cm]{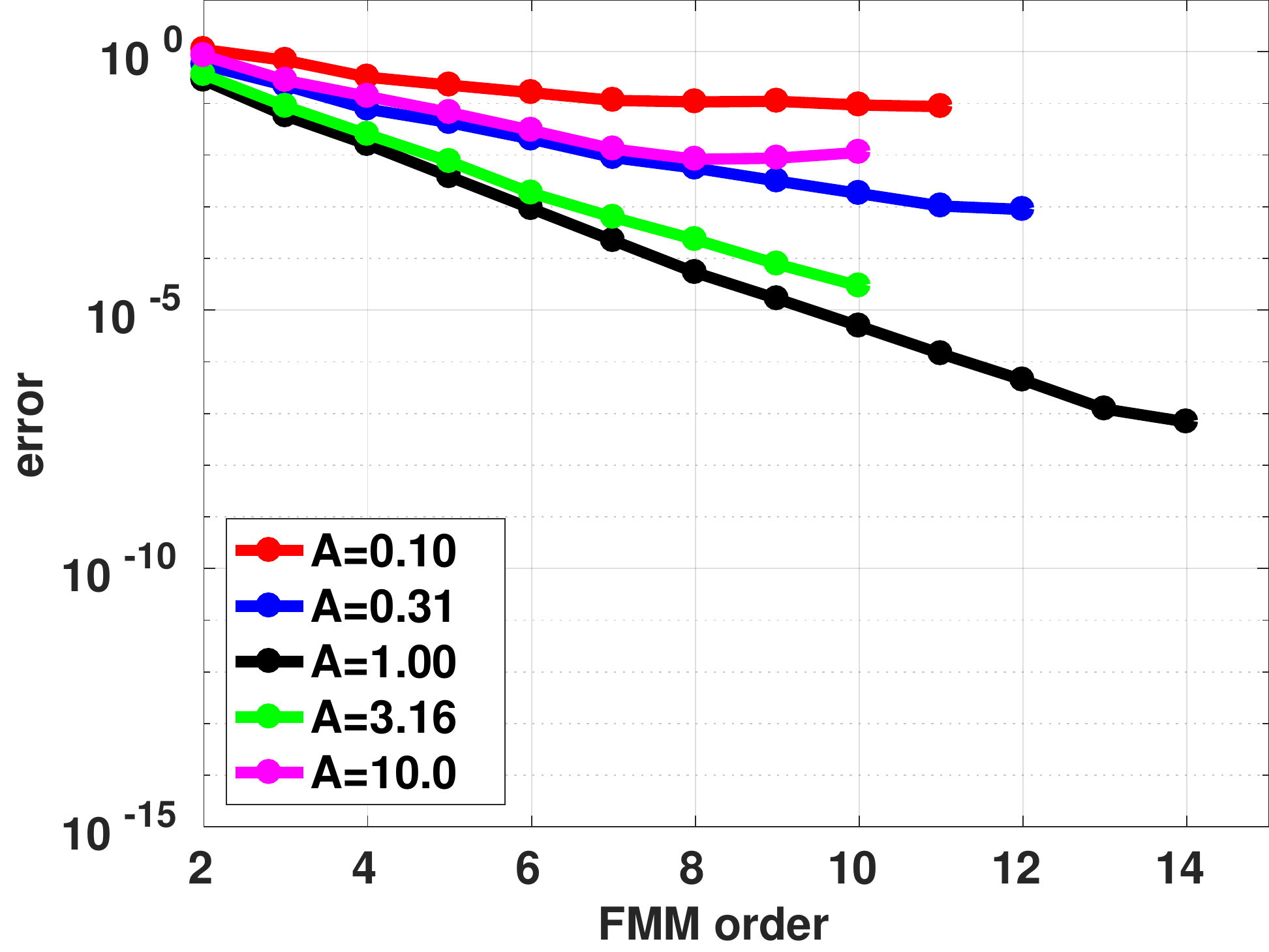}}
\caption{FMM convergence in anisotropic elastic media. The Taylor FMM and the Spherical FMM are not applicable in this case. In (a) and (c), the Chebyshev FMM was stopped at order 7 due to its large pre-computation storage. In (b) and (d), the Lagrange FMM was stopped when the error stopped decreasing.}
\label{fig:ConvergenceAniso}
\end{figure}

In anisotropic elastic media, convergence results for five different
anisotropic ratios: $A=0.1, 0.31, 1.0, 3.16, 10$ are shown in
\autoref{fig:ConvergenceAniso} ($A=1$ is equivalent to isotropic
elasticity). The Taylor FMM and the Spherical FMM are not applicable
in this case. Results in (a) and (b) verify the correctness of our
implementation. Results in \autoref{fig:ConvergenceAniso}(c) and
\autoref{fig:ConvergenceAniso}(d) correspond to the multiple-loops
configuration. As the figure shows, the convergence of the Chebyshev FMM and the Lagrange FMM slows down when the media
become more and more anisotropic, i.e., when the anisotropic ratio $A$
approaches $0.1$ or $10$. In addition, convergence results between the
Chebyshev FMM and the Lagrange FMM are similar up to order 7. With a
limited amount of computer memory, the Lagrange FMM can reach smaller errors than
the Chebyshev FMM when $A=1, 0.31, 3.16$.

\subsection{FMM running time}\label{subsec:timing}
To compare the running time among the four methods, we employ the multiple-loops configuration in \autoref{fig:BigLoop} with isotropic elasticity and anisotropic elasticity ($A=0.1, 0.31, 1.0, 3.16, 10$). Note that the number of FMM cells is fixed as $8 \times 8 \times 8$ in all experiments. We investigate varying the number of FMM cells based on the number of dislocation segments to achieve the smallest running time in \autoref{subsec:linear}.

\begin{figure}[h]
\centering
\subfigure[FMM error vs.\ running time]{\includegraphics[width=6.7cm]{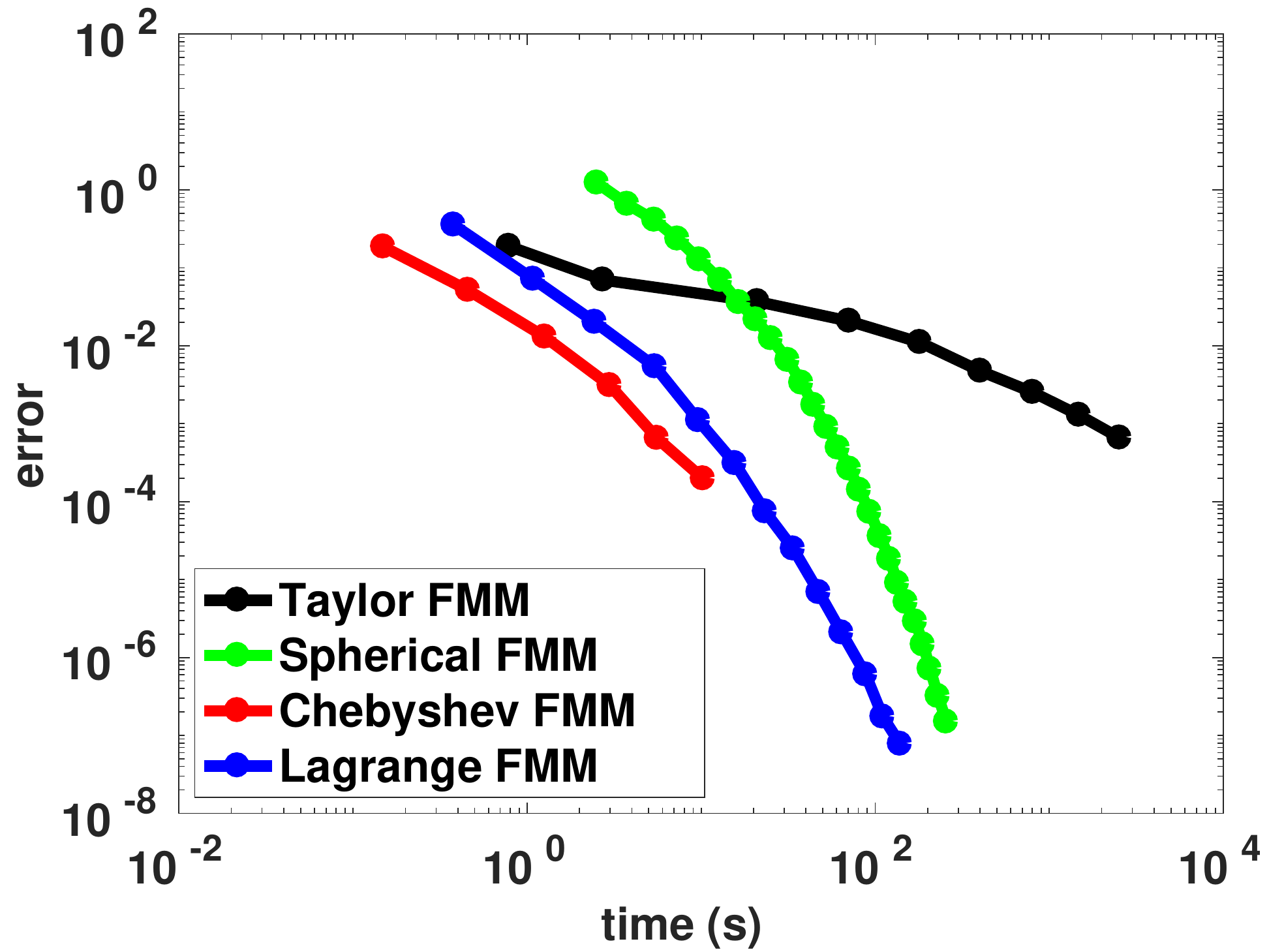}}
\subfigure[Running time vs.\ FMM order]{\includegraphics[width=6.7cm]{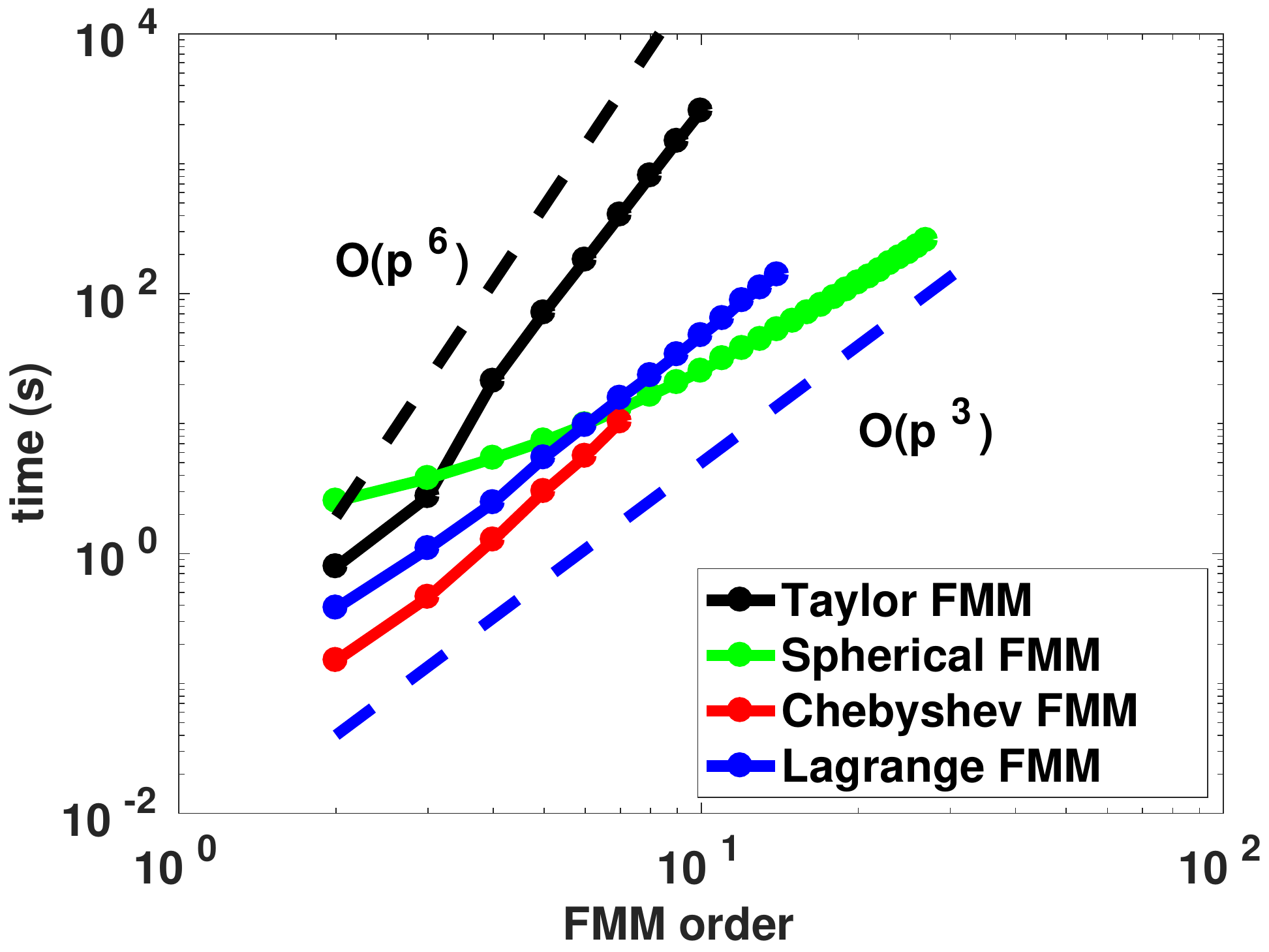}}
\caption{FMM running time in isotropic elastic media. The problem configuration is shown in \autoref{fig:BigLoop}, and all experiments used a fixed $8 \times 8 \times 8$ FMM cells.
The Taylor FMM was stopped at order 10 due to its large running time (42 minutes/run at order 10).
The Chebyshev FMM was stopped at order 7 due to its large pre-computation cost (1.7 GB at order 7).
The Lagrange FMM was stopped at order 14 when the error stopped decreasing.}
\label{fig:FMMsIso}
\begin{center}
\hspace*{\fill}
\subfigure[Chebyshev FMM]{\includegraphics[width=6.7cm]{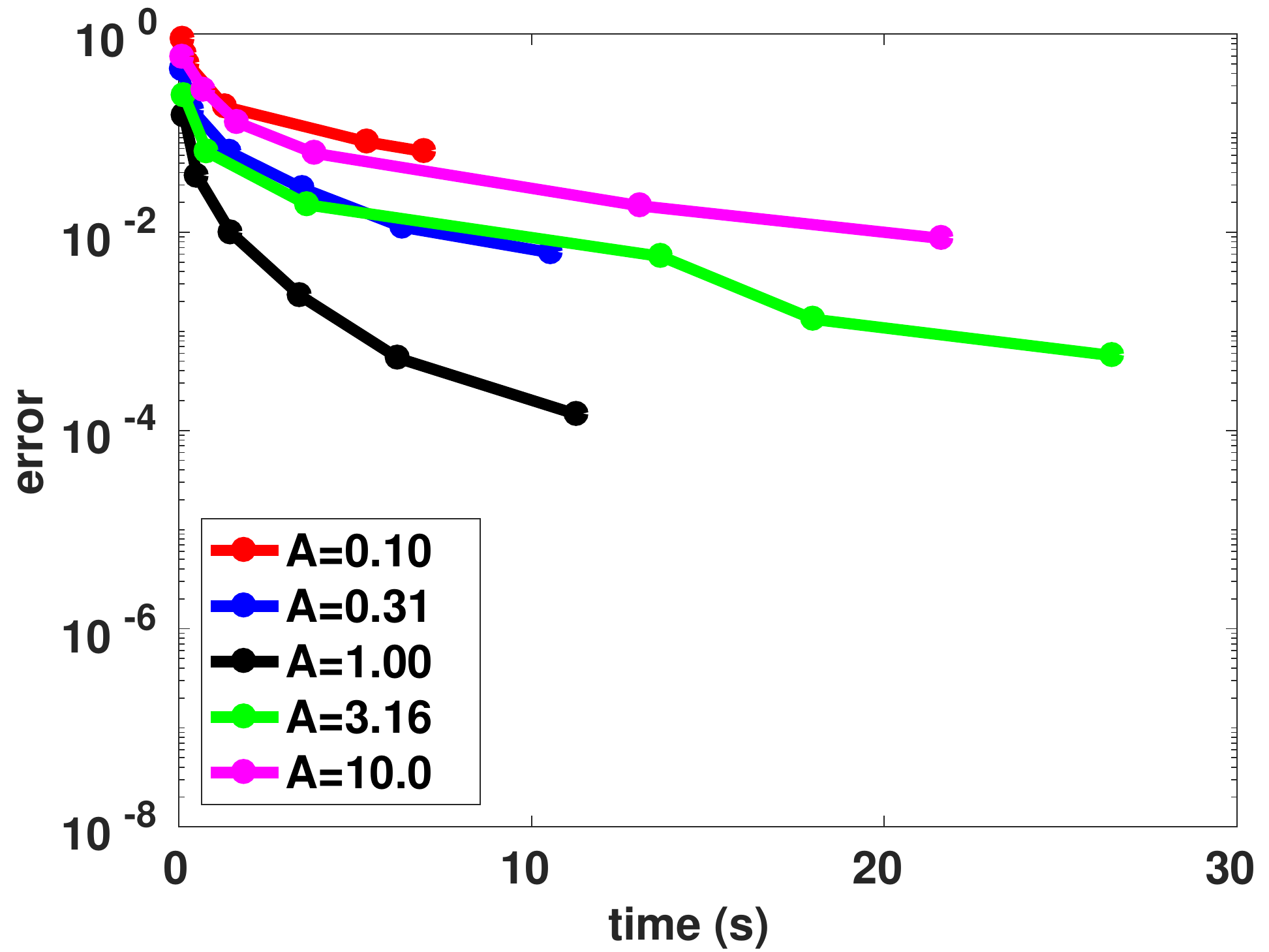}}
\hfill
\subfigure[Lagrange FMM]{\includegraphics[width=6.7cm]{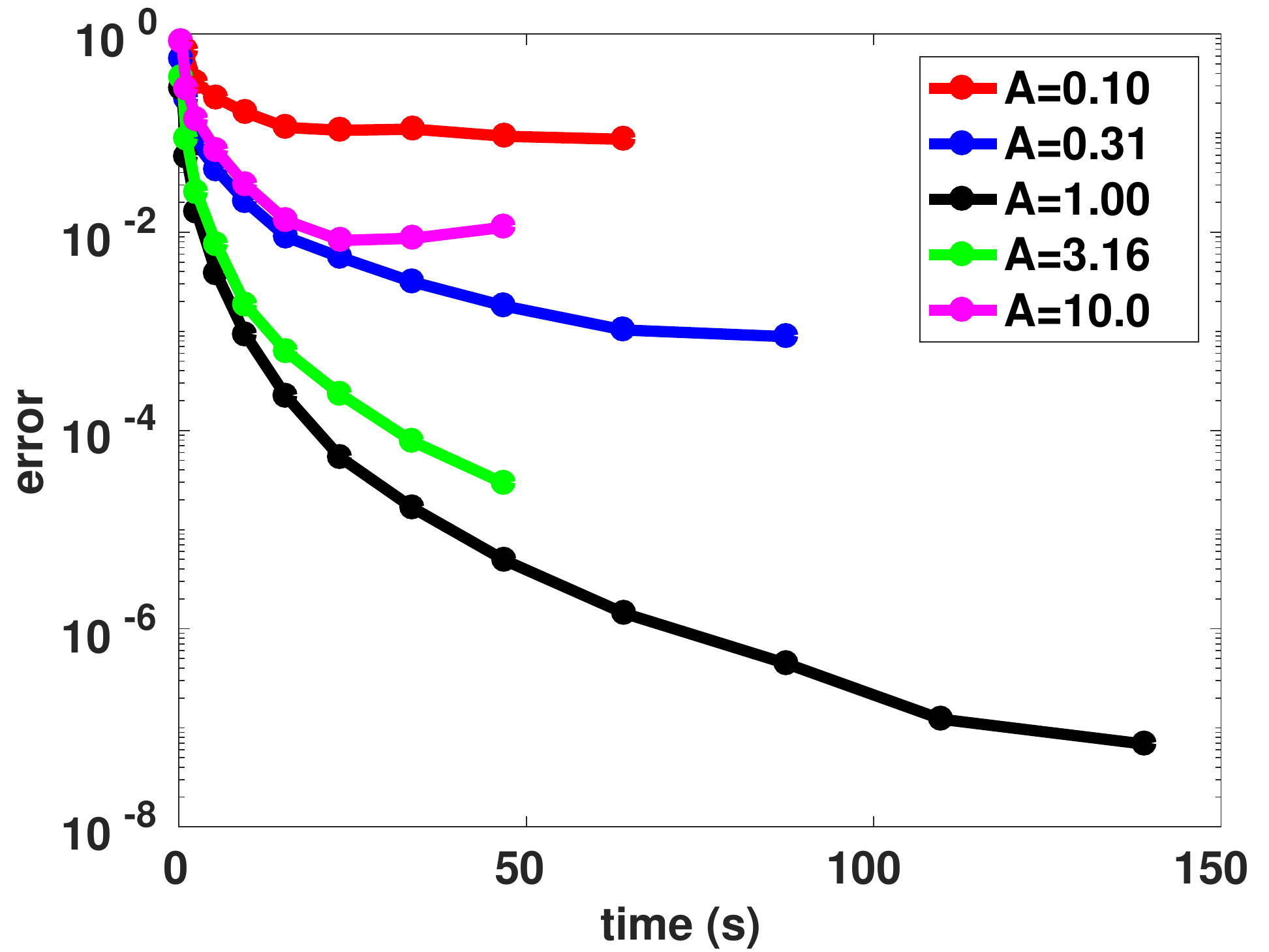}}
\caption{FMM running time in anisotropic elastic media. The problem configuration is shown in \autoref{fig:BigLoop}, and all experiments used a fixed $8 \times 8 \times 8$ FMM cells.
The Taylor FMM and the Spherical FMM are not applicable in this case.
The Chebyshev FMM was stopped at order 7 due to its large pre-computation storage.
The Lagrange FMM was stopped when the error stopped decreasing.}
\label{fig:FMMsAniso}
\end{center}
\end{figure}

In isotropic elastic media, the running time is shown in \autoref{fig:FMMsIso}. In \autoref{fig:FMMsIso}(a), the less time a method takes to attain a prescribed calculation error, the more efficient it is. For example, to reach an error of $10^{-2}$, the Chebyshev FMM takes only about a second, which is the smallest among the four methods. Therefore, with a prescribed error of $10^{-2}$, the Chebyshev FMM is the most efficient method in isotropic elastic media. Following the same logic, we conclude that for DD simulations in isotropic elastic media, 
\begin{enumerate}
\item when the prescribed error is larger than $10^{-4}$, the Chebyshev FMM is the most efficient method; \item when the prescribed error is smaller than $10^{-4}$, the Lagrange FMM is the most efficient method.
\end{enumerate}

\autoref{fig:FMMsIso}(b) shows the computational complexity of the four methods as the FMM order increases. (See \autoref{table:fmm_cost} for theoretical estimates.) As the figure shows, the Taylor FMM and the Lagrange FMM scales as $O(p^6)$ and $O(p^3)$, respectively. The Spherical FMM also converges to a complexity of $O(p^3)$, and it involves a smaller constant than that of the Lagrange FMM. Empirically, the computational complexity of the Chebyshev FMM is about $O(p^5)$.

\begin{figure}  
\begin{center}
\hspace*{\fill}
\subfigure[$A=0.10$]{\includegraphics[width=6.7cm]{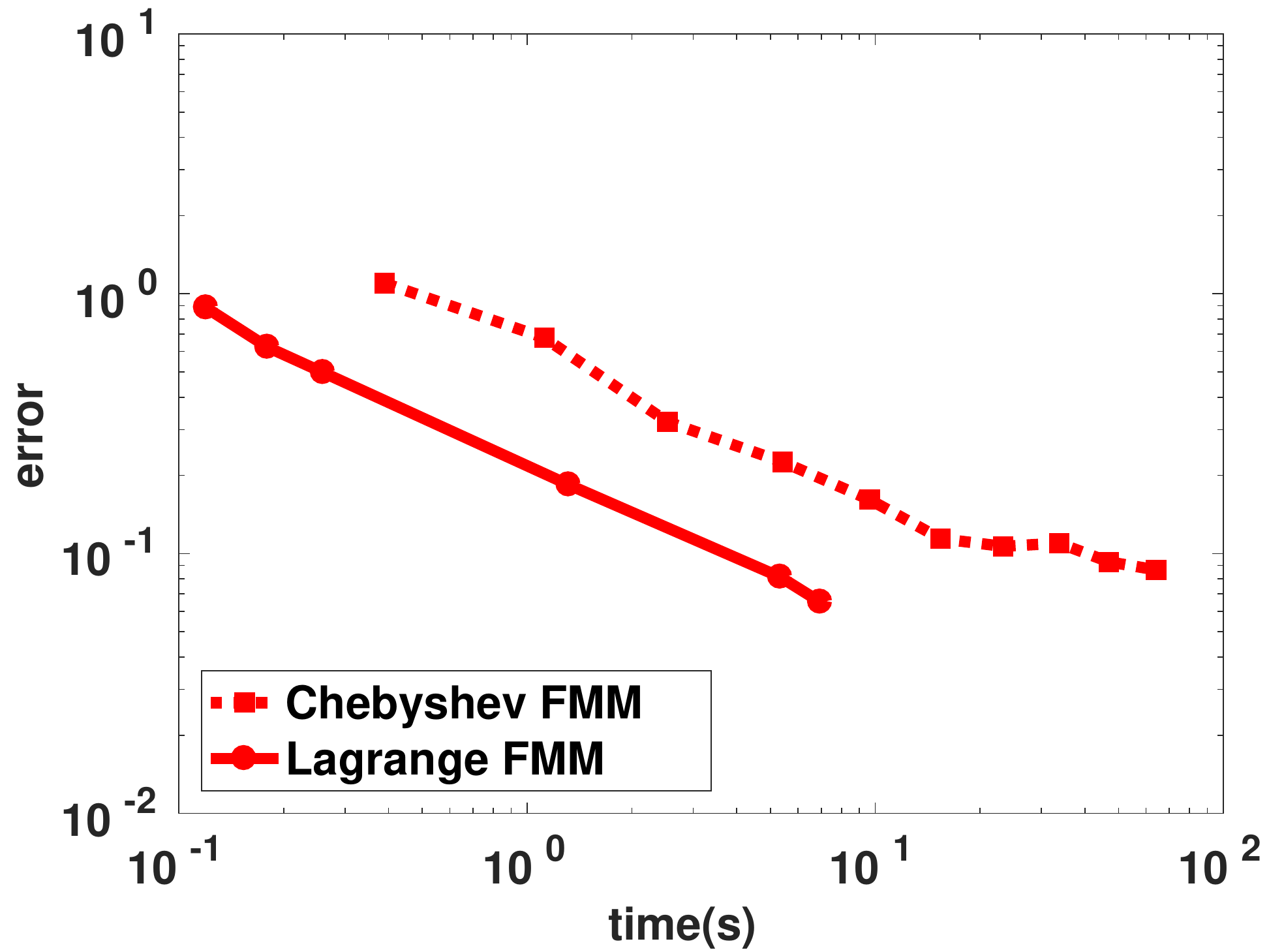}}
\hfill
\subfigure[$A=0.31$]{\includegraphics[width=6.7cm]{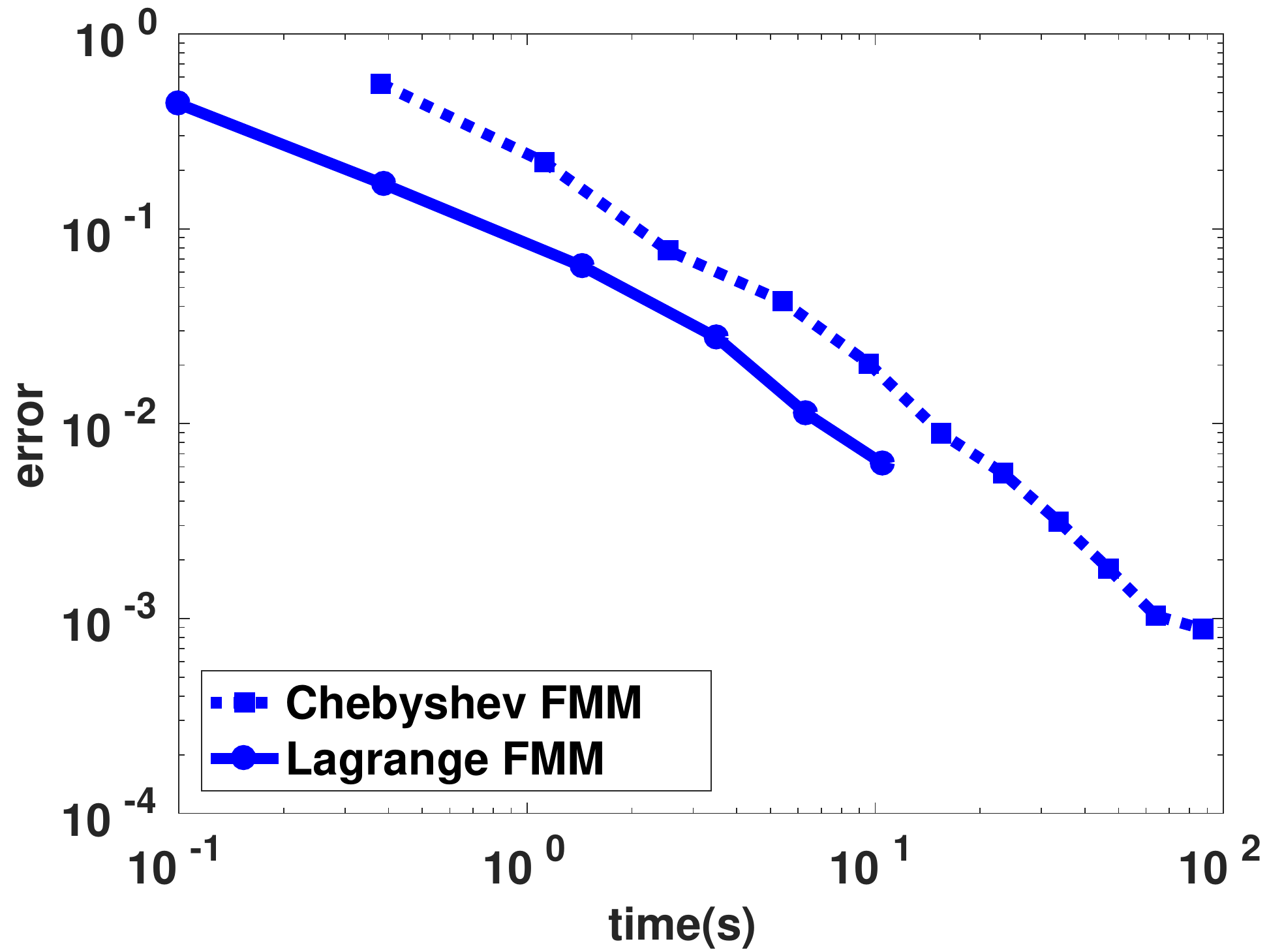}}
\hspace*{\fill}
\subfigure[$A=1.00$]{\includegraphics[width=6.7cm]{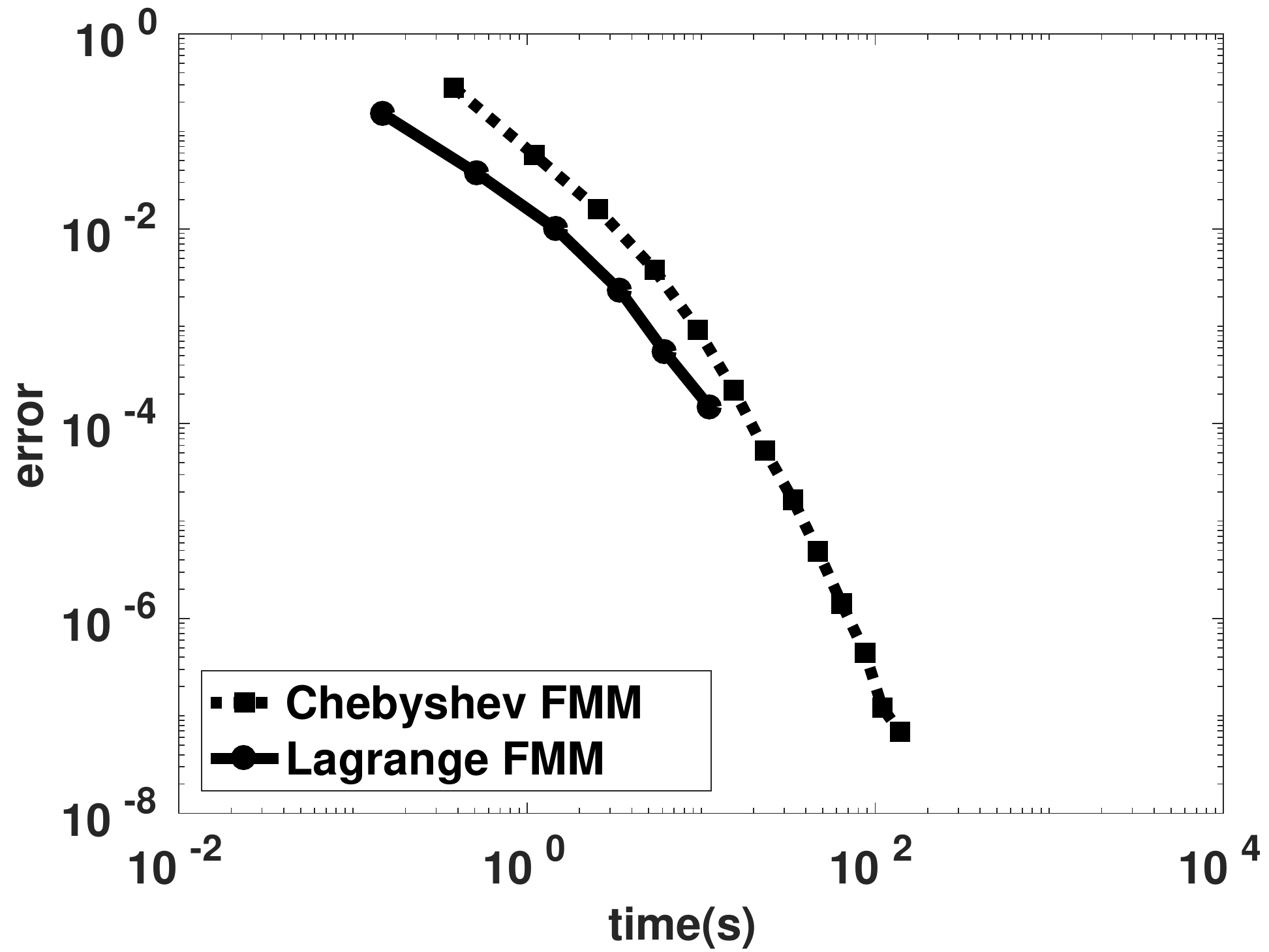}}
\hfill
\subfigure[$A=3.16$]{\includegraphics[width=6.7cm]{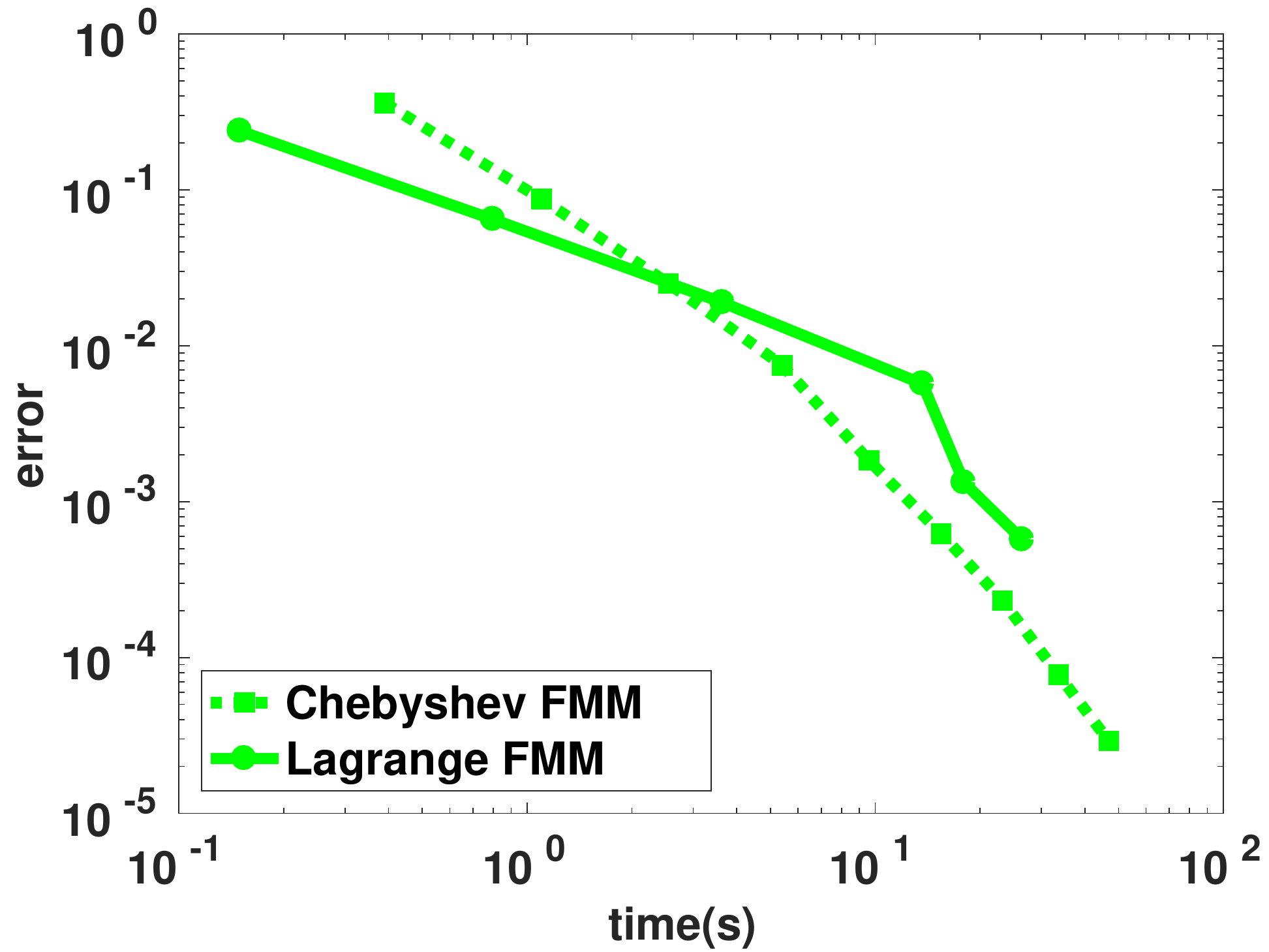}}
\centering
\subfigure[$A=10.0$]{\includegraphics[width=6.7cm]{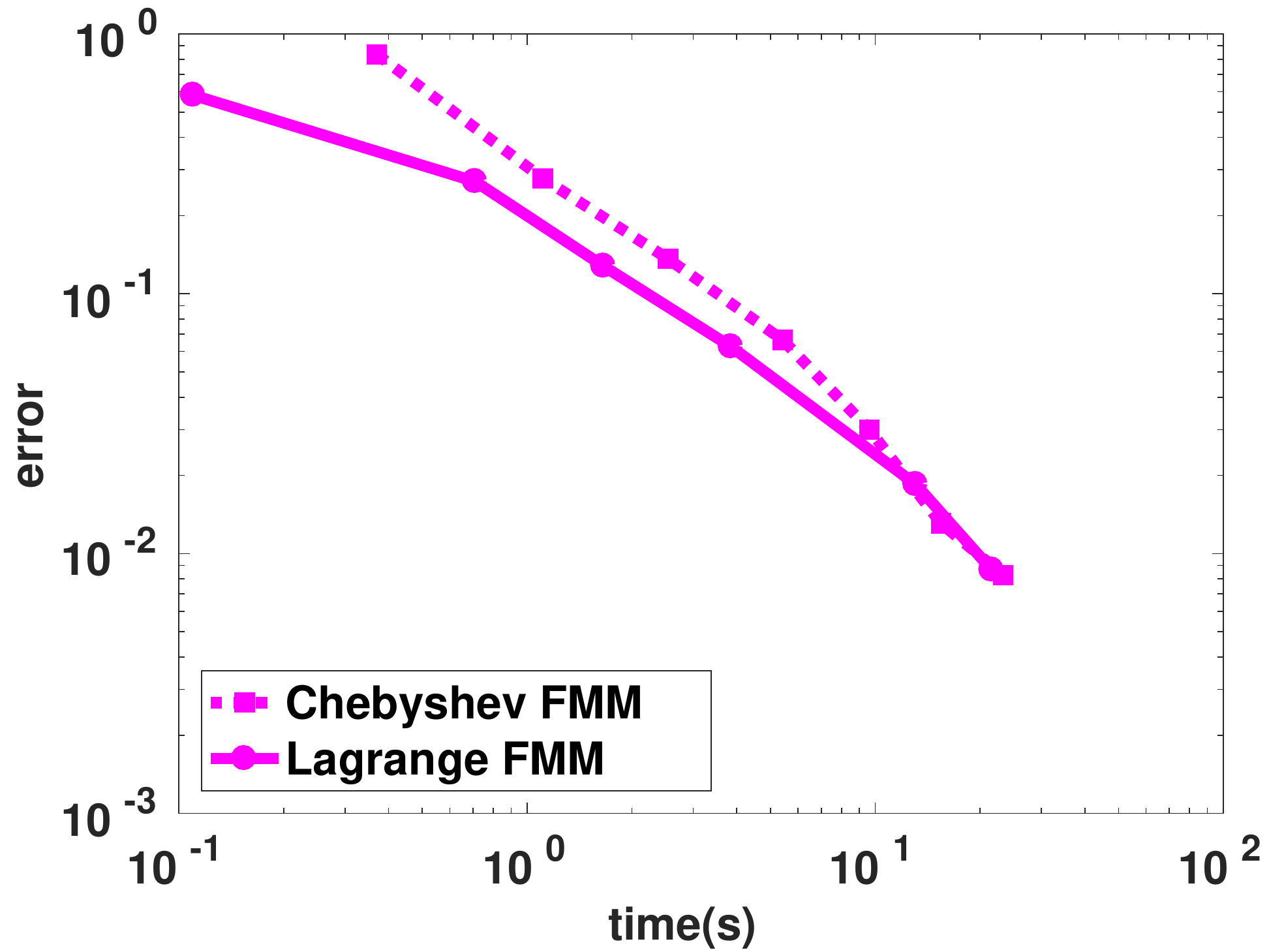}}
\caption{FMM running time in anisotropic elastic media. The results are exactly the same as in \autoref{fig:FMMsAniso}, but plotted for each anisotropic ratio to compare the Chebyshev FMM and the Lagrange FMM.}
\label{fig:FMMsAnisoRatio}
\end{center}
\end{figure}

In anisotropic elastic media, the running time is shown in \autoref{fig:FMMsAniso} and \autoref{fig:FMMsAnisoRatio} for five anisotropic ratios: $A=0.1$, $0.31$, $1.0$, $3.16$, $10$ ($A=1$ is equivalent to isotropic elasticity). \autoref{fig:FMMsAnisoRatio} compares the efficiency of the Chebyshev FMM and the Lagrange FMM. Note that the costs of near-field interactions between neighboring cells were subtracted intensionally from these timing results for the following two reasons. First, the subtraction of near-field costs does not affect the comparison between the Chebyshev FMM and the Lagrange FMM because the costs are the same for both methods. Second, as shown in \autoref{fig:FMMsAniso}(b), the subtraction of near-field costs shows that when the FMM order is fixed, the running time of the Lagrange FMM is the same for different anisotropic ratios. The reason is that the cost of evaluating the anisotropic kernel functions with different anisotropic ratios affects only the costs of pre-computing M2L translation operators and evaluating near-field interactions. Unlike in the Lagrange FMM, M2L translation costs in the Chebyshev FMM are different for distinct anisotropic ratios due to using different SVD compression ratios.

As in the isotropic case, the more time a method takes to reach a prescribed calculation error, the less efficient it is. As the figure shows,
\begin{enumerate}
\item when $A=0.1$, the Chebyshev FMM is always more efficient than the Lagrange FMM; 
\item when $A=0.31$ and $1.0$, the Chebyshev FMM is more efficient than the Lagrange FMM for errors that the Chebyshev FMM can achieve. With limited computer memory, the Lagrange FMM can reach smaller errors than the Chebyshev FMM; 
\item when $A=3.16$, the Lagrange FMM is not as efficient as the Chebyshev FMM for an error larger than $0.02$, but becomes more efficient for errors smaller than $0.02$. With limited computer memory, the Lagrange FMM can reach smaller errors than the Chebyshev FMM; 
\item when $A=10$, the Lagrange FMM is not as efficient as the Chebyshev FMM initially, but becomes comparable for an error of $0.01$.
\end{enumerate}

\subsection{Optimal running time}\label{subsec:linear}

%

We present the optimal/minimum running time, which is a function of the number of dislocation segments, with respect to the number of FMM cells. Given the number of dislocation segments, the FMM running time is a function of the number of FMM cells -- as the number of FMM cells increases, the cost for evaluating near-field interactions decreases, whereas the cost for evaluating far-field interactions increases. Therefore, the number of FMM cells need to be chosen appropriately to balance the near-field and far-field costs in order to achieve the minimum FMM running time.

We compute the optimal running time following two steps. First, we fix the number of FMM cells and compute the running time as a function of the number of dislocation segments based on the costs of near-field and far-field translations, which can be measured accurately. Second, for a given number of dislocation segments, we find the minimum among all running time corresponding to different number of FMM cells. A detailed mathematical derivation of the optimal running time is given in Appendix B.

\begin{figure}[b]
\begin{center}
\includegraphics[width=6.7cm]{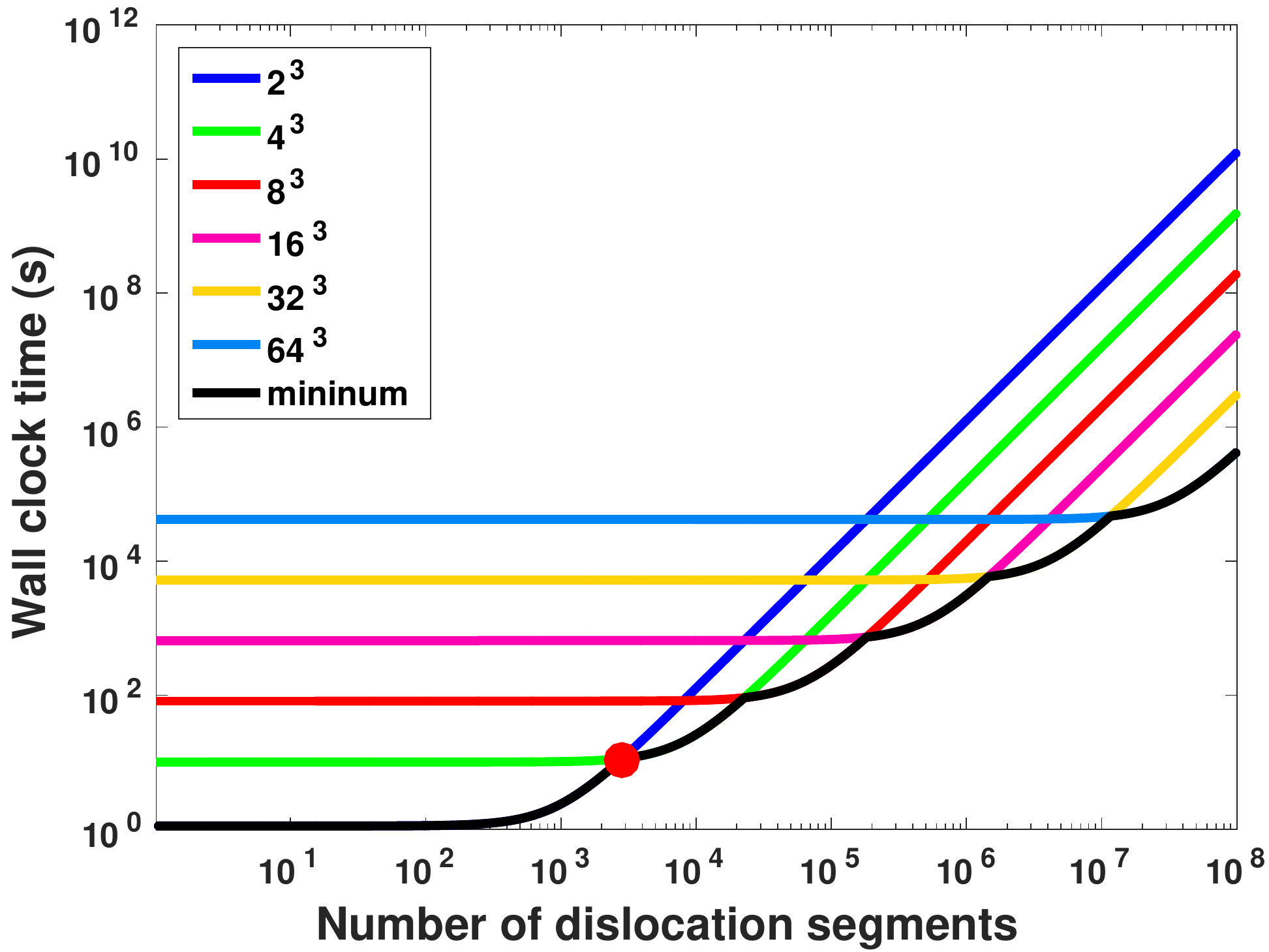}
\caption{Optimal running time of the 4-th order Taylor FMM. Every color stands for the running time obtained with a specific number of FMM cells, and the black curve represents the optimal running time. The red dot shows that the minimum running time is about $10$ seconds for $3,000$ dislocation segments.}
\label{fig:CostExplain}
\end{center}
\end{figure}

We use the $4$-th order Taylor FMM as an example to illustrate how we obtained the optimal running time. As shown in \autoref{fig:CostExplain}, when the number of FMM cells is fixed, the FMM running time is initially dominated by the constant cost for evaluating far-field translations. As the number of dislocation segments increases, the quadratic cost for evaluating near-field interaction starts to dominate the total running time. In the figure, the black curve corresponds to the optimal running time, the minimum of all curves.

\begin{figure}[b]
\begin{center}
\hspace*{\fill}
\subfigure[Error: $10^{-1}$]{\includegraphics[width=6.7cm]{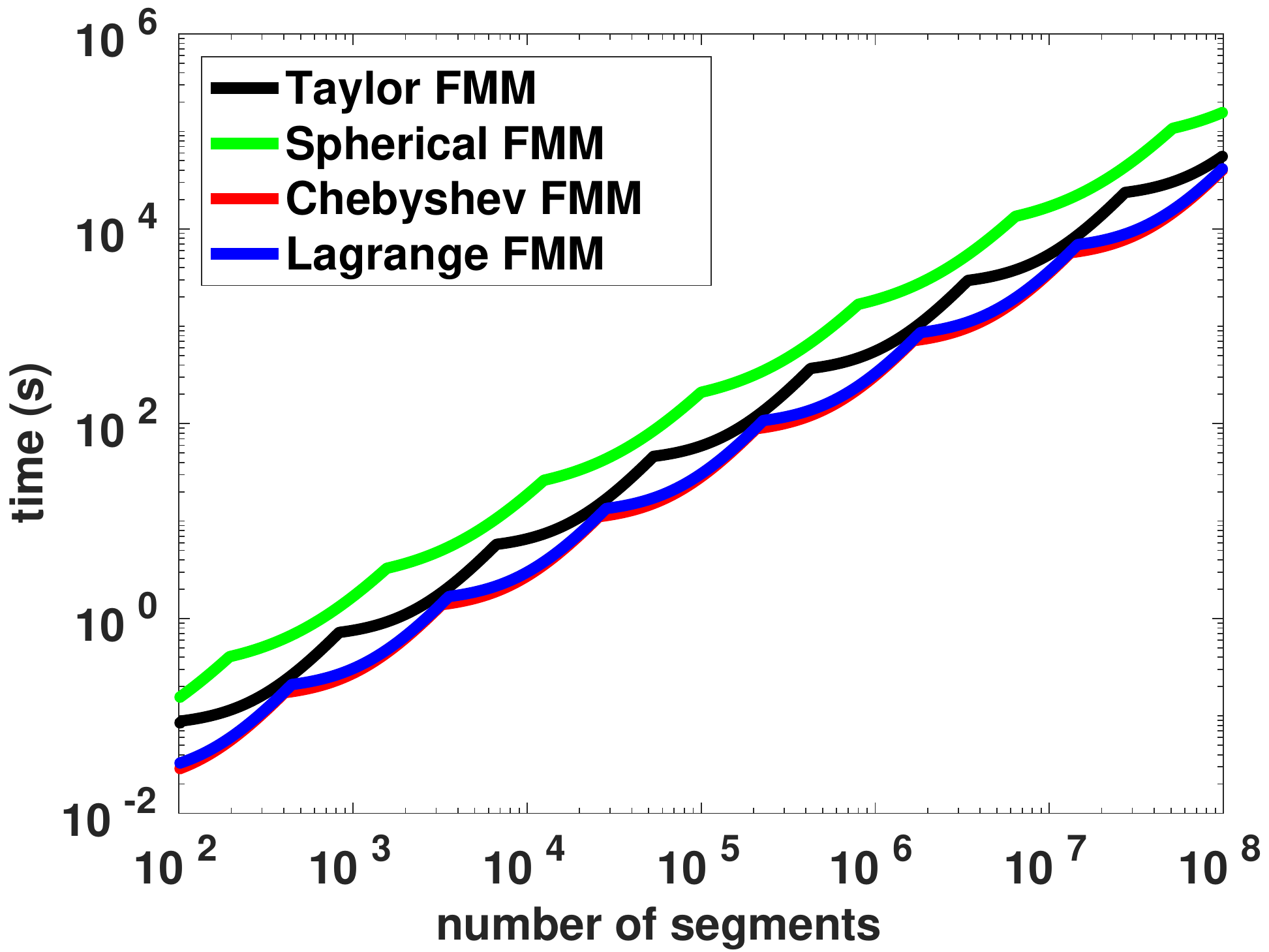}}
\hfill
\subfigure[Error: $10^{-2}$]{\includegraphics[width=6.7cm]{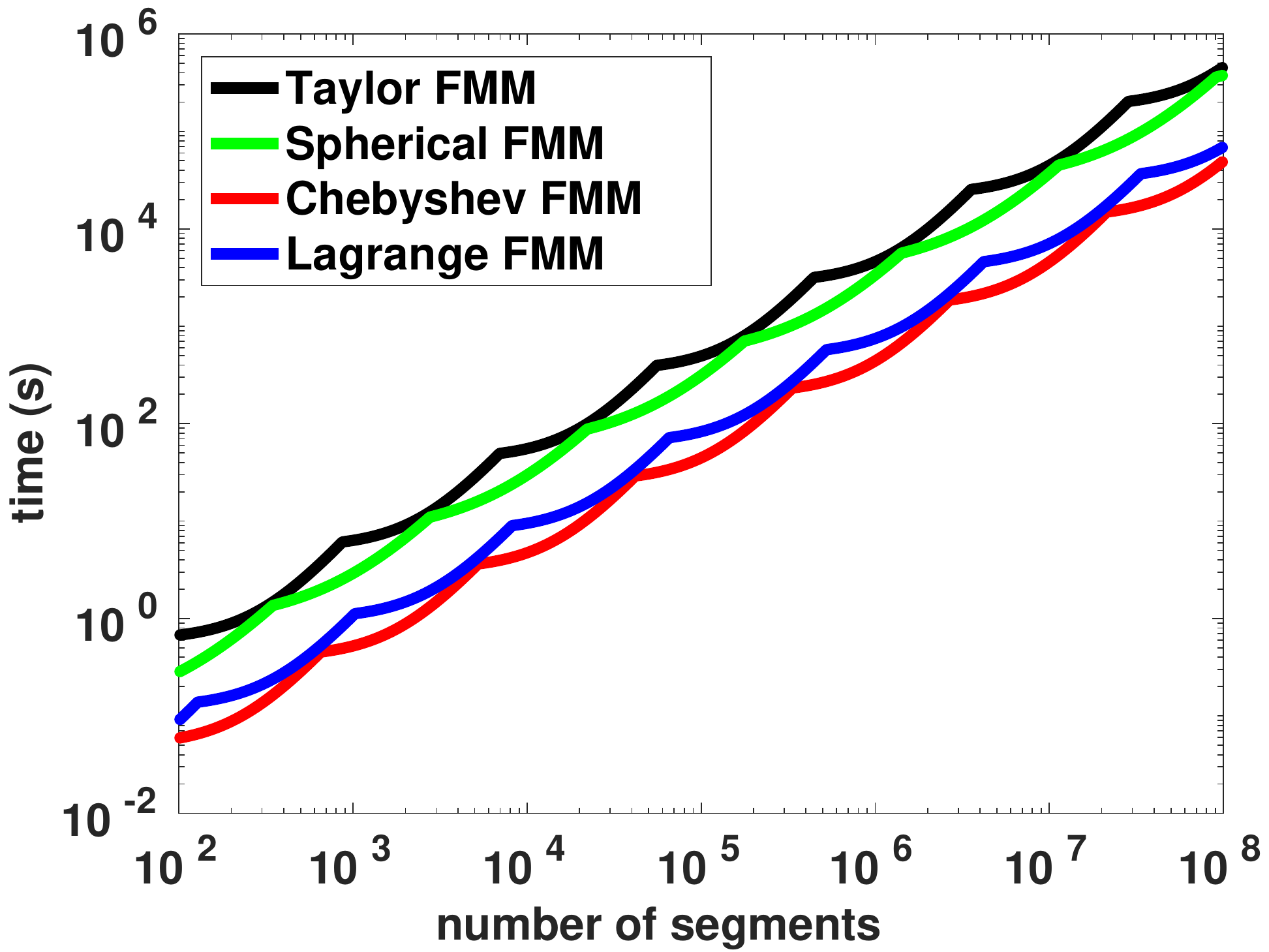}}
\hspace*{\fill}
\subfigure[Error: $10^{-4}$]{\includegraphics[width=6.7cm]{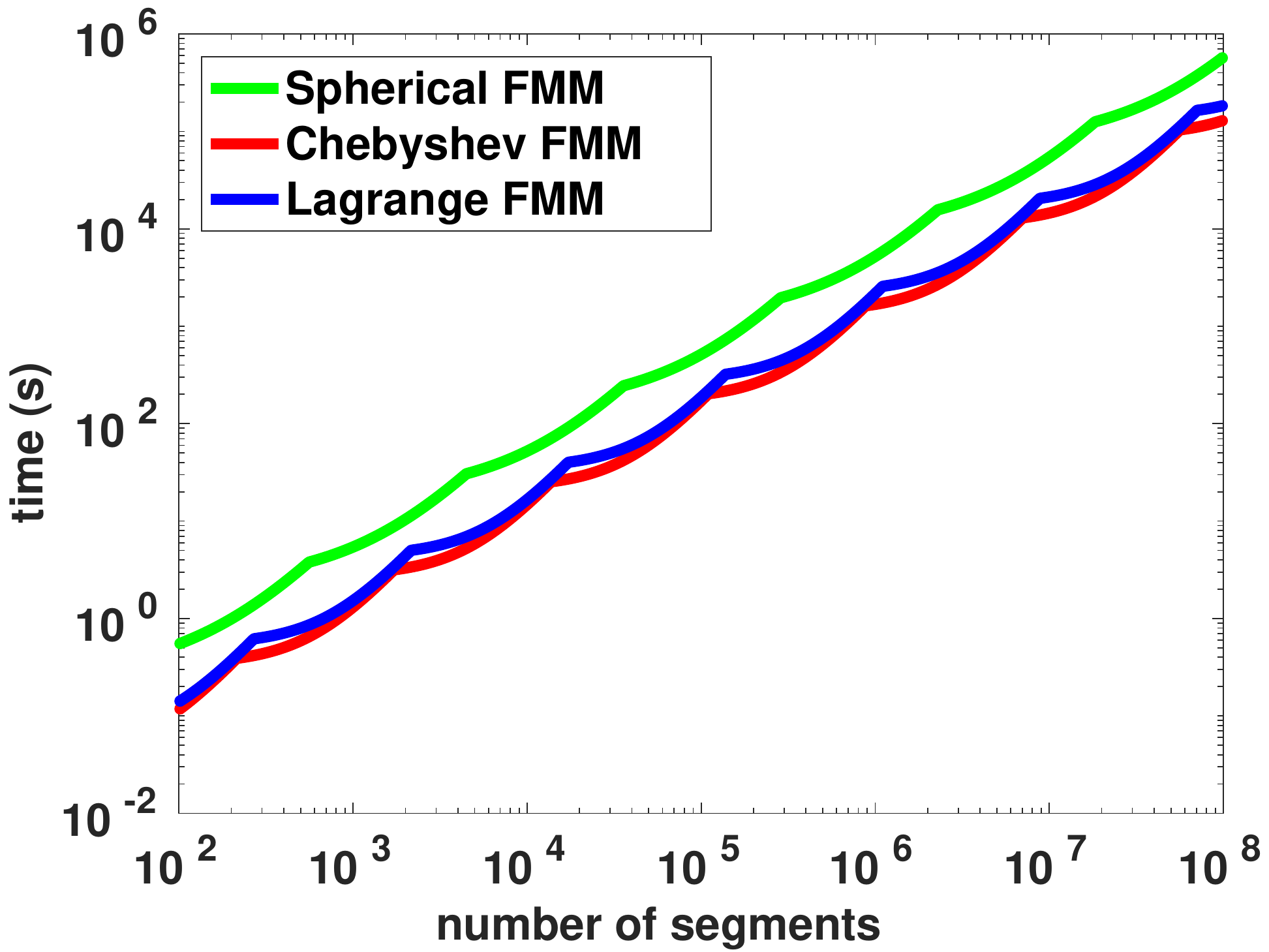}}
\hfill
\subfigure[Error: $10^{-7}$]{\includegraphics[width=6.7cm]{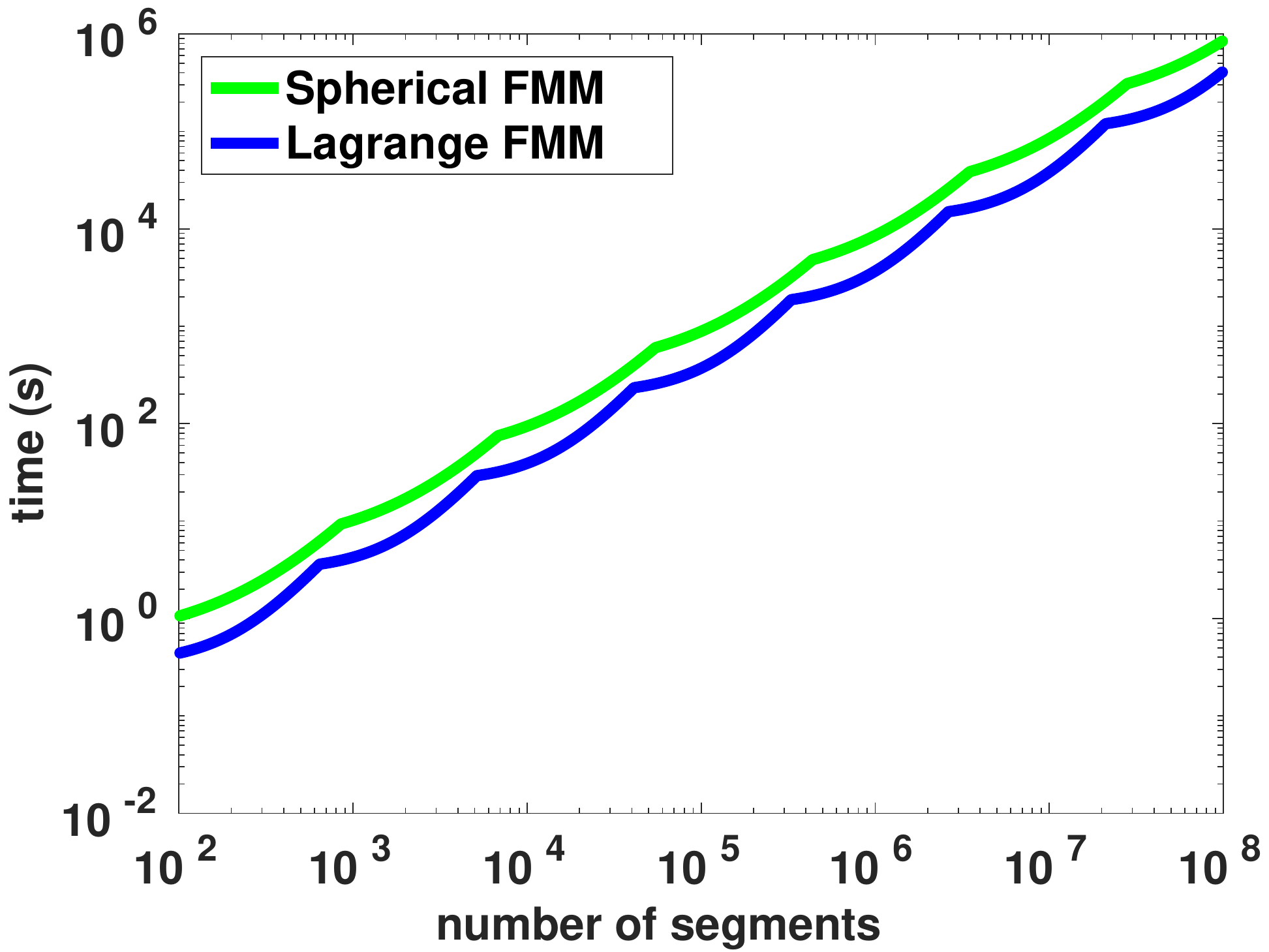}}
\hspace*{\fill}
\caption{Optimal running time (i.e., we optimize the number of FMM cells) of the four methods for different prescribed calculation errors. In (c) and (d), the Taylor FMM is not shown due to its large running time (42 minutes/run at order 10). In (d), the Chebyshev FMM is not shown due to its large pre-computation storage (1.7 GB at order 7).}
\label{fig:CostComp}
\end{center}
\end{figure}  

\begin{figure}[htbp]
\begin{center}
\hspace*{\fill}
\subfigure[`Chebyshev': $A=0.10$]{\includegraphics[width=6.5cm]{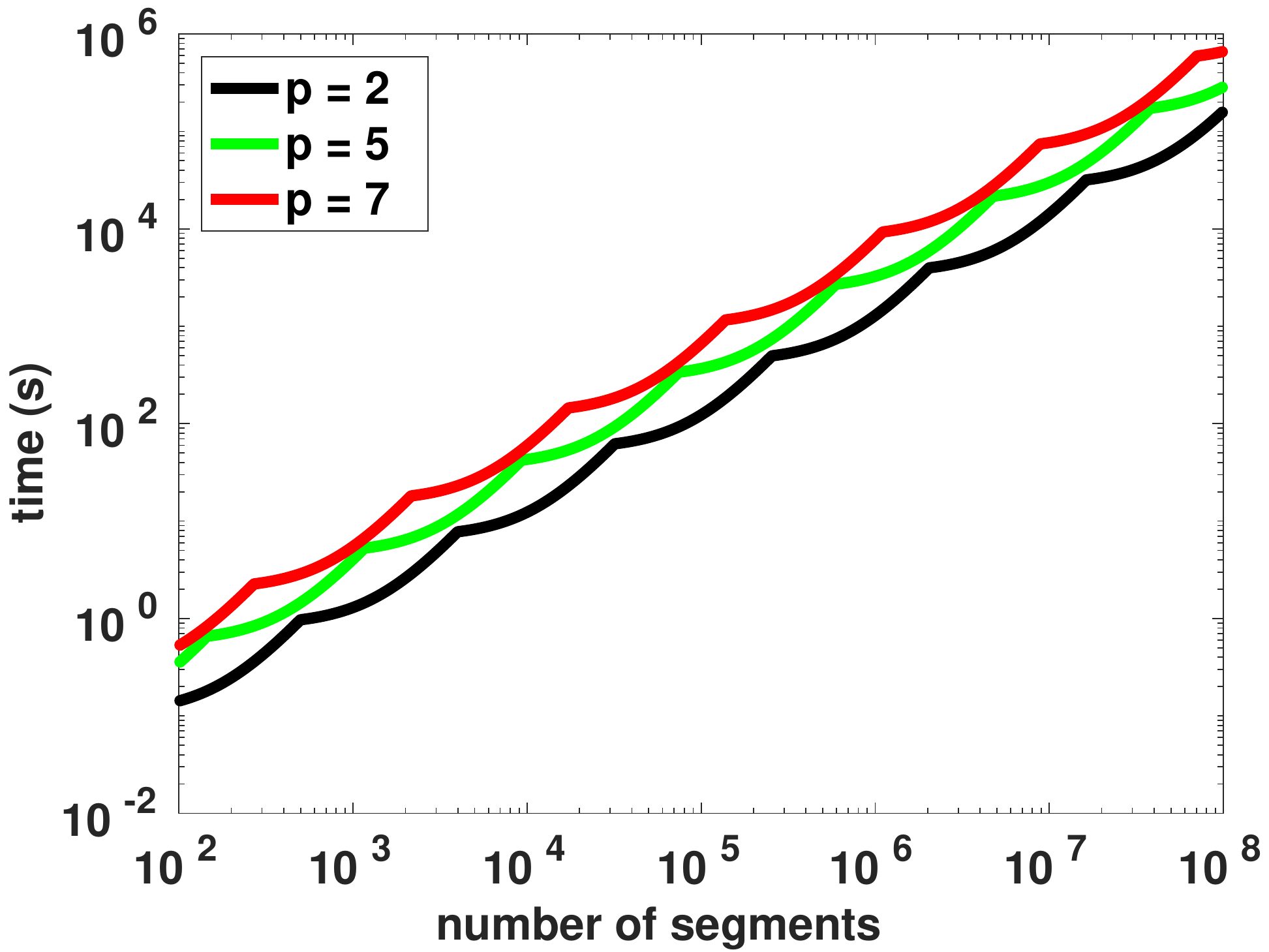}}
\hfill
\subfigure[`Chebyshev': $A=0.31$]{\includegraphics[width=6.5cm]{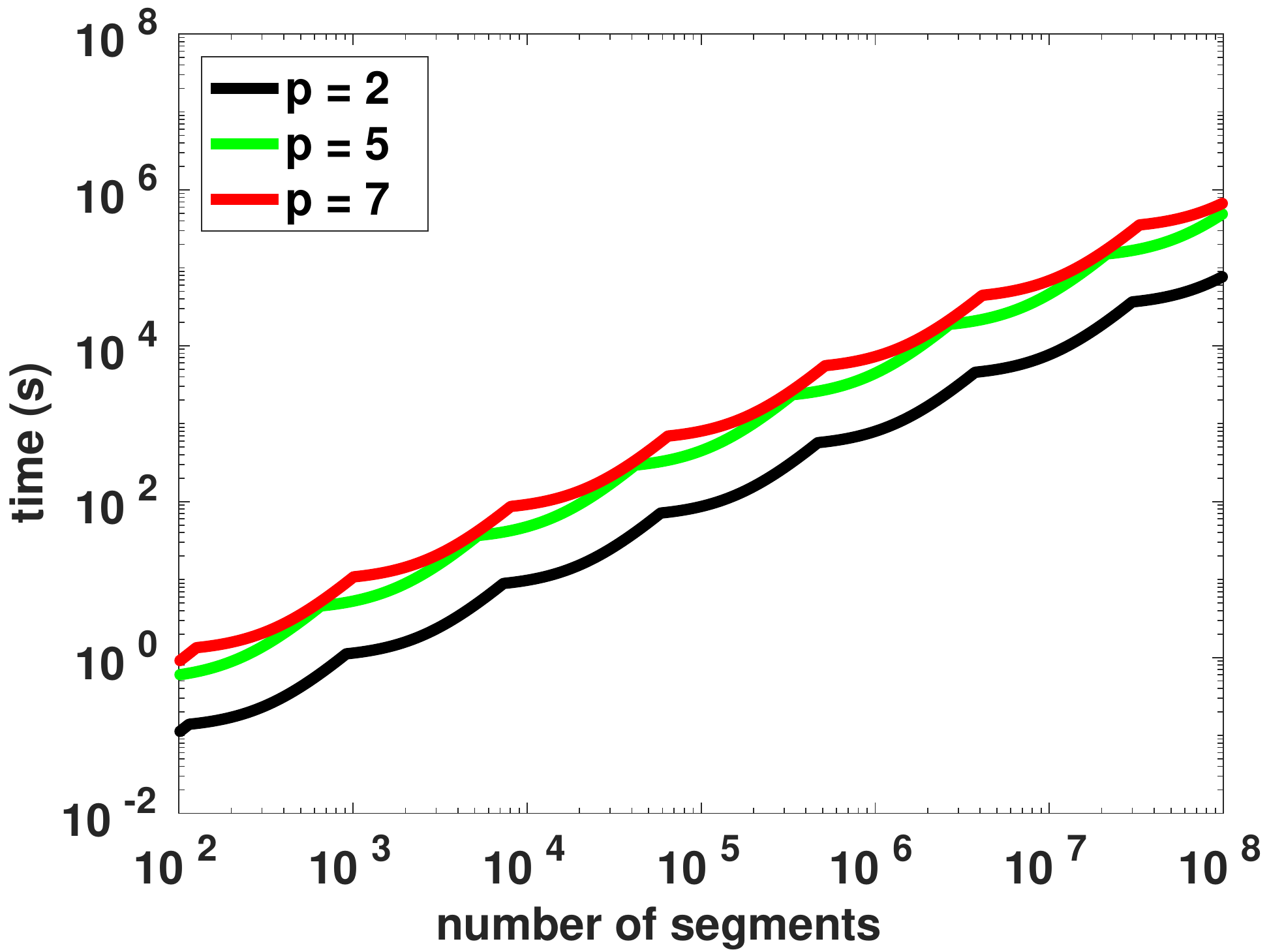}}
\hspace*{\fill}
\subfigure[`Chebyshev': $A=3.16$]{\includegraphics[width=6.5cm]{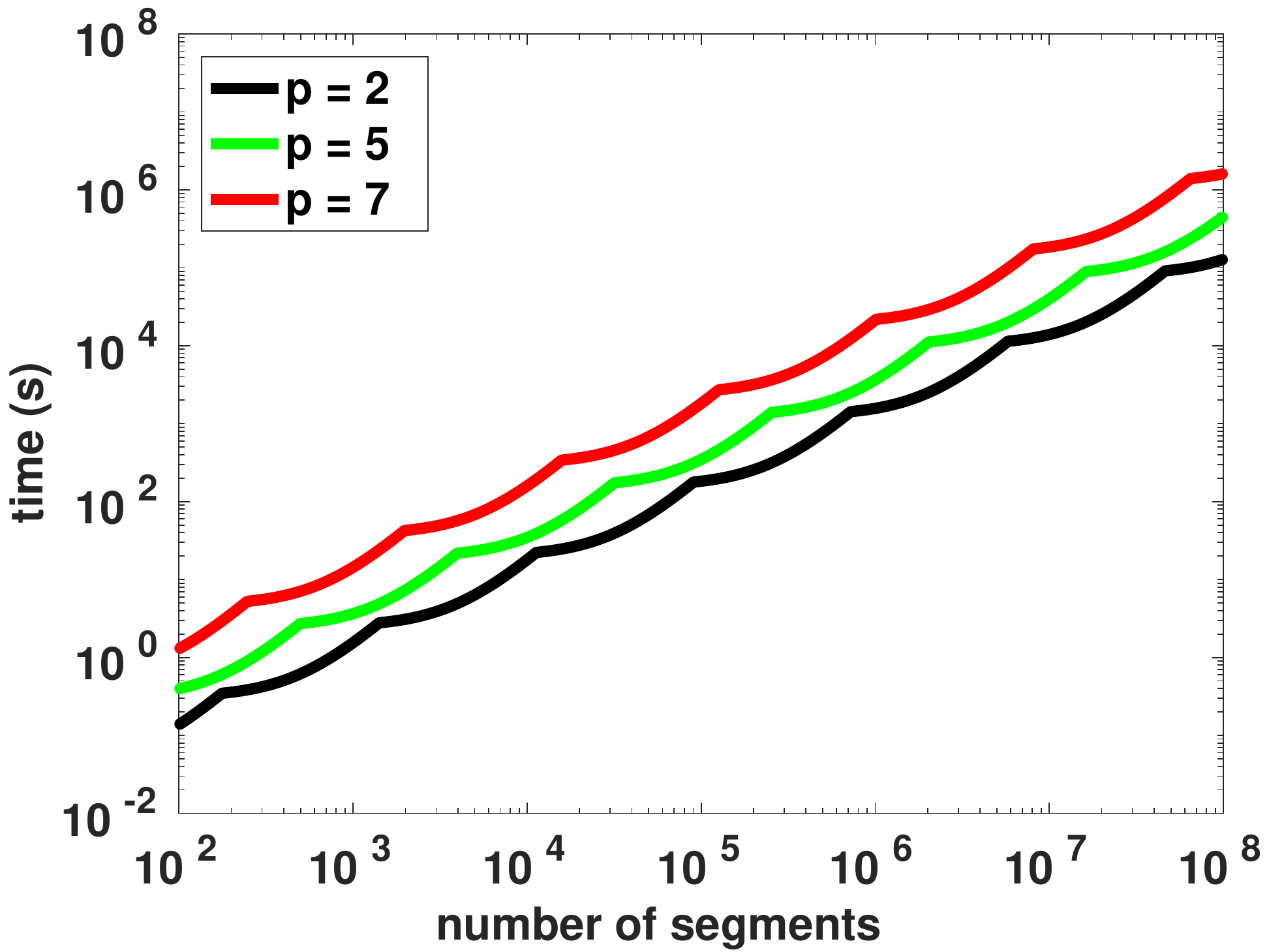}}
\hfill
\subfigure[`Chebyshev': $A=10.0$]{\includegraphics[width=6.5cm]{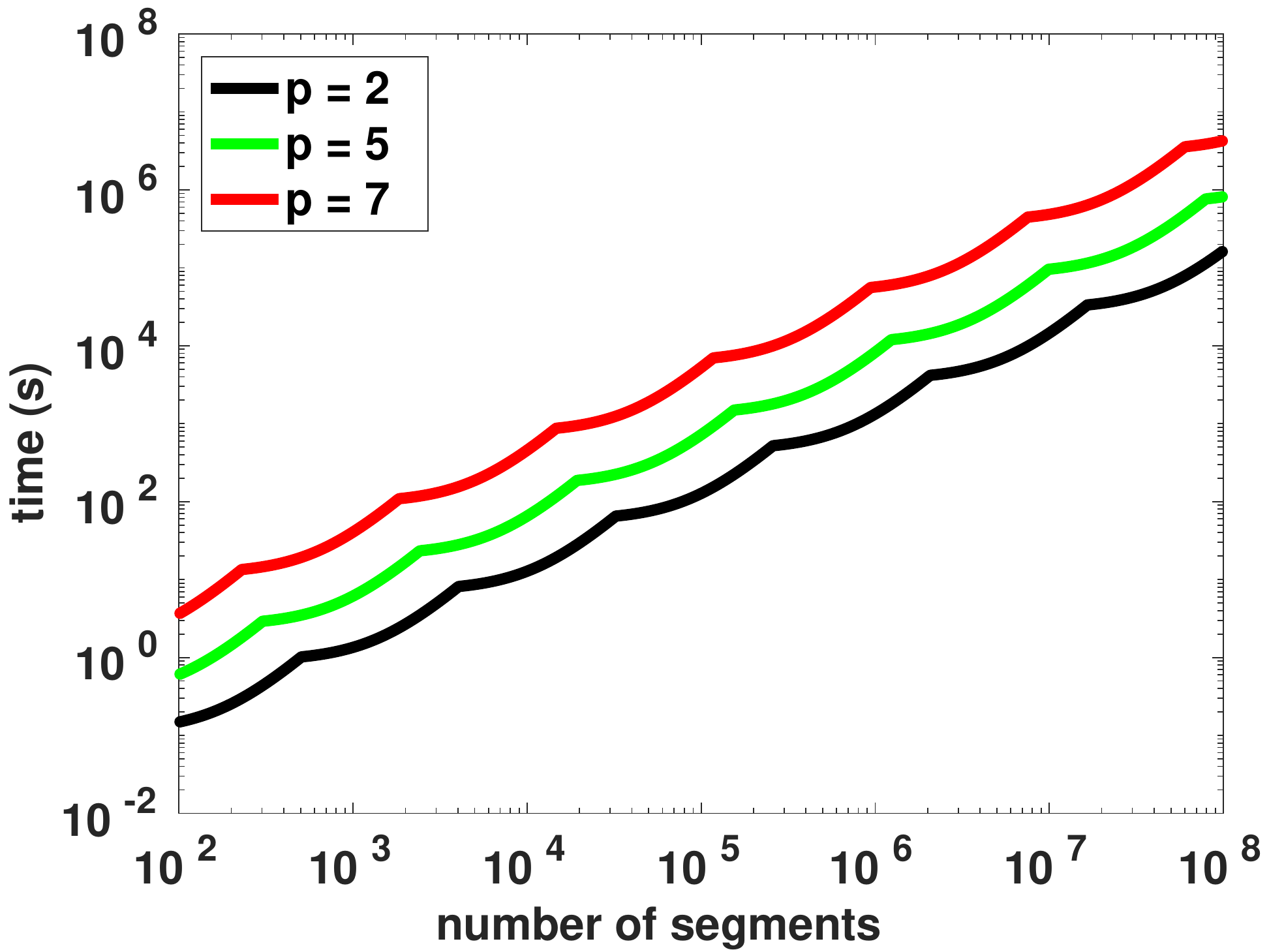}}
\hspace*{\fill}
\subfigure[`Lagrange': $A=0.31$ and $3.16$]{\includegraphics[width=6.5cm]{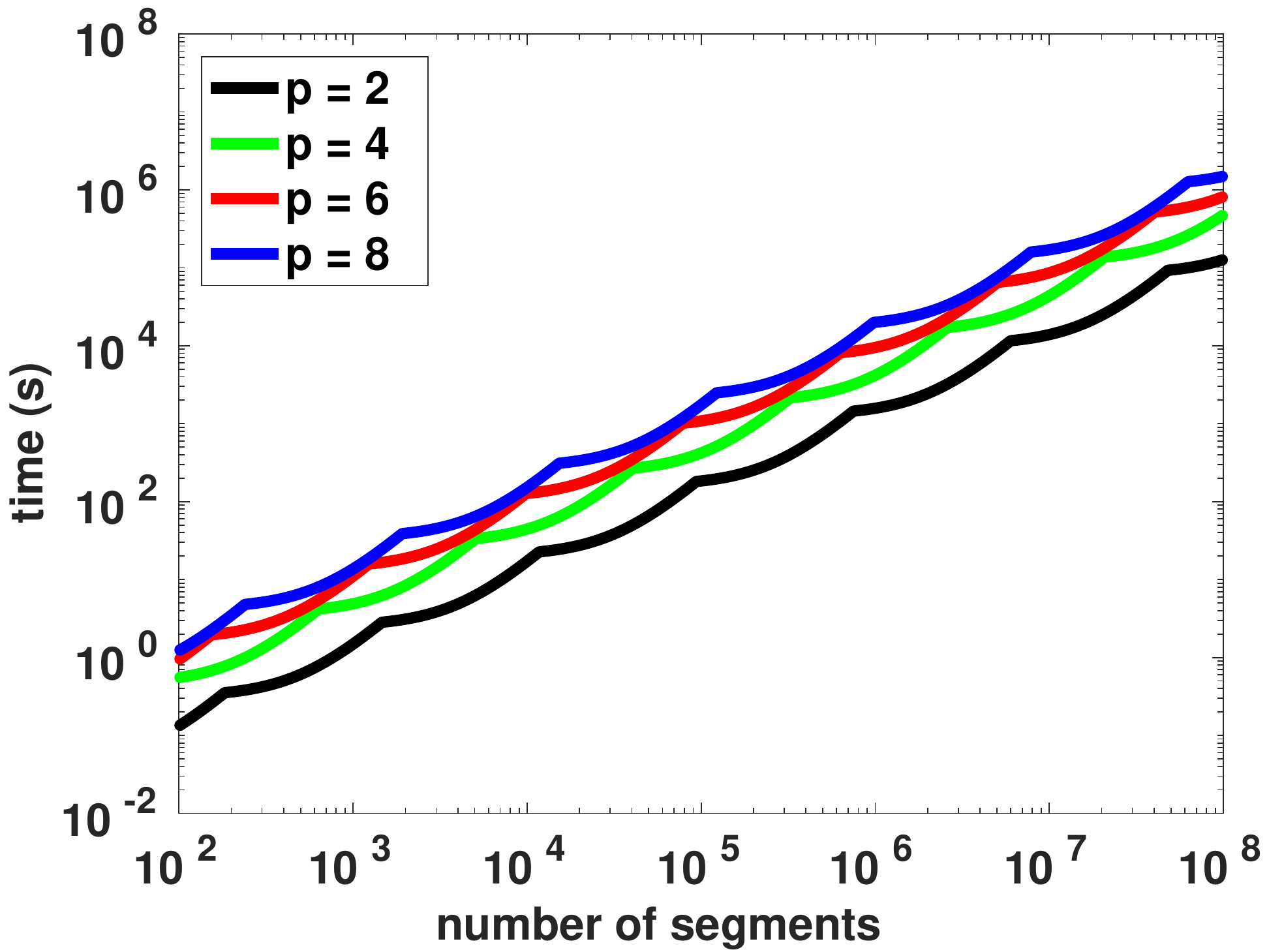}}
\hfill
\subfigure[`Lagrange': $A=0.10$ and $10.0$]{\includegraphics[width=6.5cm]{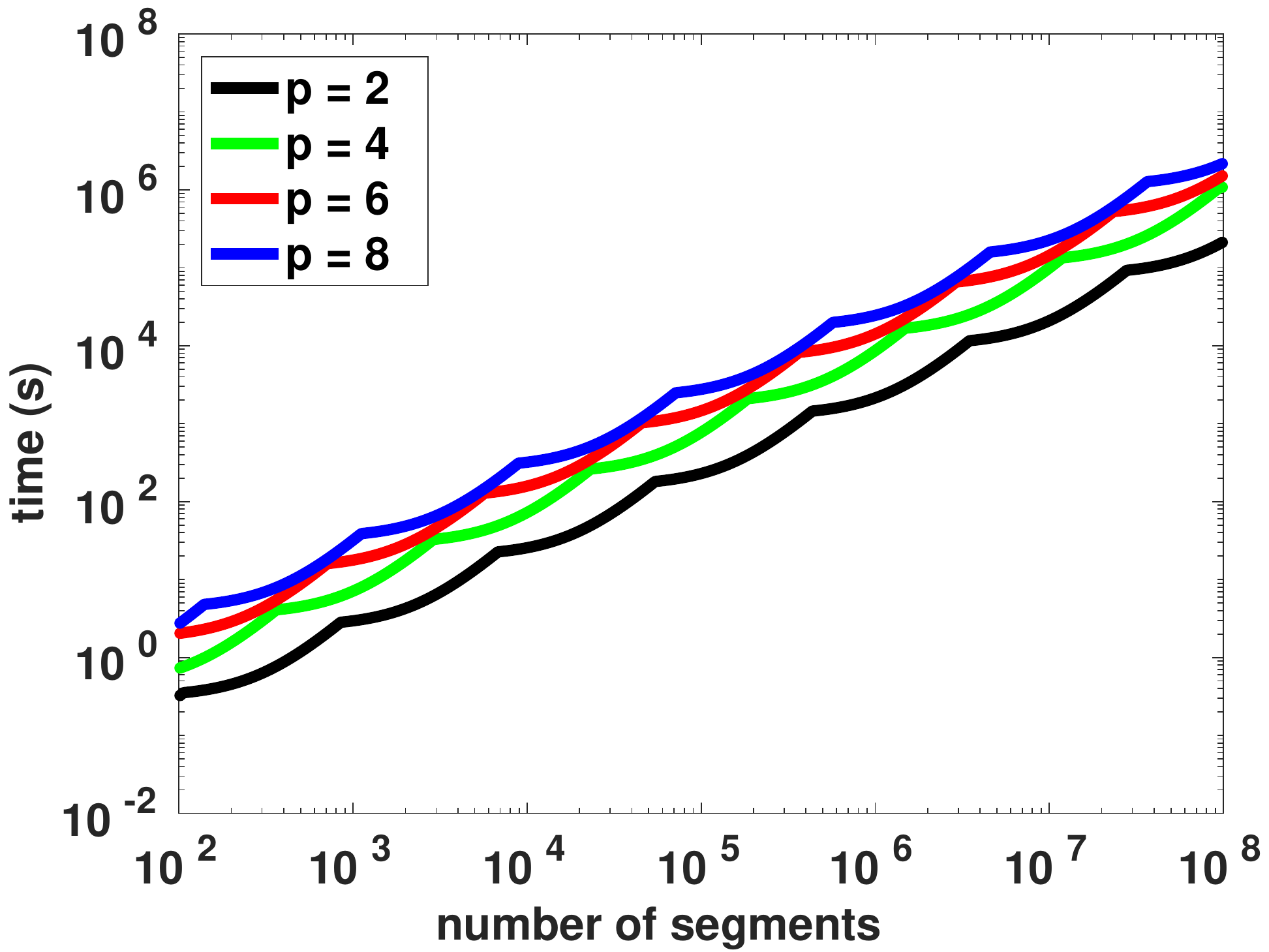}}
\caption{Optimal running time (i.e., we optimize the number of FMM cells) of the Chebyshev FMM and the Lagrange FMM in anisotropic elastic media. 
}
\label{fig:CostAniso}
\end{center}
\end{figure}

In isotropic elastic media, the optimal running time of the four methods is shown in \autoref{fig:CostComp}. The FMM order of each method is chosen such that the stress calculation error reaches four different prescribed errors: $10^{-1}$, $10^{-2}$, $10^{-3}$ and $10^{-7}$. \autoref{fig:CostComp} shows the following results. First, the Taylor FMM is competitive when the prescribed error is around $0.1$. The optimal calculation time of the Taylor FMM can be around the smallest calculation time among the four methods (at the intersection of the black, the blue and the red curve). Second, the Spherical FMM is able to converge to as small as $10^{-7}$ errors. Third, the Chebyshev FMM delivers the optimal calculation time when the prescribed error is larger than $10^{-4}$. Last, the Lagrange FMM is close to being optimal and it is able to converge to as small as $10^{-7}$ errors. 

In anisotropic elastic media, the optimal running time of the Chebyshev FMM and the Lagrange FMM is shown in \autoref{fig:CostAniso}. The Taylor FMM and the Spherical FMM are not applicable in the anisotropic case. For the Lagrange FMM, the results corresponding to $A=0.31$ and $A=3.16$ are the same in \autoref{fig:CostAniso}(e), and the reason is that we used the same $q_{\max}$ in calculating the kernel function for $A=0.31$ and $A=3.16$. The same reasoning holds for the results corresponding to $A=0.1$ and $A=10.0$ in \autoref{fig:CostAniso}(f). In the Chebyshev FMM, M2L translation costs are different for the five anisotropic ratios due to using different SVD compression ratios.

As \autoref{fig:CostAniso} shows, two curves corresponding to different FMM orders may intersect with each other, which implies that the increasing cost of more accurate far-field translation are compensated by using a larger number of FMM cells.

\subsection{Parallel Scalability} \label{subsec:parallel}
\begin{figure}[h]
\begin{center}
\hspace*{\fill}
\subfigure[Isotropic elastic dislocation network]
{\includegraphics[width=6cm, trim= 1cm 6cm 0cm 6cm]{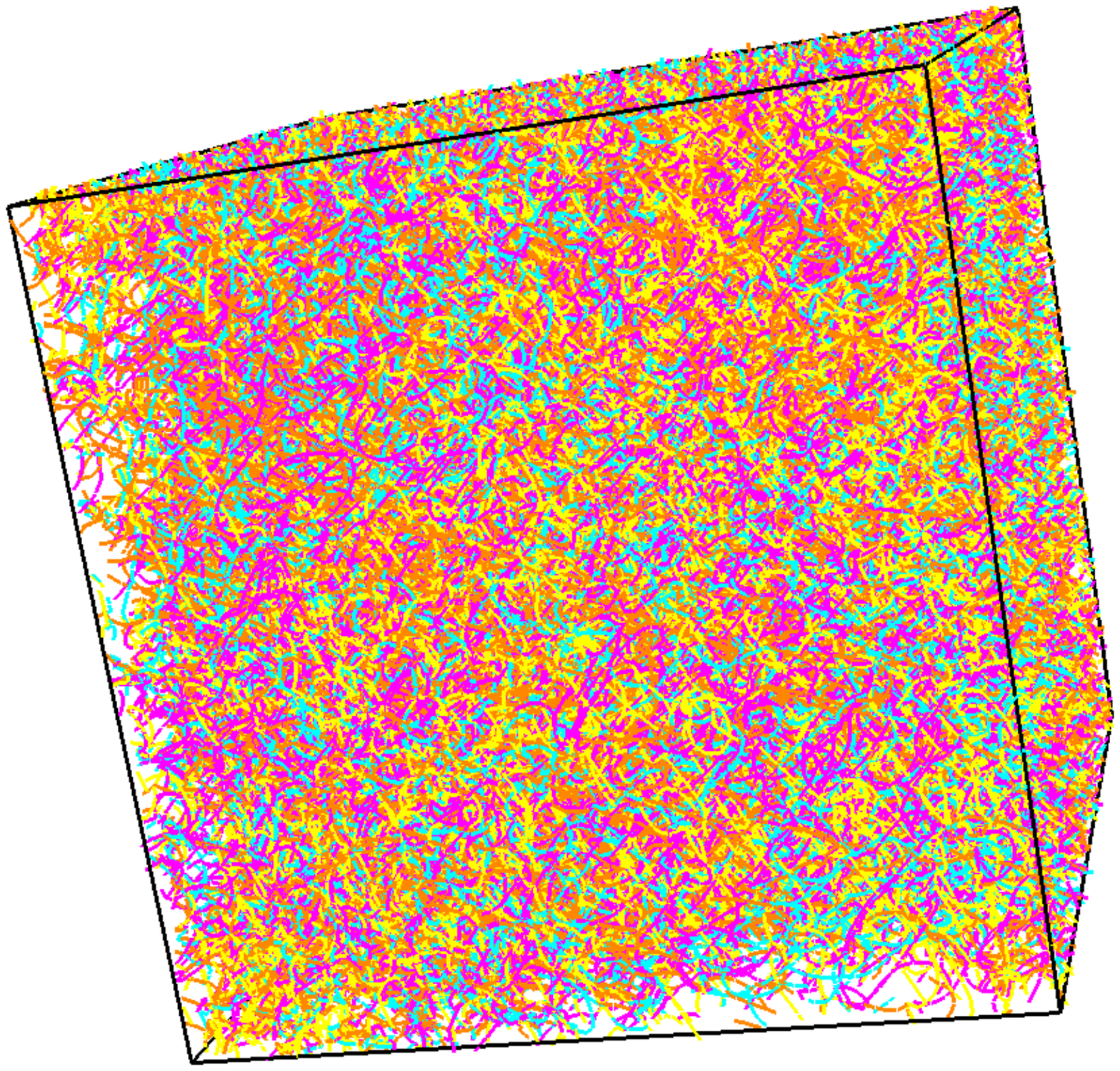}}
\hfill
\subfigure[Anisotropic elastic dislocation network]
{\includegraphics[width=6cm, trim= 1cm 6cm 0cm 6cm]{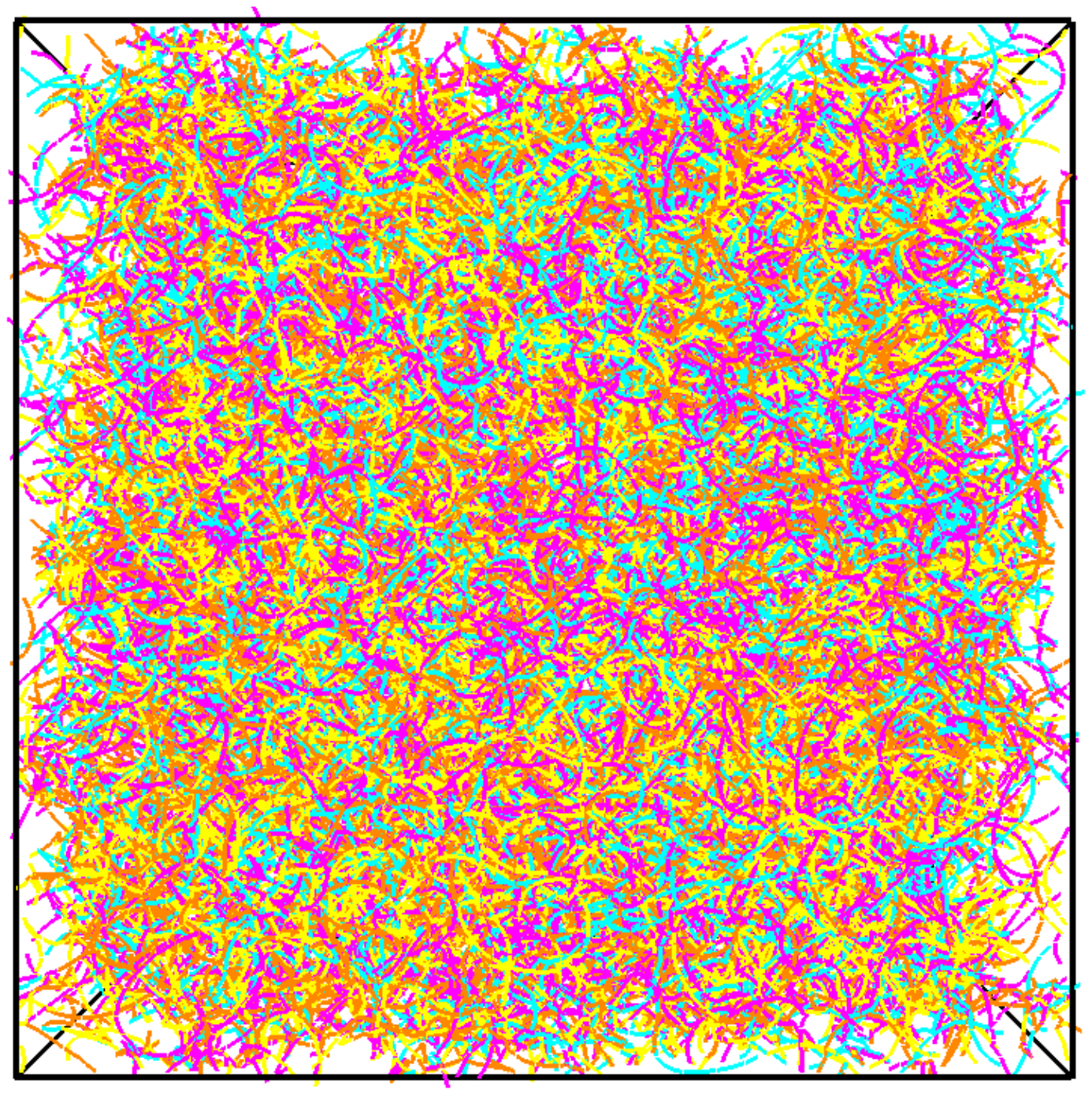}}
\hspace*{\fill}
\caption{Two dislocation networks used in parallel experiments: (a) an isotropic elastic dislocation network with about $460,000$ dislocation segments and (b) an anisotropic elastic dislocation network with about $160,000$ dislocation segments.}
\label{fig:DDdata}
\end{center}
\end{figure}

We present parallel scalability of the FMM by running ParaDiS with our FMM algorithms on distributed-memory machines. We experimented on practically large-scale dislocation networks, where the sequential force calculation took more than $10,000$ seconds, and we show the speedup factors on up to $4,096$ cores ($256$ compute nodes). The framework of distributed-memory parallel FMM was introduced in~\cite{arsenlis2007enabling} for DD simulations. \autoref{fig:DDdata} shows the two dislocation networks used in our simulations.

We demonstrate parallel scalability of the FMM using the Lagrange FMM with an interpolation order of $4$ for force calculation. For the isotropic elastic dislocation network, the calculation accuracy is about $10^{-2}$. For the anisotropic elastic dislocation network, $q_{\max}=6$ was used, which corresponds to an accuracy of about $10^{-2}$ for $A=3.16$. We used $32\times 32 \times 32$ FMM cells in the experiments to achieve the optimal running time.

\begin{figure}[h]
\begin{center}
\includegraphics[width=6.7cm]{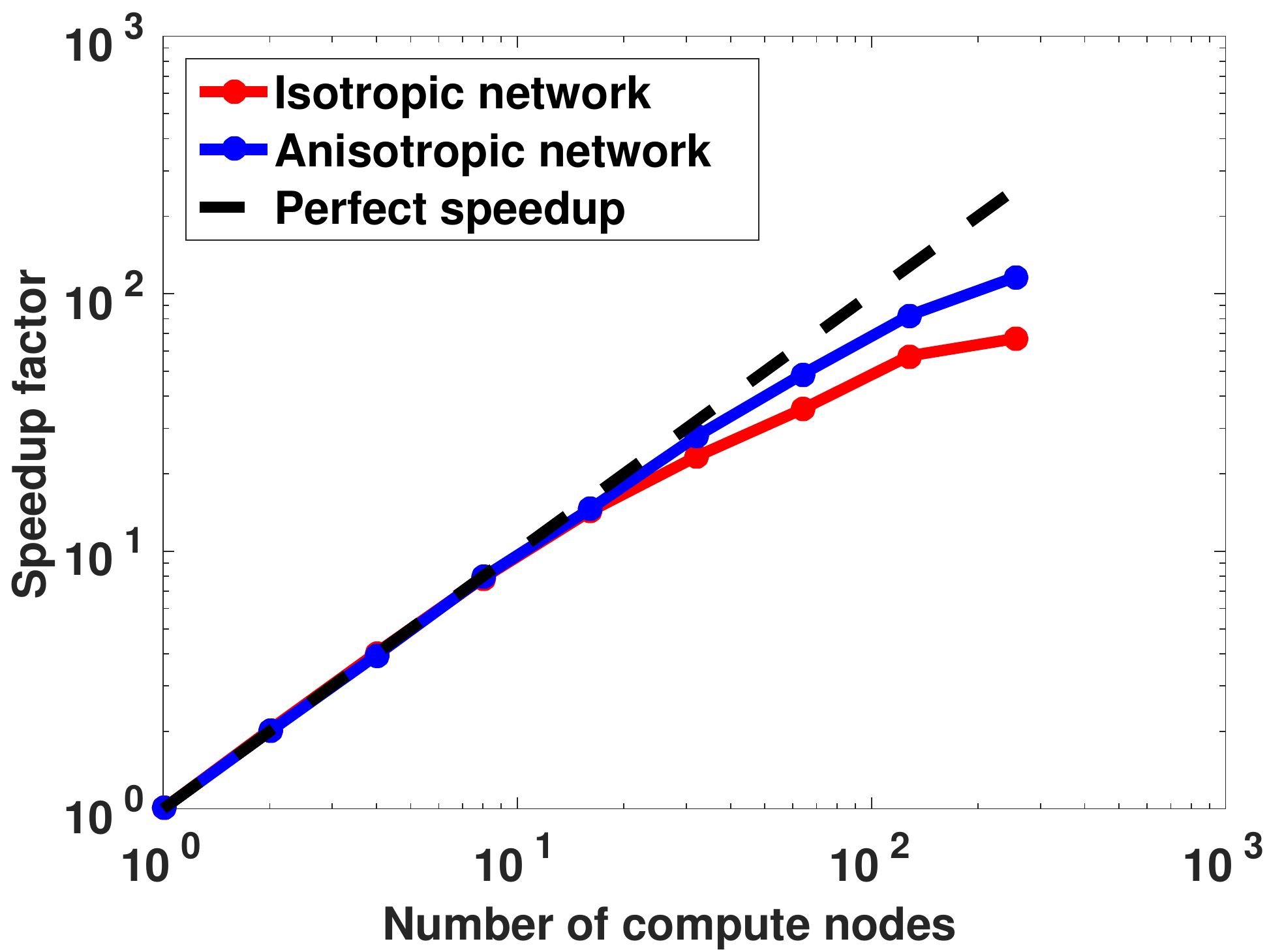}
\caption{Speedup factors of force calculation with the Lagrange FMM on up to $4,096$ cores ($256$ compute nodes).}
\label{fig:StrongScaling}
\end{center}
\end{figure}

\autoref{fig:StrongScaling} shows the speedup on up to $4,096$ cores. As the figure shows, the speedup of the force calculation is nearly perfect on $256$ cores ($16$ compute nodes). After that, the increasing rate slows down due to the decrease of computation and the increase of communication among compute nodes. On $4,096$ cores, the speedups are about $70$ and $100$, respectively, for the force calculation in an isotropic and anisotropic elastic dislocation network (the baseline is the running time on one compute node).

\section{Conclusions \label{sec:conclusion}}

This paper has presented a study of four different FMMs applied to DD simulations: the Taylor FMM, the Spherical FMM, the Chebyshev FMM, and the Lagrange FMM. In isotropic elastic media, we benchmarked and compared the four methods with respect to their convergence and running time. The Chebyshev FMM and the Lagrange FMM exhibit faster convergence than the Taylor FMM and the Spherical FMM. In terms of the computational complexity, the Lagrange FMM and the Spherical FMM both require $O(p^3)$ work, where $p$ is the FMM order. The Chebyshev FMM requires about $O(p^5)$ work, and the Taylor FMM requires $O(p^6)$ work.

In anisotropic elastic media, we have also presented benchmarks and comparisons of the Chebyshev FMM and the Lagrange FMM. The other two methods, the Taylor FMM and the Spherical FMM, are difficult to derive because the Green's function does not have an analytic closed form. In contrast, kernel-independent methods, such as the Chebyshev FMM and the Lagrange FMM, can be applied as long as kernel evaluations are available.

The Chebyshev FMM and the Lagrange FMM can be used for general kernel functions that are not highly oscillatory. The Chebyshev FMM has a relatively large memory footprint, whereas the Lagrange FMM was designed to have a reduced memory cost and fast M2L translations. Both methods have great potential to be optimized. For example, if the symmetries of the three-dimensional space and the kernel function are exploited, the number of unique M2L translation operators can be reduced to 16 \cite{messner2012optimized}, which leads to a reduce of the memory footprint by $316/16 \approx 20$x .

The results we presented in this paper are obtained from running the dislocation dynamics simulation code ParaDiS \cite{arsenlis2007enabling}. We used double-precision for all of our experiments, and a speedup of 2x in CPU time and a reduction of 2x in memory cost can be achieved if single-precision is used.

\section*{Acknowledgments}
C. Chen and E. Darve were partially supported by the Department of Energy National Nuclear Security Administration under Award Number DE-NA0002373-1; Chen would also like to thank M. McDevitt for useful comments on early drafts of this manuscript. This work was performed under the auspices of the U.S. Department of Energy by Lawrence Livermore National Laboratory under Contract DE-AC52-07NA27344. Release number LLNL-JRNL-742868.

\section*{Appendix A: proof of Theorem 1}

\begin{theorem-non} 
In the $p$-th order Lagrange FMM, every M2L operator $\bm{T}_{A \rightarrow B}$ is a three-level block-Toeplitz matrix with at most $(2p-1)^3$ unique entries, and applying $\bm{T}_{A \rightarrow B}$ to a vector takes $O(p^3\log(p))$ work. 
\end{theorem-non}
{Definition: A one-level block-Toeplitz matrix is simply a Toeplitz matrix having constant entries along its diagonals, and a $k$-level block-Toeplitz matrix is a matrix that has constant blocks along the diagonals and in which every block is a $(k-1)$-level block-Toeplitz matrix.}

\begin{proof}
In the Lagrange FMM, the interpolation nodes $\bxl$ and $\bym$ lie on equally spaced grids in box $A$ and box $B$, respectively. Therefore, the map from vector ${\bm l} - {\bm m} = (l_1-m_1, l_2-m_2, l_3-m_3)$ to vector $\bxl - \bym = (\bar{x}_{l_1} - \bar{y}_{m_1}, \bar{x}_{l_2} - \bar{y}_{m_2}, \bar{x}_{l_3} - \bar{y}_{m_3})$ is one-to-one for $l_i, m_i = 1, 2, \ldots, p$.

Since the vector ${\bm l} - {\bm m}$ takes only $(2p-1)^3$ unique values, the vector $\bxl - \bym$ also has $(2p-1)^3$ unique values. As a result, it is easy to see from \autoref{eq:m2l} that the M2L operator $\bm{T}_{A \rightarrow B}$ has at most $(2p-1)^3$ unique entries. 

We can show that the M2L translation operator $\bm{T}_{A \rightarrow B}$ in the one/two/three-dimensional Euclidean space is a one/two/three-level block-Toeplitz matrix with an inductive argument. Consider the base case, i.e., $\bm{T}_{A \rightarrow B}$ in one dimension, where $\bxl = \bar{x}_{l_1}$ and $\bym = \bar{y}_{m_1}$. Since ${\bm l} - {\bm m} = (l_1-m_1) = ((l_1+1) - (m_1+1))$, we know that two consecutive entries along a diagonal have the same value, i.e., ${\bm K}(\bar{x}_{l_1} - \bar{y}_{m_1}) = {\bm K}(\bar{x}_{l_1 + 1} - \bar{y}_{m_1 + 1})$ for $l_1, m_1 = 1, 2, \ldots, p-1$. Therefore, $\bm{T}_{A \rightarrow B}$ is a Toeplitz matrix in one dimension. 

In two dimensions, $\bxl = (\bar{x}_{l_1}, \bar{x}_{l_2})$ and $\bym = (\bar{y}_{m_1}, \bar{y}_{m_2})$. $\bm{T}_{A \rightarrow B}$ has size $p^2$ by $p^2$ and can be viewed as a $p$ by $p$ block matrix, where every block is $p$ by $p$. Since every block corresponds to a one-dimensional M2L translation operator, it is a Toeplitz matrix. In addition, two consecutive blocks along a diagonal have the same value because ${\bm l} - {\bm m} = (l_1-m_1, l_2-m_2) = (l_1 - m_1, (l_2+1) - (m_2+1))$. Therefore,  $\bm{T}_{A \rightarrow B}$ is a two-level block Toeplitz matrix in two dimensions.

Following the same logic as in two dimensions, $\bm{T}_{A \rightarrow B}$ in three dimensions can be viewed as a $p$ by $p$ block matrix where every block is $p^2$ by $p^2$. Since every $p^2$ by $p^2$ block corresponds to a two-dimensional M2L translation operator, it is a two-level block Toeplitz matrix. In addition, two consecutive blocks along a diagonal have the same value because ${\bm l} - {\bm m} = (l_1-m_1, l_2-m_2, l_3-m_3) = (l_1 - m_1, l_2 - m_2, (l_3+1) - (m_3+1))$. Therefore,  $\bm{T}_{A \rightarrow B}$ is a three-level block Toeplitz matrix in three dimensions.

The application of $\bm{T}_{A \rightarrow B}$ to a vector, i.e., applying a three-level block-Toeplitz matrix to a vector, can be accelerated with the FFT, which requires only $O(p^3 \log(p))$ work using the algorithm in \cite{lee1986fast}.
\end{proof}

\section*{Appendix B: derivation of the optimal FMM running time}

We derive the optimal FMM running time, which is a function of the number of dislocation segments, with respect to the number of FMM cells. Let $N_s$ be the number of dislocation {\textbf s}egments and $N_c$ be the number of FMM {\textbf c}ells. Assume the dislocation segments uniformly spread over the simulation domain, and thus every FMM cell has $N_s / N_c$ segments on average. The cost of evaluating near-field and far-field interactions are, $27 \, C_{n} \cdot (N_s / N_c)^2 \cdot N_c$ (27 neighboring cells) and $189 \, C_{f} \cdot N_c$ (189 cells in the far-field), where $C_{n}$ and $C_{f}$ are constants corresponding to one {\textbf n}ear-field translation cost per pair of dislocation segments and one {\textbf f}ar-field translation (M2L translation) cost per FMM cell, respectively. Therefore, the total calculation time is
\[
{\rm cost}(N_s, N_c) = 189 \, C_{f} N_c + 27 \,  C_{n} N_s^2 / N_c
\]
where the two constants -- $C_{n}$ and $C_{f}$ can be measured accurately.
For practical DD simulations, assuming that the number of dislocation segments can be estimated, we need to choose $N_c$, the number of FMM cells, to minimize the total cost. Mathematically, the minimum is 
\[
{\rm cost}(N_s) = \min_{N_c} {\rm cost}(N_s, N_c) \approx 143 \sqrt{C_f \cdot C_n} \,\, N_s
\]
which is obtained at $N_c \approx \sqrt{C_n / 7 \, C_f} \,\, N_s$ ($\sqrt{7 \, C_f / C_n }$ segments per cell).

Combing with the M2L translation cost in \autoref{table:fmm_cost}, we have the optimal running time of the four methods as follows.

\begin{table}[h]
\caption{Optimal FMM running time with respect to the number of FMM cells, i.e., a function of the number of dislocation segments. Note that the near-field translation cost $C_n$ is the same for the four methods.}
\centering
\begin{tabular}{c c c c c} \toprule
                               & Taylor            & Spherical             & Chebyshev               & Lagrange   \\ \toprule
M2L translation cost  & $O(p^6)$          & $O(p^3)$              & $O(\alpha^2_p \, p^6)$   & $O(p^3 \log(p))$ \\

cost($N_s$) in units of $143 \sqrt{C_n}$   & $O(p^3) N_s$  &  $O(p^{3/2}) N_s$  & $O(p^{3}) N_s$   &  $O(\sqrt{p^3 \log(p)}) N_s$ \\ \bottomrule
\end{tabular}
\end{table}

\bibliographystyle{model1-num-names}
\bibliography{fmm_paper.bib}

\begin{thebibliography}{48}
\expandafter\ifx\csname natexlab\endcsname\relax\def\natexlab#1{#1}\fi
\providecommand{\bibinfo}[2]{#2}
\ifx\xfnm\relax \def\xfnm[#1]{\unskip,\space#1}\fi
\bibitem[{Hirth and Lothe(1982)}]{hirth1982theory}
\bibinfo{author}{J.~P. Hirth}, \bibinfo{author}{J.~Lothe},
  \bibinfo{title}{Theory of dislocations}, \bibinfo{publisher}{John
  Wiley$\backslash$ \& Sons}, \bibinfo{year}{1982}.
\bibitem[{Barton et~al.(2011)Barton, Bernier, Becker, Arsenlis, Cavallo,
  Marian, Rhee, Park, Remington, and Olson}]{barton2011multiscale}
\bibinfo{author}{N.~Barton}, \bibinfo{author}{J.~Bernier},
  \bibinfo{author}{R.~Becker}, \bibinfo{author}{A.~Arsenlis},
  \bibinfo{author}{R.~Cavallo}, \bibinfo{author}{J.~Marian},
  \bibinfo{author}{M.~Rhee}, \bibinfo{author}{H.-S. Park},
  \bibinfo{author}{B.~Remington}, \bibinfo{author}{R.~Olson},
\newblock \bibinfo{title}{A multiscale strength model for extreme loading
  conditions},
\newblock \bibinfo{journal}{Journal of applied physics} \bibinfo{volume}{109}
  (\bibinfo{year}{2011}) \bibinfo{pages}{073501}.
\bibitem[{Greer et~al.(2008)Greer, Weinberger, and Cai}]{greer2008comparing}
\bibinfo{author}{J.~R. Greer}, \bibinfo{author}{C.~R. Weinberger},
  \bibinfo{author}{W.~Cai},
\newblock \bibinfo{title}{Comparing the strength of fcc and bcc sub-micrometer
  pillars: Compression experiments and dislocation dynamics simulations},
\newblock \bibinfo{journal}{Materials Science and Engineering: A}
  \bibinfo{volume}{493} (\bibinfo{year}{2008}) \bibinfo{pages}{21--25}.
\bibitem[{Arsenlis et~al.(2012)Arsenlis, Rhee, Hommes, Cook, and
  Marian}]{arsenlis2012dislocation}
\bibinfo{author}{A.~Arsenlis}, \bibinfo{author}{M.~Rhee},
  \bibinfo{author}{G.~Hommes}, \bibinfo{author}{R.~Cook},
  \bibinfo{author}{J.~Marian},
\newblock \bibinfo{title}{A dislocation dynamics study of the transition from
  homogeneous to heterogeneous deformation in irradiated body-centered cubic
  iron},
\newblock \bibinfo{journal}{Acta Materialia} \bibinfo{volume}{60}
  (\bibinfo{year}{2012}) \bibinfo{pages}{3748--3757}.
\bibitem[{Kutka(1998)}]{kutka1998observations}
\bibinfo{author}{R.~V. Kutka}, \bibinfo{title}{Observations on the kinetics of
  relaxation in epitaxial films grown on conventional and compliant substrates:
  A continuum simulation of dislocation glide near an interface},
  \bibinfo{year}{1998}.
\bibitem[{Schwarz(1999)}]{schwarz1999simulation}
\bibinfo{author}{K.~Schwarz},
\newblock \bibinfo{title}{Simulation of dislocations on the mesoscopic scale.
  i. methods and examples},
\newblock \bibinfo{journal}{Journal of Applied Physics} \bibinfo{volume}{85}
  (\bibinfo{year}{1999}) \bibinfo{pages}{108--119}.
\bibitem[{Weygand et~al.(2002)Weygand, Friedman, Van~der Giessen, and
  Needleman}]{weygand2002aspects}
\bibinfo{author}{D.~Weygand}, \bibinfo{author}{L.~Friedman},
  \bibinfo{author}{E.~Van~der Giessen}, \bibinfo{author}{A.~Needleman},
\newblock \bibinfo{title}{Aspects of boundary-value problem solutions with
  three-dimensional dislocation dynamics},
\newblock \bibinfo{journal}{Modelling and Simulation in Materials Science and
  Engineering} \bibinfo{volume}{10} (\bibinfo{year}{2002})
  \bibinfo{pages}{437}.
\bibitem[{Kubin et~al.(1992)Kubin, Canova, Condat, Devincre, Pontikis, and
  Br{\'e}chet}]{kubin1992dislocation}
\bibinfo{author}{L.~P. Kubin}, \bibinfo{author}{G.~Canova},
  \bibinfo{author}{M.~Condat}, \bibinfo{author}{B.~Devincre},
  \bibinfo{author}{V.~Pontikis}, \bibinfo{author}{Y.~Br{\'e}chet},
\newblock \bibinfo{title}{Dislocation microstructures and plastic flow: a 3d
  simulation},
\newblock in: \bibinfo{booktitle}{Solid State Phenomena},
  volume~\bibinfo{volume}{23}, \bibinfo{organization}{Trans Tech Publ}, pp.
  \bibinfo{pages}{455--472}.
\bibitem[{Devincre(1996)}]{devincre1996meso}
\bibinfo{author}{B.~Devincre},
\newblock \bibinfo{title}{Meso-scale simulation of the dislocation dynamics},
\newblock \bibinfo{journal}{NATO ASI Series E Applied Sciences-Advanced Study
  Institute} \bibinfo{volume}{308} (\bibinfo{year}{1996})
  \bibinfo{pages}{309--324}.
\bibitem[{Devincre and Kubin(1997)}]{devincre1997mesoscopic}
\bibinfo{author}{B.~Devincre}, \bibinfo{author}{L.~Kubin},
\newblock \bibinfo{title}{Mesoscopic simulations of dislocations and
  plasticity},
\newblock \bibinfo{journal}{Materials Science and Engineering: A}
  \bibinfo{volume}{234} (\bibinfo{year}{1997}) \bibinfo{pages}{8--14}.
\bibitem[{Gulluoglu et~al.(1989)Gulluoglu, Srolovitz, LeSar, and
  Lomdahl}]{gulluoglu1989dislocation}
\bibinfo{author}{A.~Gulluoglu}, \bibinfo{author}{D.~Srolovitz},
  \bibinfo{author}{R.~LeSar}, \bibinfo{author}{P.~Lomdahl},
\newblock \bibinfo{title}{Dislocation distributions in two dimensions},
\newblock \bibinfo{journal}{Scripta metallurgica} \bibinfo{volume}{23}
  (\bibinfo{year}{1989}) \bibinfo{pages}{1347--1352}.
\bibitem[{Wang and LeSar(1995)}]{wang1995n}
\bibinfo{author}{H.~Wang}, \bibinfo{author}{R.~LeSar},
\newblock \bibinfo{title}{O(n) algorithm for dislocation dynamics},
\newblock \bibinfo{journal}{Philosophical Magazine A} \bibinfo{volume}{71}
  (\bibinfo{year}{1995}) \bibinfo{pages}{149--164}.
\bibitem[{Zbib et~al.(1998)Zbib, Rhee, and Hirth}]{zbib1998plastic}
\bibinfo{author}{H.~M. Zbib}, \bibinfo{author}{M.~Rhee}, \bibinfo{author}{J.~P.
  Hirth},
\newblock \bibinfo{title}{On plastic deformation and the dynamics of 3d
  dislocations},
\newblock \bibinfo{journal}{International Journal of Mechanical Sciences}
  \bibinfo{volume}{40} (\bibinfo{year}{1998}) \bibinfo{pages}{113--127}.
\bibitem[{Wang et~al.(2004)Wang, Ghoniem, and LeSar}]{wang2004multipole}
\bibinfo{author}{Z.~Wang}, \bibinfo{author}{N.~Ghoniem},
  \bibinfo{author}{R.~LeSar},
\newblock \bibinfo{title}{Multipole representation of the elastic field of
  dislocation ensembles},
\newblock \bibinfo{journal}{Physical Review B} \bibinfo{volume}{69}
  (\bibinfo{year}{2004}) \bibinfo{pages}{174102}.
\bibitem[{Wang et~al.(2006)Wang, Ghoniem, Swaminarayan, and
  LeSar}]{wang2006parallel}
\bibinfo{author}{Z.~Wang}, \bibinfo{author}{N.~Ghoniem},
  \bibinfo{author}{S.~Swaminarayan}, \bibinfo{author}{R.~LeSar},
\newblock \bibinfo{title}{A parallel algorithm for 3d dislocation dynamics},
\newblock \bibinfo{journal}{Journal of computational physics}
  \bibinfo{volume}{219} (\bibinfo{year}{2006}) \bibinfo{pages}{608--621}.
\bibitem[{Zhao et~al.(2010)Zhao, Huang, and Xiang}]{zhao2010new}
\bibinfo{author}{D.~Zhao}, \bibinfo{author}{J.~Huang},
  \bibinfo{author}{Y.~Xiang},
\newblock \bibinfo{title}{A new version fast multipole method for evaluating
  the stress field of dislocation ensembles},
\newblock \bibinfo{journal}{Modelling and Simulation in Materials Science and
  Engineering} \bibinfo{volume}{18} (\bibinfo{year}{2010})
  \bibinfo{pages}{045006}.
\bibitem[{LeSar and Rickman(2002)}]{lesar2002multipole}
\bibinfo{author}{R.~LeSar}, \bibinfo{author}{J.~Rickman},
\newblock \bibinfo{title}{Multipole expansion of dislocation interactions:
  application to discrete dislocations},
\newblock \bibinfo{journal}{Physical Review B} \bibinfo{volume}{65}
  (\bibinfo{year}{2002}) \bibinfo{pages}{144110}.
\bibitem[{Arsenlis et~al.(2007)Arsenlis, Cai, Tang, Rhee, Oppelstrup, Hommes,
  Pierce, and Bulatov}]{arsenlis2007enabling}
\bibinfo{author}{A.~Arsenlis}, \bibinfo{author}{W.~Cai},
  \bibinfo{author}{M.~Tang}, \bibinfo{author}{M.~Rhee},
  \bibinfo{author}{T.~Oppelstrup}, \bibinfo{author}{G.~Hommes},
  \bibinfo{author}{T.~G. Pierce}, \bibinfo{author}{V.~V. Bulatov},
\newblock \bibinfo{title}{Enabling strain hardening simulations with
  dislocation dynamics},
\newblock \bibinfo{journal}{Modelling and Simulation in Materials Science and
  Engineering} \bibinfo{volume}{15} (\bibinfo{year}{2007})
  \bibinfo{pages}{553}.
\bibitem[{Yin et~al.(2012)Yin, Barnett, Fitzgerald, and Cai}]{yin2012computing}
\bibinfo{author}{J.~Yin}, \bibinfo{author}{D.~Barnett},
  \bibinfo{author}{S.~Fitzgerald}, \bibinfo{author}{W.~Cai},
\newblock \bibinfo{title}{Computing dislocation stress fields in anisotropic
  elastic media using fast multipole expansions},
\newblock \bibinfo{journal}{Modelling and Simulation in Materials Science and
  Engineering} \bibinfo{volume}{20} (\bibinfo{year}{2012})
  \bibinfo{pages}{045015}.
\bibitem[{Greengard and Rokhlin(1987)}]{greengard1987fast}
\bibinfo{author}{L.~Greengard}, \bibinfo{author}{V.~Rokhlin},
\newblock \bibinfo{title}{A fast algorithm for particle simulations},
\newblock \bibinfo{journal}{Journal of computational physics}
  \bibinfo{volume}{73} (\bibinfo{year}{1987}) \bibinfo{pages}{325--348}.
\bibitem[{Greengard and Rokhlin(1997)}]{greengard1997new}
\bibinfo{author}{L.~Greengard}, \bibinfo{author}{V.~Rokhlin},
\newblock \bibinfo{title}{A new version of the fast multipole method for the
  laplace equation in three dimensions},
\newblock \bibinfo{journal}{Acta numerica} \bibinfo{volume}{6}
  (\bibinfo{year}{1997}) \bibinfo{pages}{229--269}.
\bibitem[{Gimbutas and Rokhlin(2003)}]{gimbutas2003generalized}
\bibinfo{author}{Z.~Gimbutas}, \bibinfo{author}{V.~Rokhlin},
\newblock \bibinfo{title}{A generalized fast multipole method for
  nonoscillatory kernels},
\newblock \bibinfo{journal}{SIAM Journal on Scientific Computing}
  \bibinfo{volume}{24} (\bibinfo{year}{2003}) \bibinfo{pages}{796--817}.
\bibitem[{Ying et~al.(2004)Ying, Biros, and Zorin}]{ying2004kernel}
\bibinfo{author}{L.~Ying}, \bibinfo{author}{G.~Biros},
  \bibinfo{author}{D.~Zorin},
\newblock \bibinfo{title}{A kernel-independent adaptive fast multipole
  algorithm in two and three dimensions},
\newblock \bibinfo{journal}{Journal of Computational Physics}
  \bibinfo{volume}{196} (\bibinfo{year}{2004}) \bibinfo{pages}{591--626}.
\bibitem[{Martinsson and Rokhlin(2007)}]{martinsson2007accelerated}
\bibinfo{author}{P.-G. Martinsson}, \bibinfo{author}{V.~Rokhlin},
\newblock \bibinfo{title}{An accelerated kernel-independent fast multipole
  method in one dimension},
\newblock \bibinfo{journal}{SIAM Journal on Scientific Computing}
  \bibinfo{volume}{29} (\bibinfo{year}{2007}) \bibinfo{pages}{1160--1178}.
\bibitem[{Fong and Darve(2009)}]{fong2009black}
\bibinfo{author}{W.~Fong}, \bibinfo{author}{E.~Darve},
\newblock \bibinfo{title}{The black-box fast multipole method},
\newblock \bibinfo{journal}{Journal of Computational Physics}
  \bibinfo{volume}{228} (\bibinfo{year}{2009}) \bibinfo{pages}{8712--8725}.
\bibitem[{Trefethen(2013)}]{trefethen2013approximation}
\bibinfo{author}{L.~N. Trefethen}, \bibinfo{title}{Approximation theory and
  approximation practice}, \bibinfo{publisher}{Siam}, \bibinfo{year}{2013}.
\bibitem[{Mura and Kinoshita(1971)}]{mura1971green}
\bibinfo{author}{T.~Mura}, \bibinfo{author}{N.~Kinoshita},
\newblock \bibinfo{title}{Green's functions for anisotropic elasticity},
\newblock \bibinfo{journal}{physica status solidi (b)} \bibinfo{volume}{47}
  (\bibinfo{year}{1971}) \bibinfo{pages}{607--618}.
\bibitem[{Cai et~al.(2006)Cai, Arsenlis, Weinberger, and Bulatov}]{cai2006non}
\bibinfo{author}{W.~Cai}, \bibinfo{author}{A.~Arsenlis}, \bibinfo{author}{C.~R.
  Weinberger}, \bibinfo{author}{V.~V. Bulatov},
\newblock \bibinfo{title}{A non-singular continuum theory of dislocations},
\newblock \bibinfo{journal}{Journal of the Mechanics and Physics of Solids}
  \bibinfo{volume}{54} (\bibinfo{year}{2006}) \bibinfo{pages}{561--587}.
\bibitem[{Aubry and Arsenlis(2013)}]{aubry2013use}
\bibinfo{author}{S.~Aubry}, \bibinfo{author}{A.~Arsenlis},
\newblock \bibinfo{title}{Use of spherical harmonics for dislocation dynamics
  in anisotropic elastic media},
\newblock \bibinfo{journal}{Modelling and Simulation in Materials Science and
  Engineering} \bibinfo{volume}{21} (\bibinfo{year}{2013})
  \bibinfo{pages}{065013}.
\bibitem[{Aubry et~al.(2013)Aubry, Fitzgerald, and Arsenlis}]{aubry2013methods}
\bibinfo{author}{S.~Aubry}, \bibinfo{author}{S.~Fitzgerald},
  \bibinfo{author}{A.~Arsenlis},
\newblock \bibinfo{title}{Methods to compute dislocation line tension energy
  and force in anisotropic elasticity},
\newblock \bibinfo{journal}{Modelling and Simulation in Materials Science and
  Engineering} \bibinfo{volume}{22} (\bibinfo{year}{2013})
  \bibinfo{pages}{015001}.
\bibitem[{Yin et~al.(2010)Yin, Barnett, and Cai}]{yin2010efficient}
\bibinfo{author}{J.~Yin}, \bibinfo{author}{D.~M. Barnett},
  \bibinfo{author}{W.~Cai},
\newblock \bibinfo{title}{Efficient computation of forces on dislocation
  segments in anisotropic elasticity},
\newblock \bibinfo{journal}{Modelling and Simulation in Materials Science and
  Engineering} \bibinfo{volume}{18} (\bibinfo{year}{2010})
  \bibinfo{pages}{045013}.
\bibitem[{Rokhlin(1990)}]{rokhlin1990rapid}
\bibinfo{author}{V.~Rokhlin},
\newblock \bibinfo{title}{Rapid solution of integral equations of scattering
  theory in two dimensions},
\newblock \bibinfo{journal}{Journal of Computational Physics}
  \bibinfo{volume}{86} (\bibinfo{year}{1990}) \bibinfo{pages}{414--439}.
\bibitem[{Rokhlin(1992)}]{rokhlin1992diagonal}
\bibinfo{author}{V.~Rokhlin}, \bibinfo{title}{Diagonal forms of translation
  operators for Helmholtz equation in three dimensions},
  \bibinfo{type}{Technical Report}, DTIC Document, \bibinfo{year}{1992}.
\bibitem[{Darve(2000{\natexlab{a}})}]{darve2000fast}
\bibinfo{author}{E.~Darve},
\newblock \bibinfo{title}{The fast multipole method: numerical implementation},
\newblock \bibinfo{journal}{Journal of Computational Physics}
  \bibinfo{volume}{160} (\bibinfo{year}{2000}{\natexlab{a}})
  \bibinfo{pages}{195--240}.
\bibitem[{Darve(2000{\natexlab{b}})}]{darve2000siam}
\bibinfo{author}{E.~Darve},
\newblock \bibinfo{title}{The fast multipole method i: error analysis and
  asymptotic complexity},
\newblock \bibinfo{journal}{SIAM Journal on Numerical Analysis}
  \bibinfo{volume}{38} (\bibinfo{year}{2000}{\natexlab{b}})
  \bibinfo{pages}{98--128}.
\bibitem[{Engquist and Ying(2007)}]{engquist2007fast}
\bibinfo{author}{B.~Engquist}, \bibinfo{author}{L.~Ying},
\newblock \bibinfo{title}{Fast directional multilevel algorithms for
  oscillatory kernels},
\newblock \bibinfo{journal}{SIAM Journal on Scientific Computing}
  \bibinfo{volume}{29} (\bibinfo{year}{2007}) \bibinfo{pages}{1710--1737}.
\bibitem[{Fu and Rodin(2000)}]{fu2000fast}
\bibinfo{author}{Y.~Fu}, \bibinfo{author}{G.~J. Rodin},
\newblock \bibinfo{title}{Fast solution method for three-dimensional stokesian
  many-particle problems},
\newblock \bibinfo{journal}{International Journal for Numerical Methods in
  Biomedical Engineering} \bibinfo{volume}{16} (\bibinfo{year}{2000})
  \bibinfo{pages}{145--149}.
\bibitem[{Dutt et~al.(1996)Dutt, Gu, and Rokhlin}]{dutt1996fast}
\bibinfo{author}{A.~Dutt}, \bibinfo{author}{M.~Gu},
  \bibinfo{author}{V.~Rokhlin},
\newblock \bibinfo{title}{Fast algorithms for polynomial interpolation,
  integration, and differentiation},
\newblock \bibinfo{journal}{SIAM Journal on Numerical Analysis}
  \bibinfo{volume}{33} (\bibinfo{year}{1996}) \bibinfo{pages}{1689--1711}.
\bibitem[{Yarvin and Rokhlin(1998)}]{yarvin1998generalized}
\bibinfo{author}{N.~Yarvin}, \bibinfo{author}{V.~Rokhlin},
\newblock \bibinfo{title}{Generalized gaussian quadratures and singular value
  decompositions of integral operators},
\newblock \bibinfo{journal}{SIAM Journal on Scientific Computing}
  \bibinfo{volume}{20} (\bibinfo{year}{1998}) \bibinfo{pages}{699--718}.
\bibitem[{Yarvin and Rokhlin(1999)}]{yarvin1999improved}
\bibinfo{author}{N.~Yarvin}, \bibinfo{author}{V.~Rokhlin},
\newblock \bibinfo{title}{An improved fast multipole algorithm for potential
  fields on the line},
\newblock \bibinfo{journal}{SIAM Journal on Numerical Analysis}
  \bibinfo{volume}{36} (\bibinfo{year}{1999}) \bibinfo{pages}{629--666}.
\bibitem[{Anderson(1992)}]{anderson1992implementation}
\bibinfo{author}{C.~R. Anderson},
\newblock \bibinfo{title}{An implementation of the fast multipole method
  without multipoles},
\newblock \bibinfo{journal}{SIAM Journal on Scientific and Statistical
  Computing} \bibinfo{volume}{13} (\bibinfo{year}{1992})
  \bibinfo{pages}{923--947}.
\bibitem[{Makino(1999)}]{makino1999yet}
\bibinfo{author}{J.~Makino},
\newblock \bibinfo{title}{Yet another fast multipole method without
  multipoles-pseudoparticle multipole method},
\newblock \bibinfo{journal}{Journal of Computational Physics}
  \bibinfo{volume}{151} (\bibinfo{year}{1999}) \bibinfo{pages}{910--920}.
\bibitem[{Cheng et~al.(2005)Cheng, Gimbutas, Martinsson, and
  Rokhlin}]{cheng2005compression}
\bibinfo{author}{H.~Cheng}, \bibinfo{author}{Z.~Gimbutas},
  \bibinfo{author}{P.-G. Martinsson}, \bibinfo{author}{V.~Rokhlin},
\newblock \bibinfo{title}{On the compression of low rank matrices},
\newblock \bibinfo{journal}{SIAM Journal on Scientific Computing}
  \bibinfo{volume}{26} (\bibinfo{year}{2005}) \bibinfo{pages}{1389--1404}.
\bibitem[{L{\'e}tourneau et~al.(2014)L{\'e}tourneau, Cecka, and
  Darve}]{letourneau2014cauchy}
\bibinfo{author}{P.-D. L{\'e}tourneau}, \bibinfo{author}{C.~Cecka},
  \bibinfo{author}{E.~Darve},
\newblock \bibinfo{title}{Cauchy fast multipole method for general analytic
  kernels},
\newblock \bibinfo{journal}{SIAM Journal on Scientific Computing}
  \bibinfo{volume}{36} (\bibinfo{year}{2014}) \bibinfo{pages}{A396--A426}.
\bibitem[{Dahlquist and Bj{\"o}rck(1974)}]{dahlquist1974numeriska}
\bibinfo{author}{G.~Dahlquist}, \bibinfo{author}{{\AA}.~Bj{\"o}rck},
  \bibinfo{title}{Numeriska metoder}, \bibinfo{publisher}{Prentice Hall},
  \bibinfo{year}{1974}.
\bibitem[{Prenter et~al.(2008)}]{prenter2008splines}
\bibinfo{author}{P.~M. Prenter}, et~al., \bibinfo{title}{Splines and
  variational methods}, \bibinfo{publisher}{Courier Corporation},
  \bibinfo{year}{2008}.
\bibitem[{Messner et~al.(2012)Messner, Bramas, Coulaud, and
  Darve}]{messner2012optimized}
\bibinfo{author}{M.~Messner}, \bibinfo{author}{B.~Bramas},
  \bibinfo{author}{O.~Coulaud}, \bibinfo{author}{E.~Darve},
\newblock \bibinfo{title}{Optimized m2l kernels for the chebyshev interpolation
  based fast multipole method},
\newblock \bibinfo{journal}{arXiv preprint arXiv:1210.7292}
  (\bibinfo{year}{2012}).
\bibitem[{Lee(1986)}]{lee1986fast}
\bibinfo{author}{D.~Lee},
\newblock \bibinfo{title}{Fast multiplication of a recursive block toeplitz
  matrix by a vector and its application},
\newblock \bibinfo{journal}{Journal of Complexity} \bibinfo{volume}{2}
  (\bibinfo{year}{1986}) \bibinfo{pages}{295--305}.

\end{thebibliography}
\end{document}